\documentclass[amsmath,amssymb,aps,showpacs,twocolumn,prd,floatfix]{revtex4-1}

\usepackage{graphicx}
\usepackage{epsfig,amsmath}
\usepackage{amssymb}
\usepackage{rotate}
\usepackage{color}
\usepackage{bm}
\usepackage{hyperref}

\newcommand{\mnras}{MNRAS}
\newcommand{\aap}{A{\&}A}

\newcommand{\jcap}{JCAP}

\newcommand{\be}{\begin{eqnarray}}
\newcommand{\ee}{\end{eqnarray}}
\newcommand{\Pnl}{P_\mathrm{nl}}
\newcommand{\Plinj}{P_\mathrm{lin}^{(j)}}
\newcommand{\Plinjp}{P_\mathrm{lin,+}^{(j)}}
\newcommand{\Plinjm}{P_\mathrm{lin,-}^{(j)}}
\newcommand{\Plinjpm}{P_\mathrm{lin,\pm}^{(j)}}
\newcommand{\Plin}{P_\mathrm{lin}}

\newcommand{\hMpci}{h\,\mathrm{Mpc}^{-1}}

\newcommand{\bfk}{{\boldsymbol{k}}}
\newcommand{\bfq}{{\boldsymbol{q}}}
\newcommand{\bfp}{{\boldsymbol{p}}}

\newcommand{\bfOmg}{{\boldsymbol{\Omega}}}
\newcommand{\dd}{{\mathrm{d}}}
\newcommand{\hiMpc}{{h^{-1}\mathrm{Mpc}}}
\newcommand{\hiGpc}{{h^{-1}\mathrm{Gpc}}}

\newcommand{\Dirac}{{\delta_{\rm D}}}
\newcommand{\dpartial}{{\delta}}

\newcommand{\tree}{{\mathrm{tree}}}
\newcommand{\reg}{{\rm reg}}

\newcommand{\oneloop}{{1\mbox{-}{\rm loop}}}
\newcommand{\twoloop}{{2\mbox{-}{\rm loop}}}

\begin{document}

\title{Moving around the cosmological parameter space: a nonlinear power spectrum reconstruction based on high-resolution cosmic responses}

\author{Takahiro Nishimichi}
\affiliation{Kavli Institute for the Physics and Mathematics of the Universe (WPI), The University of Tokyo Institutes for Advanced Study, The University of Tokyo, 5-1-5 Kashiwanoha, Kashiwa, Chiba 277-8583, Japan}

\author{Francis Bernardeau}
\affiliation{UPMC - CNRS, UMR7095, Institut d'Astrophysique de Paris, F-75014, Paris, France}
\affiliation{CEA - CNRS, URA 2306, Institut de Physique Th\'eorique, F-91191 Gif-sur-Yvette, France}

\author{Atsushi Taruya}
\affiliation{Center for Gravitational Physics, Yukawa Institute for Theoretical Physics, Kyoto University, Kyoto 606-8502, Japan}
\affiliation{Kavli Institute for the Physics and Mathematics of the Universe (WPI), The University of Tokyo Institutes for Advanced Study, The University of Tokyo, 5-1-5 Kashiwanoha, Kashiwa, Chiba 277-8583, Japan}

\begin{abstract}
We present numerical measurements of the power spectrum response function 
of the gravitational growth of cosmic structures, defined as the functional 
derivative of the nonlinear spectrum with respect to the linear counterpart, 
based on $1,400$ cosmological simulations. We develop a simple analytical 
model based on a regularization of the standard perturbative calculation.
Using the model prediction, we show that this function gives a 
natural way to interpolate the nonlinear power spectrum over cosmological 
parameter space from single or multi-step interpolations. We demonstrate that once an 
accurate numerical spectrum template is available for one (or a small number of) cosmological 
model(s), it doubles  the range in $k$ for which percent level accuracy can be obtained 
even for large change in the cosmological parameters.
The python package \texttt{RESPRESSO} we developed to make those predictions is publicly available.
\end{abstract}

\maketitle

\section{Introduction}
\label{sec:introduction}

In light of ultimate future observational projects that will map 
the cosmic web in detail over an unprecedentedly large cosmic volume, 
such as Euclid~\citep{EUCLID} or LSST~\citep{LSST},
theoretical model templates must be provided with 
extreme accuracy that meets the statistical error 
level of the observed data. 
Substantial theoretical efforts have been made both 
with analytical and numerical approaches in recent years. 

One important aspect of such theoretical studies is the ability to predict 
the dependence of the target statistics on the cosmological parameters, not
just a careful calibration of the model prediction for one specific parameter set.
Since numerical simulations are generally more computationally expensive, 
one often develops an analytical model for the statistical quantity of 
interest based on the perturbative expansion, and then use numerical simulations 
only supplementary at a small number of parameter sets to verify the accuracy 
of the analytical model.
A parameter inference is then performed by confronting the model with observational 
data after having the accuracy under control by tests with simulations.

Alternatively, one might fully rely on simulations and interpolate over 
the cosmological parameter space, either by developing a fitting formula
or by employing a non-parametric machine-learning approaches. 
It has been shown that halofit formula \cite{2003MNRAS.341.1311S} provides a $\sim5\%$ level prediction
of the nonlinear power spectrum of the matter density field after a careful calibration \cite{2012ApJ...761..152T},
and Gaussian process can make a smooth and accurate interpolation 
when combined with efficient sampling schemes 
(e.g. \cite{Coyote1,Coyote2,Coyote3}). However, the latter approach
was supplemented with analytical models on large scales, 
on which the statistical error-control is rather demanding.
Furthermore, simulation-based emulation gets more and more 
expensive when one considers a higher dimensional parameter 
space~\footnote{For instance, the accuracy level of such approach 
is degraded to $\sim4\%$ in an eight dimensional parameter space 
with $36$ sampling points \cite{MiraTitan2}. 
This is compared to $\sim1\%$ accuracy up to $k\sim 1\hMpci$ with 
$37$ sampling points in five-dimensional subspace \cite{Coyote3}.}.
Thus analytical approaches are still important and complementary to 
numerical techniques in this context.

Very recently, the fundamental limitations of the perturbation theory (PT) treatment have been highlighted. In particular, the Eulerian PT schemes developed so far are based on the the continuity, Euler and Poisson equations in an expanding Universe in the Newtonian limit, which are derived from the cosmological Vlasov-Poisson equation under the assumptions of irrotational and single-stream flow. With this treatment, the standard PT solves the time evolution of the cosmic density and velocity fields order by order (see \cite{Bernardeau02} for a review). The fundamental assumptions are, however, eventually violated at late times on small scales where the cosmic matter flow experiences shell crossings. One solid way to account for the small-scale dynamics is to go back to a fundamental description, i.e., the Vlasov-Poisson system, and to develop a proper treatment beyond the shell crossings \cite{Taruya17}. Alternatively, effective field theory approaches are developed to supplement the standard PT calculation with non-perturbative corrections to match to simulation measurements \cite{Baumann12,Carrasco12,Hertzberg14,Baldauf15}.

In this context, the 
response function has been introduced recently in 
\cite{Bernardeau:2014lr}. 
It describes how the large-scale structure of the universe, seen as a system obeying 
an intricate nonlinear evolution, is responding to a small change in the initial 
conditions. The notion of response function is transverse in physics and can be encountered in a variety of systems. Here we focus on its use to 
the density power spectrum which measures the amplitude of the density fluctuations as a function of scale (or more precisely of wavelength in a Fourier decomposition). The latter is indeed identified as the key ingredient that can be used to describe to a large extent the statistical properties of the cosmic field. The large-scale structure of the universe can then be seen as a system that transforms a linear density spectrum - the amplitude of the density fluctuations when they are small - into a nonlinear density spectrum. 
The theory we are then developing consists in measuring and computing how the nonlinear power spectrum in the final state is responding to a small change in the initial density spectrum. 

It was discussed in \cite{Bernardeau:2014lr} that the apparent breakdown of the perturbative 
predictions, such as the three-loop order predictions of the standard PT shown in \cite{Blas:2013qy}, 
is closely related to the too strong response of large-scale modes to small-scale modes, where 
the latter is in the strongly nonlinear regime. This was further confirmed by the numerical experiments in \cite{Nishimichi16} based on an ensemble of simulations with slightly different linear power spectra~\footnote{See also \cite{Little91,Neyrinck13} for earlier numerical studies on the mode transfer in the context of large-scale structure cosmology.}. 
Such mode transfer is observed to be strongly suppressed in fully nonlinear 
numerical simulations. We need a way to regularize such mode transfer to have a well-behaved theory.

The first goal of this study is to perform a larger set of cosmological simulations to study
fine structures of the response function. We discuss detailed mode-coupling structure
based on the simulation results and give it a physical interpretation based on 
analytical calculations. We propose a phenomenological model built from analytical results 
that smoothly interpolates the response
function in different regimes, including the suppressed mode transfer from small to 
large scales and that matches to the simulation measurement.

Besides giving a direct insight to the mode-coupling structure, 
the interest of such a function in more practical situations is at least two-fold: it allows to estimate the nonlinear power spectrum for a model close enough to another model for which the nonlinear spectrum is known and it allows to estimate the covariance matrix at one-loop order as shown in \cite{Mohammed17}. The final aim of this study is to put forward the former possibility and develop a code to realize this idea. Assisted by a well-calibrated simulation template for a fiducial cosmological model, it paves the way to give a quick and accurate prediction of the cosmological-parameter dependence to be reasonably implemented in the standard Markov-chain Monte Carlo technique to constrain the model parameters. The code we developed is available as a python package, which we call \texttt{RESPRESSO}, and can be found at 
\url{http://www-utap.phys.s.u-tokyo.ac.jp/~nishimichi/public_codes/respresso/index.html}.

The paper is organized as follows. We first describe our numerical 
experiments and discuss the features seen in the measured response function in Section~\ref{sec:numerical}. We then give physical 
interpretations to the numerical results by confronting with analytical
models in different regimes in Section~\ref{sec:interp}. A proposed model, that 
gives a good match to the simulation data over different scales, 
is presented there. We then apply the response function to reconstruct
the nonlinear power spectrum for different cosmological models starting
from a well-calibrated simulation template for a fiducial cosmological model
in Section~\ref{sec:recon}. We finally conclude in 
Section~\ref{sec:conclusions}. 
We show a more detailed derivation of the analytical response function in Appendix~\ref{appendix:derivation_PT}.

\section{Response function from $N$-body simulations}
\label{sec:numerical}

As we mentioned in Sec.~\ref{sec:introduction}, the response function introduced in \cite{Bernardeau:2014lr} characterizes the nonlinear mode-coupling between the Fourier modes through the nonlinear evolution of the large-scale structure. To be more precise, it specifically quantifies the variation of the nonlinear power spectrum at redshift $z$, $\delta P(k;\,z)$,  with respect to a small initial disturbance added in the initial or linearly extrapolated power spectrum, $\delta P_{\rm lin}(q; z)$, through
\be
\delta P(k;z)= \int d\ln q \,K(k,\,q;\,z)\,\delta P_{\rm lin}(q;\,z)
\label{eq:response_def1}
\ee
The function $K$ is the response function. One can alternatively write it as
\be
K(k,\,q;\,z)= q \,\frac{\delta\,P(k;\,z)}{\delta\,P_{\rm lin}(q;\,z)}.
\label{eq:response_def2}
\ee

\begin{table*}[htb]
  \centering
  \caption{\label{tab:cosmos}Cosmological parameters for our simulations. We show the number of particles per dimension $N^{1/3}$, box size $L$ in $\hiMpc$, the matter $\Omega_\mathrm{m} = 1-\Omega_\Lambda$ and baryon $\Omega_\mathrm{b}$ density parameter, normalized Hubble parameter $h$, the amplitude of the primordial scalar perturbation $A_\mathrm{s}/10^9$ at the pivot scale $k_0 = 0.05\mathrm{Mpc}^{-1}$ and its tilt $n_\mathrm{s}$. We also show the number of realizations and if or not we adopt the Angulo-Pontzen method to suppress the cosmic variance.}
  \begin{tabular}{ l | c  c  c  c  c  c  c  c  c  c |} \hline
    Name & $N^{1/3}$ & box size & $\Omega_\mathrm{m}$ & $\Omega_\mathrm{b}$ & $h$ & $A_{\mathrm s}$ & $n_{\mathrm s}$ & realizations & Angulo-Pontzen \\ \hline \hline
	low-res & $512$ & $1024$ & $0.279$ & $0.0461$ & $0.701$ & $2.19$ & $0.960$ & $1400$ & No. \\ \hline
	PL15 & $2048$ & $2048$ & $0.316$ & $0.0492$ & $0.673$ & $2.21$ & $0.965$ & $10$ & Yes.\\
	WM3 & $2048$ & $2048$ & $0.234$ & $0.0410$ & $0.734$ & $2.09$ & $0.961$ & $2$ & Yes.\\
	WM5 & $2048$ & $2048$ & $0.279$ & $0.0461$ & $0.701$ & $2.19$ & $0.960$ & $2$ & Yes.\\
   low-ns & $2048$ & $2048$ & $0.316$ & $0.0492$ & $0.673$ & $2.29$ & $0.915$ & $2$ & Yes.\\
   high-ns & $2048$ & $2048$ & $0.316$ & $0.0492$ & $0.673$ & $2.12$ & $1.015$ & $2$ & Yes.\\
	EXT015 & $2048$ & $2048$ & $0.15$ & $0.0492$ & $0.673$ & $14.74$\footnotemark[1] & $0.965$ & $2$ & Yes.\\
	EXT045 & $2048$ & $2048$ & $0.45$ & $0.0492$ & $0.673$ & $0.71$\footnotemark[1] & $0.965$ & $2$ & Yes.\\ \hline
  \end{tabular}
  \footnotetext[1]{The normalization is chosen such that the combination $\sigma_8\Omega_\mathrm{m}^{0.5}$ equals to $0.4$ to roughly match to the recent lensing observations\cite{Kilbinger13,Kwan17,Hildebrandt17}.}
\end{table*}

In this section, we present updated numerical results of response function based on a larger set of cosmological $N$-body simulations.  Following Ref.~\cite{Nishimichi16},
the measurement is performed on the density fields obtained from $N$-body simulations using the discretized estimator:
\be
\hat{K}_{i,j}\Plinj = \frac{\Pnl^{(i)}[\Plinjp]-\Pnl^{(i)}[\Plinjm]}{\left(\ln{\Plinjp}-\ln{\Plinjm}\right)\Delta q/q},
\label{eq:response_estimator}
\ee
where the subscripts $i$ and $j$ stand for the wavenumber bin in which the band-averaged power is measured.
We add either positive or negative perturbation to the linear power spectrum $\Plin$ at the $j$-th bin to have the spectra, $\Plinjpm$. 
Following the previous study, we adopt $\pm1\%$ of the original amplitude for these perturbed spectra.
We assume that the evolution of cosmic structures is fully determined by the function $\Plin$ extrapolated to the same epoch (i.e., we ignore the history-dependence of the structure growth, but such effect might be important on strongly nonlinear scales; see e.g., \cite{Mead17}).
Then, the resultant nonlinear power spectra in the numerator, $\Pnl^{(i)}$ at the $i$-th wavenumber bin, are given as functionals of the perturbed linear spectra, $\Plinjpm$.

It is clear from Eq.~(\ref{eq:response_estimator}) that we need (at least) two simulations to measure the function $K(k,q)$ 
at \textit{each} $q$ bin. Thus, a large number of simulations are generally required to determine the shape of this function in
a fine binning on $q$. High-resolution simulations with sufficiently large volume, such as the one presented later in Sec.~\ref{subsec:highres}, are idealistic but 
we here limit ourselves to the simulations with $N=512^3$ particles in periodic cubes with the comoving size of $L_\mathrm{box}=1\,\hiGpc$, and the eventual accuracy of our procedure will be
discussed later with a smaller number of high-resolution simulations in terms of the recovered nonlinear power spectra.

We set the bin width to be $\Delta q = 0.005\,\hMpci$ and cover the wave mode range up to $\sim 1.5\,\hMpci$.
We perform $N$-body simulations for all the $100$ bins in the interval of $[0.005\hMpci, 0.505\hMpci)$, but we sample only every $20$ bins beyond $q\sim0.5\hMpci$. This makes the total number of sampled bins to be $110$. 
For each of the $100$ lower wavenumber bins, we perform $5$ independent random realizations with both positively and negatively perturbed spectra, $\Plinjpm$, while we create $20$ realizations for each of the coarsely sampled high 
wavenumber range to reduce the statistical error. We indeed had to use a larger number of realizations for the high-$q$ data to obtain a clear signal of the response function
because of the substantial suppression of response on small scales. We eventually have $1,400$ simulations in total at our disposal, counting all the simulations with positive and negative perturbations in the linear spectrum at different wavenumber bins.

All these simulations assume a flat-$\Lambda$CDM cosmological model with the parameters derived by
five-year observation of WMAP \cite{WMAP5}. We compute the linear matter transfer function
using the linear Boltzmann solver \texttt{CAMB} \cite{CAMB}.
We then give displacements and velocities to the simulation particles located in a regular lattice based 
on the second-order Lagrangian perturbation theory \cite{Scoccimarro98,Crocce06a} implemented 
in a parallel code by \cite{Nishimichi09,Valageas11a}. 
The initial redshift is set to $z_\mathrm{in}=15$. This corresponds to the time when the rms displacement is $\sim25\%$ of the mean inter-particle separation. A higher starting redshift does not improve the convergence of the nonlinear power spectrum for this set of simulations (compared to the higher resolution simulations described later). This is because the initial displacement of the particles become smaller for a higher initial redshift, and the relative impact of the artificial force arising from the grid pattern gets larger, leading to a biased result in the linear growing mode \cite{Garrison16}. 
We employ \texttt{Gadget2} \cite{GADGET2} to simulate the subsequent gravitational evolution of the 
particle distribution. The gravitational softening length is set to be $5\%$ of the mean inter-particle
separation. Finally, we measure the power spectra using a standard FFT-based method with cloud-in-cell (CIC) density estimate on $1024^3$ grid points.

\begin{figure*}[!ht]
   \centering
   \includegraphics[width=15.1cm]{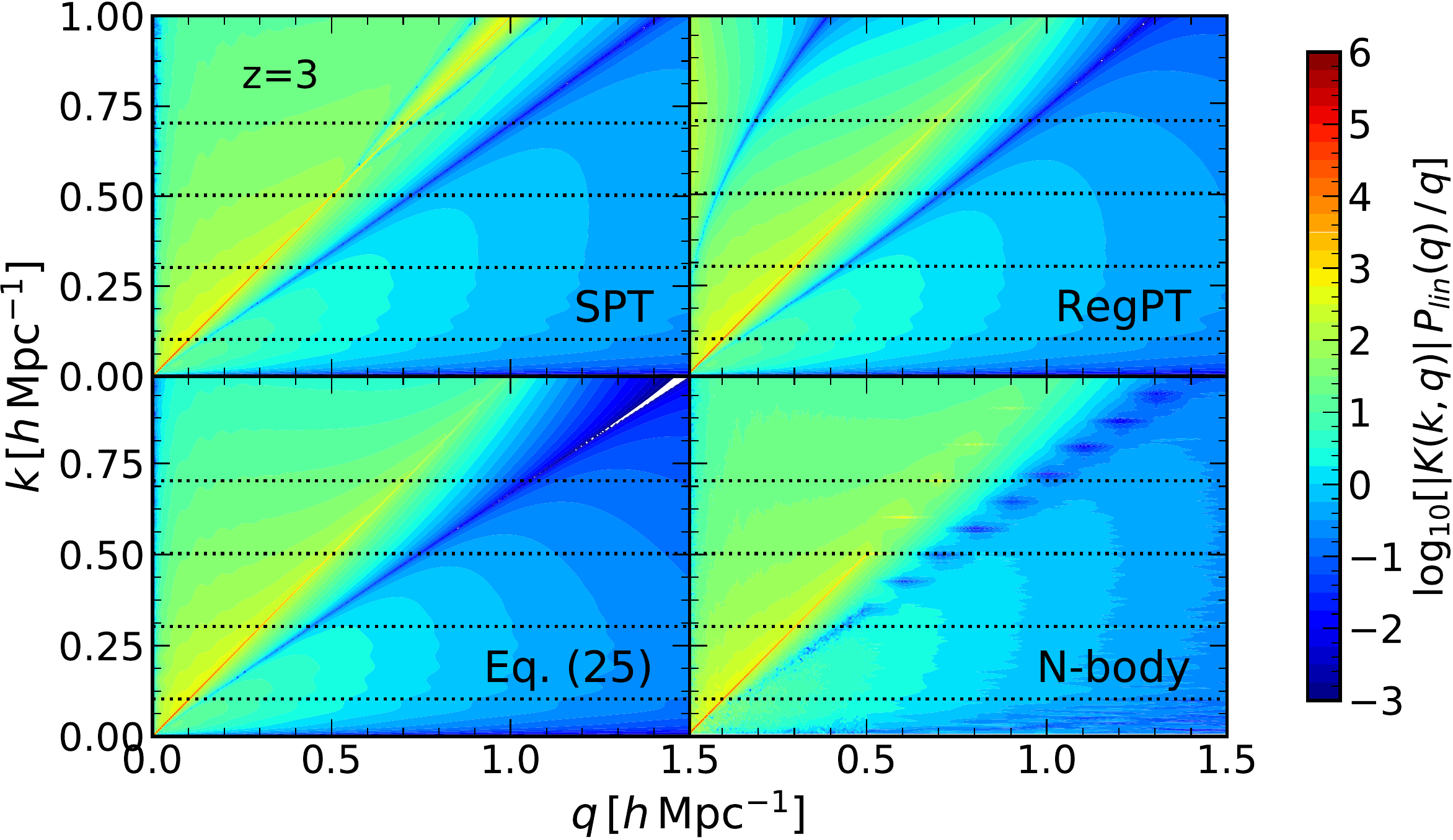}
   \includegraphics[width=15.1cm]{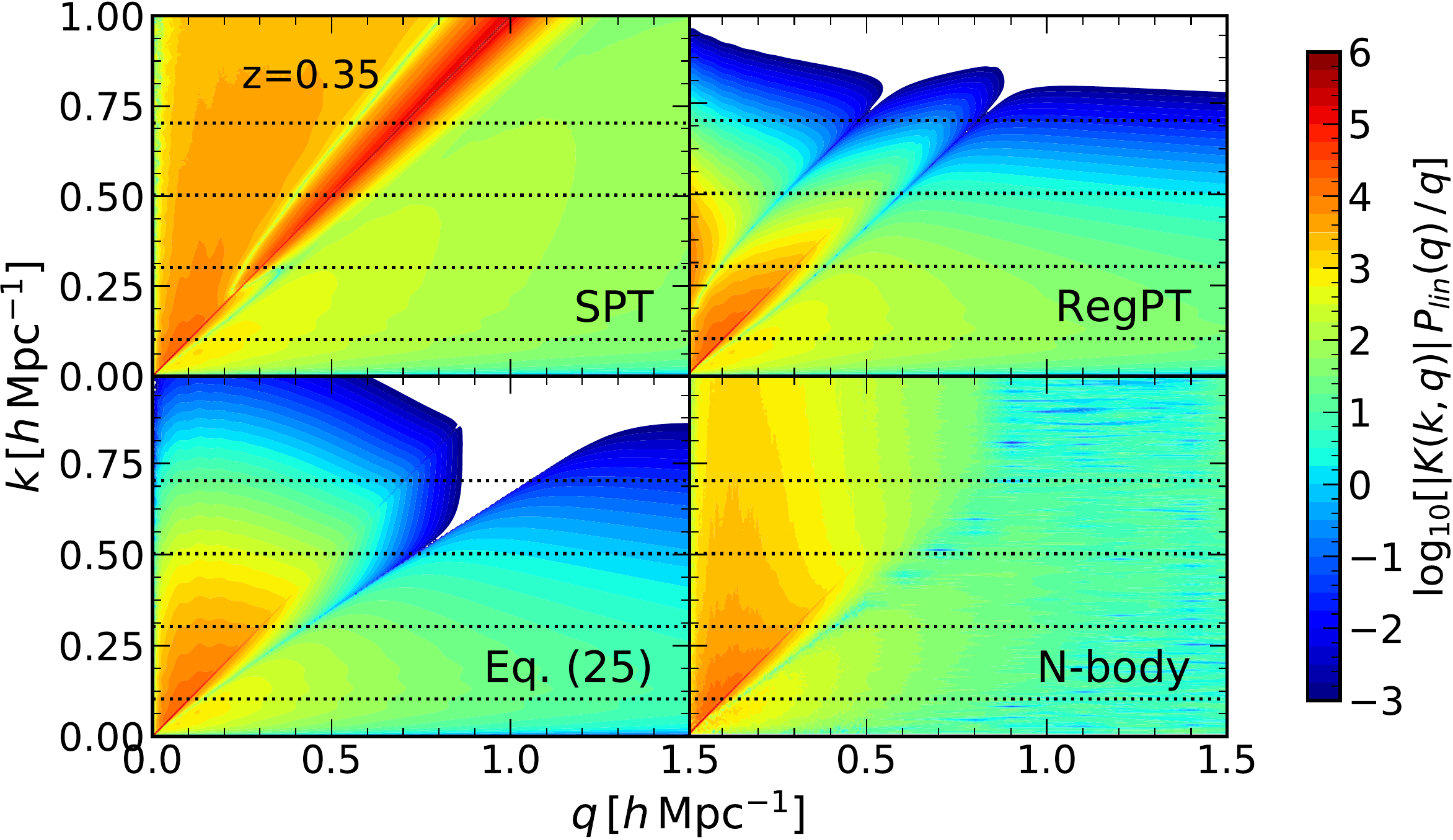}
   \caption{Response function from different approaches at high (upper: $z=3$) and low (lower: $z=0.35$) redshifts. We show standard perturbation theory (top left), RegPT (top right), our new regularized prescription (bottom left), all of which are computed at the two-loop order, and the numerical simulations (bottom right). The horizontal dotted lines mark constant $k$ values along which we present the $q$-dependence of the function in Fig.~\ref{fig:response_qdep}.}
   \label{fig:response_2d}
\end{figure*}

The measured response function is shown in the lower right panel of Fig.~\ref{fig:response_2d} at $z=3$ (top) and $z=0.35$ (bottom). We show the absolute value of the combination $K(k,q)\Plin(q)/q$ with the color-coded amplitude in logarithmic scale. At both redshifts, the $K(k,q)$ function is positive in the upper-left half region corresponding to $q<k$. At $q=k$, it exhibits a peak structure, and then the amplitude rapidly drops towards larger $q$ (for a fixed $k$) to cross zero. The zero crossing points thus appear in a diagonal line slightly tilted to the $q>k$ side. We also show different analytical prescriptions for the function $K(k,q)$ in the other three panels. We leave a detailed explanation for these prescriptions to the next section.

\begin{figure*}[!ht]
   \centering
   \includegraphics[width=7.3cm]{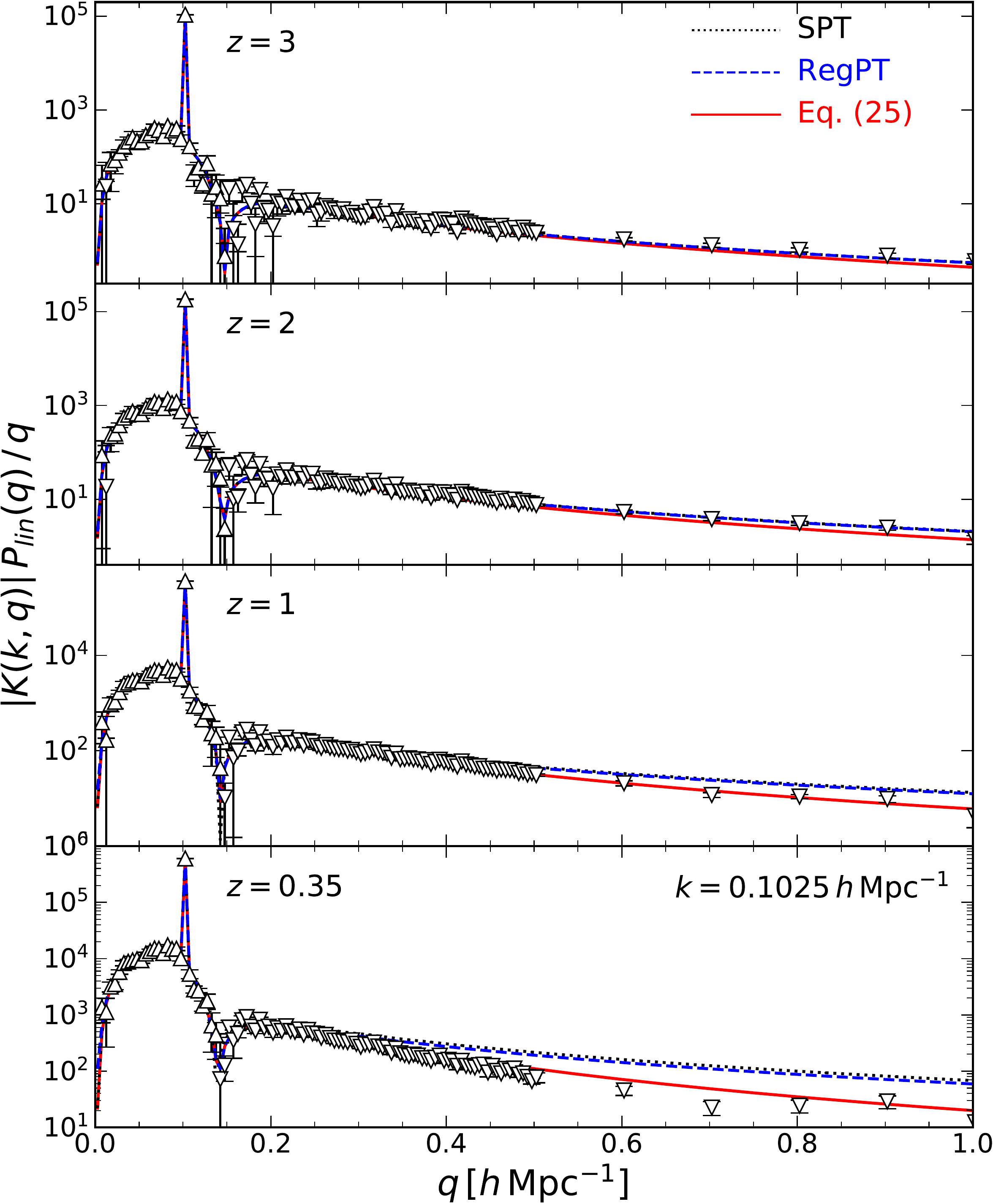}
   \includegraphics[width=7.3cm]{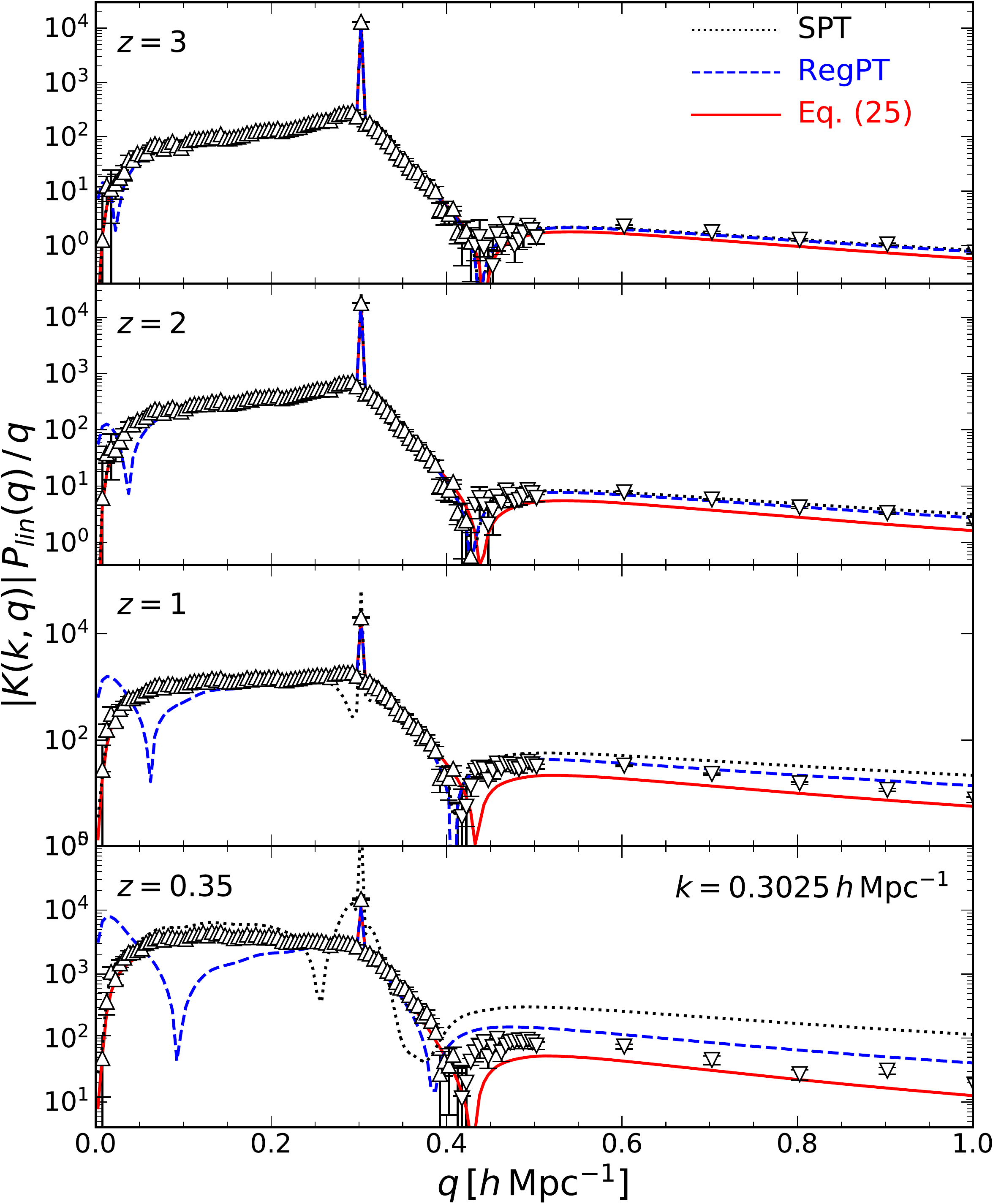}
   \includegraphics[width=7.3cm]{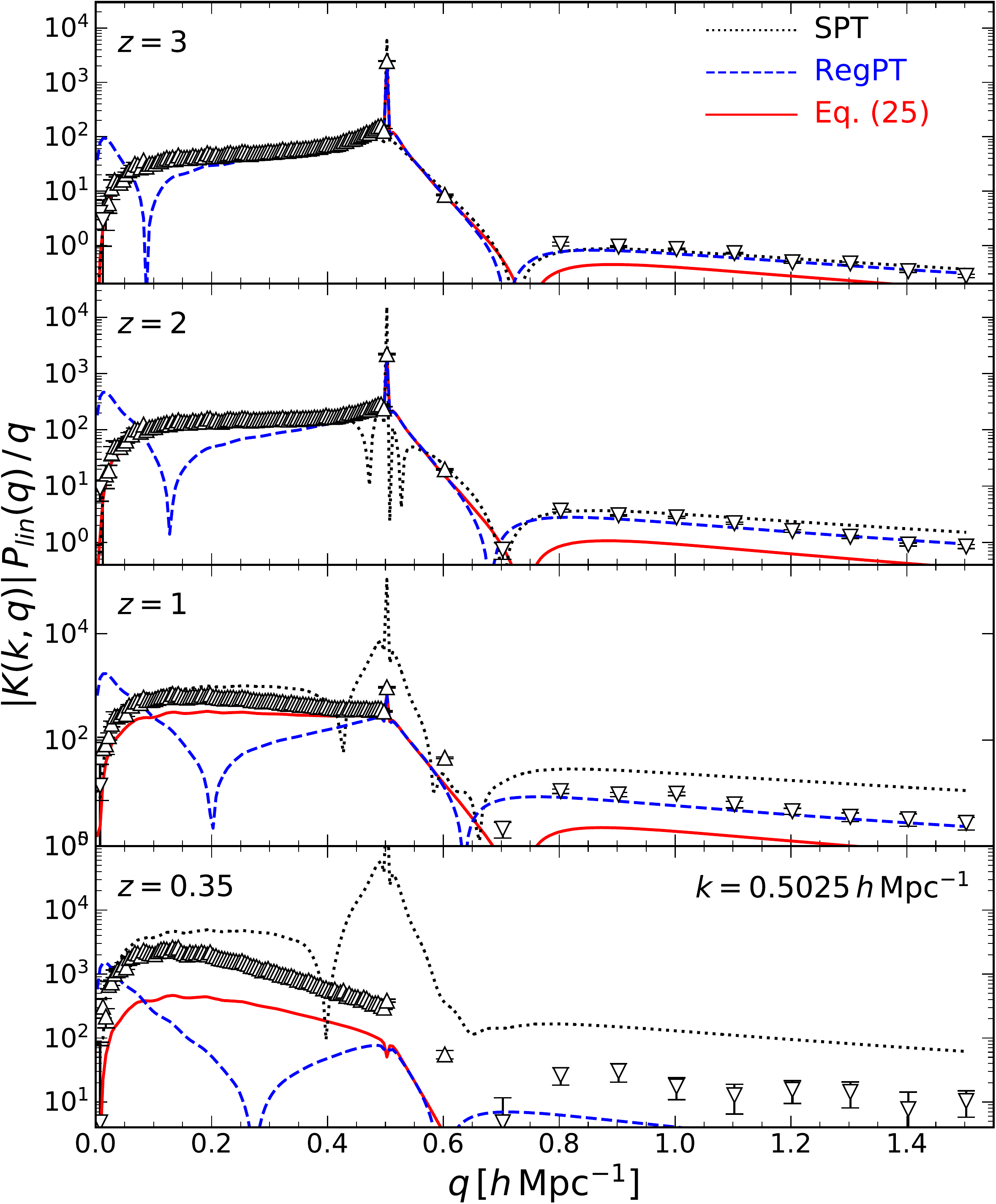}
   \includegraphics[width=7.3cm]{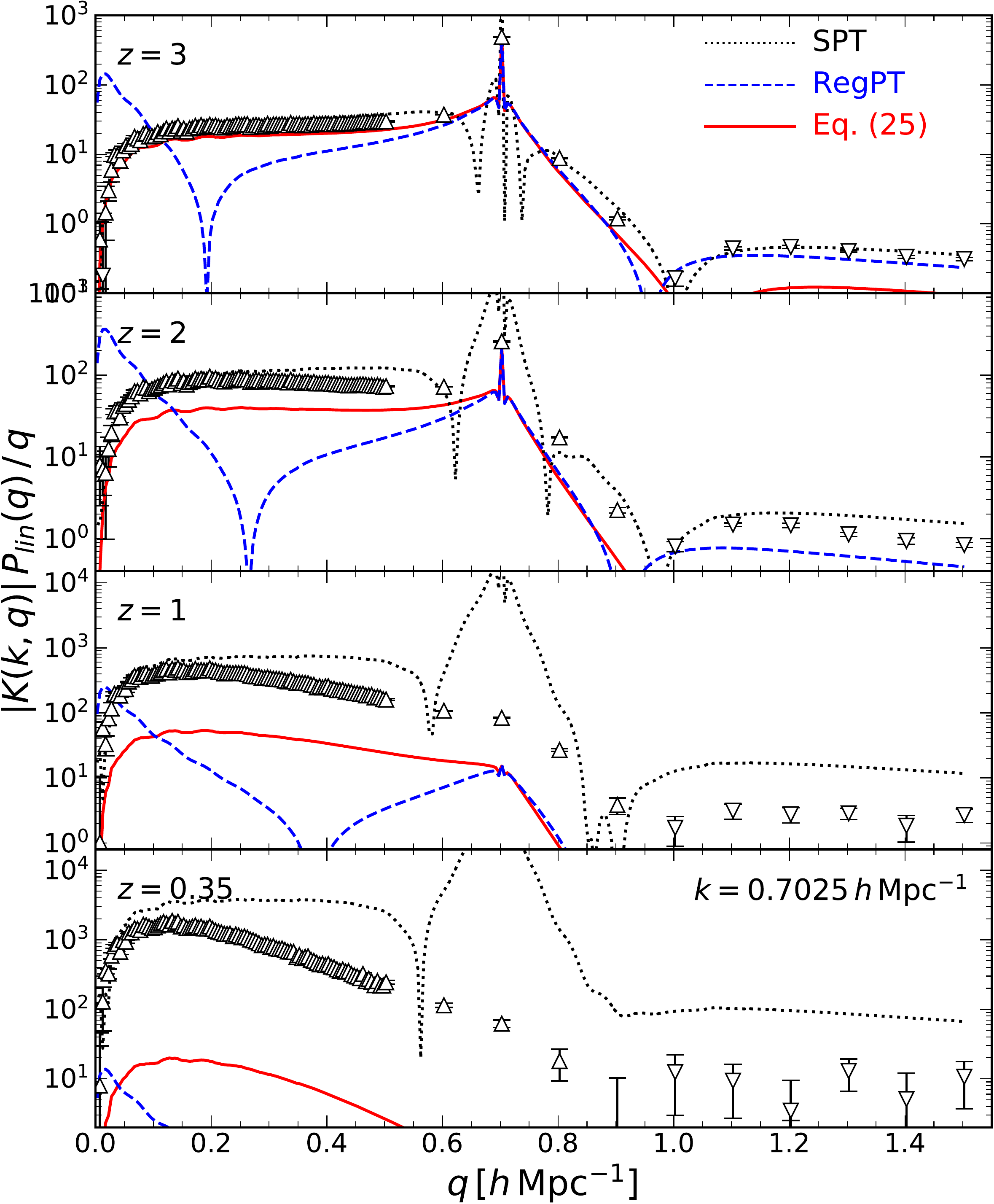}
   \caption{Response function as a function of wavenumber $q$ for various fixed $k$ values and at different redshifts as indicated in the panels. Simulation data are shown by triangles with error bars (upward triangles for positive values, and downward triangles for negative values of $K(k,q)$). Different analytical predictions are also shown: standard perturbation theory (dotted), RegPT (dashed) and a new hybrid model~(\ref{eq:our_model}) proposed in this paper (solid). Data points are sparse on $q>0.5\hMpci$ simply because of our simulation design.}
   \label{fig:response_qdep}
\end{figure*}

The behavior of the function $K(k,q)$ at some fixed values of $k$ is shown in Fig.~\ref{fig:response_qdep}. These $k$ values correspond to the locations of the horizontal dotted lines in Fig.~\ref{fig:response_2d}. Here, a positive (negative) value of $K(k,q)$ is shown by an upward (downward) triangle symbol with an error bar (simulation data). The three analytical predictions are overplotted by different lines. The overall trend of the simulation data is accurately followed by the analytical curves especially on low $k$ at high redshifts including the peak structure at $k=q$ and the location of a change of sign after the peak. At the other limit, any of the three analytical models cannot explain the simulation data at all on high $k$ at low redshifts. In this strongly nonlinear and non-perturbative regime, the function measured from the simulations show rather simple structure without a clear peak.

In what follows, we will focus mostly on the weakly nonlinear regime where perturbative calculations work well ($k\sim0.3\hMpci$, in most cases). We will discuss different features seen on different regime of wavenumber $q$ for $k$ fixed in this regime in more detail. We will investigate how well different analytical calculations explain these features exhibited in the simulation data. We will also discuss briefly where such calculations eventually breakdown.

\section{Response function from perturbation theory}
\label{sec:interp}

In this section, we present analytical calculations of the response function based on perturbation theory (PT). The results are confronted with the response function measured from $N$-body simulations. As we will see below, the predictions made with the standard and resummed PT treatments do not perfectly match the simulation results, but in several different regimes, they quantitatively explain the measured results of response function. We discuss the reasons for these, and then propose a simple PT model that incorporates all the necessary ingredients to quantitatively explain the overall trends without introducing free parameters. In Sec.~\ref{subsec:PT_LSS}, we begin by briefly reviewing the perturbation theory of large-scale structure, focusing on two specific treatments. We then present the analytic expression of response function in Sec.~\ref{subsec:response_PT}. After a detailed comparison of PT predictions with $N$-body simulations in Sec.~\ref{subsec:comparison}, a simple PT model is proposed in Sec.~\ref{subsec:our_model}. We briefly discuss the validity range of the model in Sec.~\ref{subsec:validity}.

\subsection{Perturbation theory of large-scale structure}
\label{subsec:PT_LSS}

\subsubsection{Standard PT (SPT)}

The perturbation theory of large-scale structure provides a systematic way to analytically compute the statistics of cosmic density and velocity fields beyond the linear regime of gravitational evolution. Regarding the linear density field as a small perturbed quantity, a systematic calculation is made under the irrotational single-stream flow approximation, and the resultant predictions are all linked to the initial fields through the nonlinear mode-coupling between different Fourier modes (see \cite{Bernardeau02} for a comprehensive review). For an adiabatic initial condition in the linear growing mode, an analytical expression for the mass density field is obtained, and is summarized in Fourier space as
\begin{align}
 &\delta(\bfk,t)=\sum_{n=1}\,\delta^{(n)}(\bfk,t)\,;\,
\nonumber\\
&\delta^{(n)}(\bfk,t) = \int \frac{d^3\bfq_1\cdots d^3\bfq_n}{(2\pi)^{3(n-1)}}\,\Dirac(\bfk-\bfq_{1\cdots n})\nonumber\\
&\qquad\qquad\quad \times F_{\rm sym}^{(n)}(\bfq_1,\cdots,\bfq_n)\,\delta_{\rm lin}(\bfq_1,t)\,\cdots\,\delta_{\rm lin}(\bfq_n,t)
\label{eq:delta_SPT}
\end{align}
with $\bfq_{1\cdots n}=\bfq_1+\cdots+\bfq_n$. The function, $F_{\rm sym}^{(n)}$, is the $n$-th order PT kernel symmetrized over the arguments, which characterizes the mode-coupling through the non-linear evolution, and only the contributions from the fastest growing mode are considered here. Then, the PT kernels are analytically constructed based on the recursion relation (e.g., \cite{Goroff86,Bernardeau02,Crocce06b}). Note that the time dependence of the higher-order density fields (i.e., $n\geq2$) is wholly encapsulated in the linear density field, $\delta_{\rm lin}$. In what follows, we will omit the time dependence from the argument, and follow the convention to evaluate all the relevant variables at the epoch of interest.

We are especially interested in the late-time nonlinear evolution of the density field, starting with tiny Gaussian fluctuations, as we see in the $N$-body simulations. In the standard PT, imposing the Gaussianity of linear density field, all the statistical quantities such as the power spectrum and bispectrum are constructed with the PT kernels, given the power spectrum of initial density field:  
\begin{equation}
\langle 
\delta_{\rm lin}(\bfk)\delta_{\rm lin}(\bfk')\rangle=(2\pi)^3\Dirac(\bfk+\bfk')P_{\rm lin}(k). 
\end{equation}
The power spectrum of the nonlinear density field can then be computed order-by-order, substituting Eq.~(\ref{eq:delta_SPT}) into its definition: 
\begin{eqnarray}
\langle\delta(\bfk)\delta(\bfk')\rangle=(2\pi)^3\Dirac(\bfk+\bfk')P(k).   
\end{eqnarray}
The resultant expression in the standard PT is summarized as 
\begin{align}
 P^{\rm SPT}(k)= P_{\rm lin}(k) + P^{\rm SPT}_{\rm 1\mbox{-}loop}(k) + P^{\rm SPT}_{\rm 2\mbox{-}loop}(k) +\cdots,  
\end{align}
where the second and third terms in the right-hand side correspond to the so-called one- and two-loop corrections, respectively, and the explicit expressions are presented in Appendix \ref{appendix:derivation_PT} [see Eqs.~(\ref{eq:pk_SPT}) with (\ref{eq:pk_SPT_1loop}) and (\ref{eq:pk_SPT_2loop})]. Below, we will compute the power spectrum at two-loop, and derive the expression for the response function up to the corresponding order.

\subsubsection{RegPT}

As we will demonstrate below and partly shown already in Ref.~\cite{Nishimichi16}, the standard PT prediction fails to capture the whole complexity of the response function. While this is partly related to the fact that the single-stream treatment, as the basis of all the existing PT formalisms, cannot properly deal with the small-scale physics such as the formation and merger of halos, it is to be noted that the standard PT itself is known to have a bad convergence property, and it produces ill-behaved higher-order corrections. One way to remedy this is to reorganize the PT expansion by introducing non-perturbative objects, expressed as partial infinite sums of terms in the standard PT expansion. Since such a treatment possesses a different mode-coupling structure, the prediction of the response function would be improved to some extent.

\begin{figure}
\centering
\includegraphics[width=7.8cm]{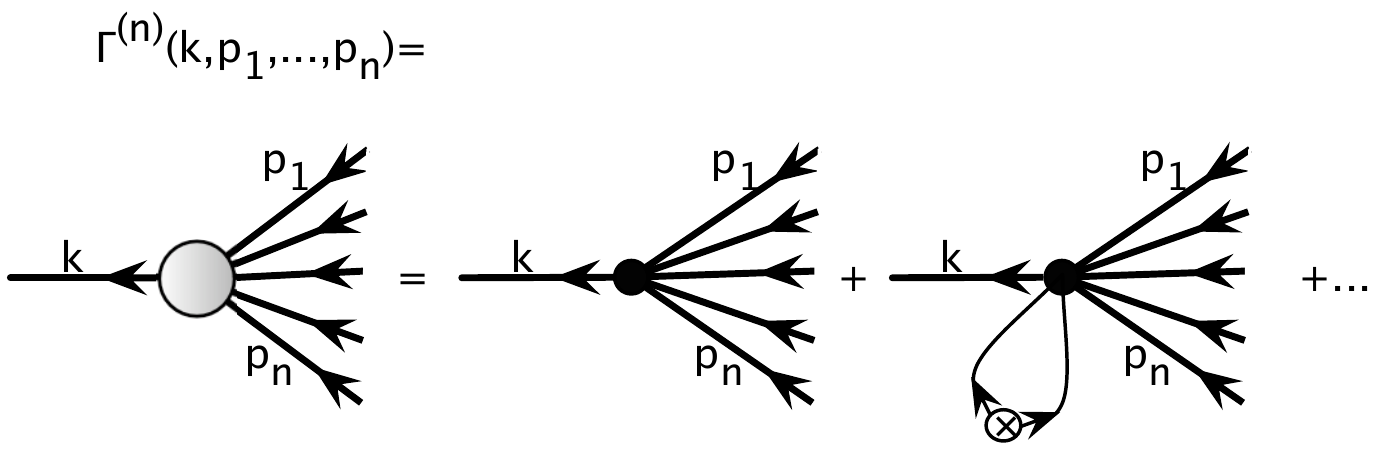} 
\caption{Diagrammatic representation of the first two terms of the multi-point
  propagator $\Gamma^{(n)}$ in the standard PT expansion. $\Gamma^{(n)}$ represents the average value of the emerging nonlinear mode $\bfk$ given $n$ initial modes in the linear regime. Here we show the first two contributions: tree-level and one-loop. Note that each object represents a collection of (topologically) different diagrams: each black dot represents a set of trees that connect respectively $n+1$ lines for the first term, $n+3$ for the second.}
\label{Gamma5}
\end{figure}

As one of the alternatives to the standard PT approach, we consider the multi-point propagator expansion proposed by Ref.~\cite{Bernardeau:2008lr}, in which the multi-point propagators are the building blocks of the expansion. They are fully non-perturbative objects defined as the ensemble average (over fluctuations in the medium) of the infinitesimal response of the evolved density field to a small initial perturbation.  More precisely, we can define the $(n+1)$-point propagator, $\Gamma^{(n)}$, through the functional derivative as
\begin{eqnarray}
\displaystyle\frac{1}{n!}\left\langle \frac{\dpartial^n \delta(\bfk)}{\dpartial\delta_{\rm lin}(\bfq_{1})\dots \dpartial\delta_{\rm lin}(\bfq_{n})}\right\rangle &\equiv 
\frac{1}{(2\pi)^{3(n-1)}} \Dirac(\bfk-\bfq_{1\cdots n})\nonumber\\
&\times\Gamma^{(n)}(\bfq_{1},\dots,\bfq_{n}).
\end{eqnarray}
In the standard PT treatment, the propagator, $\Gamma^{(n)}$, is expanded into an infinite series starting with the leading-order term, $F_{\rm sym}^{(n)}$, and one systematically obtains the higher-order corrections from the $(n+2p)$-th order density field with a positive integer $p>0$ \cite{Taruya:2012qy}:
\begin{widetext}
\begin{align}
& \Gamma^{(n)}(\bfk_1,\cdots,\bfk_n)=F_{\rm sym}^{(n)}(\bfk_1,\cdots,\bfk_n)+
\sum_{p=1}^\infty\,\Gamma^{(n)}_{p\mbox{-}{\rm loop}}(\bfk_1,\cdots,\bfk_n)\,;
\nonumber\\
&\Gamma^{(n)}_{p\mbox{-}{\rm loop}}(\bfk_1,\cdots,\bfk_n)=c^{(n)}_m
\int\frac{d^3\bfq_1\cdots d^3\bfq_p}{(2\pi)^{3p}}
 F_{\rm sym}^{(n+2p)}(\bfq_1,-\bfq_1,\cdots,\bfq_p,-\bfq_p,\bfk_1,\cdots,\bfk_n)\,P_{\rm lin}(q_1)\cdots P_{\rm lin}(q_p).
\label{eq:def_Gamma^n_p-loop}
\end{align}
\end{widetext}
Here, the coefficient $c^{(n)}_m$ is given by $_{2m+n}C_n\,(2m-1)!!$ with $_{2m+n}C_n$ being the binomial coefficient. A diagrammatic representation of this expansion can be found in Fig.~\ref{Gamma5}. 

Taking advantage of these forms, it is then possible to show explicitly that 
the power spectrum $P(k)$ can be expressed in terms of the multi-point propagators $\Gamma^{(n)}(\bfq_{1},\dots,\bfq_{n})$ as,
\begin{eqnarray}
&&P(k) = \sum_{n=1}^{\infty}\,n!\,\int
\frac{\dd^{3}\bfq_{1}\dots\dd^{3}\bfq_{n}}{(2\pi)^{3(n-1)}}\,\Dirac(\bfk-\bfq_{1\cdots n})\nonumber\\
&&\qquad\quad\times\Bigl\{\Gamma^{(n)}(\bfq_{1},\dots,\bfq_{n})\Bigr\}^2\,
\Plin(q_{1})\cdots\Plin(q_{n}).
\label{eq:power_gamma}
\end{eqnarray}
This is an important and nontrivial result presented in Ref.~\cite{Bernardeau:2008lr} that shows how the perturbative series can be reorganized.

In order to obtain an improved PT prediction with Eq.~(\ref{eq:power_gamma}), one has to develop a model of $\Gamma^{(n)}$ that has a non-perturbative property. Indeed, such a property can be exploited by taking the high-$k$ limit at each order, and then systematically summing up all the terms at this limit. As a result, the multi-point propagators are shown to behave as follows:
\begin{equation}
\Gamma^{(n)}(\bfk_{1},\dots,\bfk_{n})\,\, \stackrel{k\to\infty}{\longrightarrow} \,\, \exp(-\alpha_k)\ \Gamma^{(n)}_{\tree}(\bfk_{1},\dots,\bfk_{n}),
\label{eq:gamma_highk}
\end{equation}
with the tree-level propagator $\Gamma_\mathrm{tree}^{(n)}$ identified with the $n$-th order PT kernel, $F_{\rm sym}^{(n)}$. Here, the quantity $\alpha_k$ is defined by
\begin{equation}
\alpha_k = \frac12k^2\sigma_{\rm d}^2\,; \quad
\sigma_{\rm d}^2=\int \frac{dq}{6\pi^2}\,P_{\rm lin}(q),
\label{eq:def_alpha}
\end{equation}
where $k=\vert\bfk_{1}+\dots+\bfk_{n}\vert$, and $\sigma_{\rm d}$ is the one-dimensional root mean square of the linear displacement field. Note that this resummation is a priori valid when incoming modes are all large.  We consider a $k$-dependent cutoff to perform the integral for $\sigma_{\rm d}$ to have a better match with simulations as advocated in Ref.~\cite{Taruya:2012qy}.

With the asymptotic property given above, one can unambiguously construct a {\it regularized} propagator that reproduces Eq.~(\ref{eq:gamma_highk}) at the high-$k$ limit as well as the standard PT results shown in Fig.~\ref{Gamma5} at the low-$k$ limit \cite{Bernardeau:2012fk}. The expressions relevant to the power spectrum calculation at the two-loop order are
\begin{eqnarray}
&&\Gamma^{(1)}_{\reg}(k) = \Bigl\{1+\alpha_k+\frac12\alpha_k^2 + \Gamma_{\oneloop}^{(1)}(k)(1+\alpha_k)\nonumber\\
&&\qquad\qquad\qquad\qquad +\,\Gamma_{\twoloop}^{(1)}(k)\Bigr\}\exp\left(-\alpha_k\right),
\label{eq:gamma1_reg}\\
&&\Gamma^{(2)}_{\reg}(\bfk_1,\bfk_2) = \Bigl\{(1+\alpha_k)F_{\rm sym}^{(2)}(\bfk_1,\bfk_2) \nonumber\\
&&\qquad\qquad\qquad\qquad + \Gamma_{\oneloop}^{(2)}(\bfk_1,\bfk_2)\Bigr\}\exp\left(-\alpha_k\right),
\label{eq:gamma2_reg}\\
&&\Gamma^{(3)}_{\reg}(\bfk_1,\bfk_2,\bfk_3) = F_{\rm sym}^{(3)}(\bfk_1,\bfk_2,\bfk_3)\exp\left(-\alpha_k\right),
\label{eq:gamma3_reg}
\end{eqnarray}
where the function $\Gamma_{p\mbox{-}{\rm loop}}^{(n)}$ is defined in Eq.~(\ref{eq:def_Gamma^n_p-loop}). In what follows, we adopt a specific implementation of the multi-point propagator expansion with regularized propagators given above, called RegPT, and following Ref.~\cite{Taruya:2012qy}, we will compute the power spectrum given in Eq.~(\ref{eq:power_gamma}). We will then derive the analytic expression for the response function based on this prescription.

\subsection{Response function at the two-loop order}
\label{subsec:response_PT}

Based on the two different PT treatments described so far, we here present the analytic form of the response function up to the two-loop order. 
According to the definition in Eq.~(\ref{eq:response_def2}),  it is rather obvious that at the lowest (i.e., linear) order in the standard PT, we have the only contribution: 
\begin{equation}
K(k,q)=q\,\Dirac(k-q),
\end{equation}
which appears non-vanishing only when $k=q$. Subsequent perturbative corrections include contributions which have a broad support over $q$ modes, and the broadband structure depends on the PT treatment. In the following, we will focus our investigation on the higher-order contributions to the response function.

\subsubsection{The Standard PT expression}

The explicit analytical form of the response function can be obtained from Eq.~(\ref{eq:pk_SPT_2loop}) based on the definition Eq.~(\ref{eq:response_def1}).  A detailed derivation is presented in Appendix \ref{appendix:derivation_PT}. The resultant expression up to the two-loop order is summarized as 
\begin{eqnarray}
K^\mathrm{SPT}(k,q)= K_\mathrm{tree}^\mathrm{SPT}(k,q) + K_\mathrm{1-loop}^\mathrm{SPT}(k,q) + K_\mathrm{2-loop}^\mathrm{SPT}(k,q),\nonumber\\
\label{eq:K_SPT}
\end{eqnarray}
with
\begin{widetext}
\begin{align}
K_\mathrm{tree}^\mathrm{SPT}(k,q) &= q\,\delta_\mathrm{D}(q-k),
\label{eq:K_SPT_tree}\\
K_\mathrm{1-loop}^\mathrm{SPT}(k,q) &= 2q\,\Gamma^{(1)}_\mathrm{1-loop}(k)\delta_\mathrm{D}(q-k) + \frac{q^3}{2\pi^2}\left[2\Plin(k)L^{(1)}(q,k) + 4X^{(2)}(q,k)\right],
\label{eq:K_SPT_1loop}\\
K_\mathrm{2-loop}^\mathrm{SPT}(k,q) &= q\left\{\left[\Gamma^{(1)}_\mathrm{1-loop}(k)\right]^2+2\,\Gamma^{(1)}_\mathrm{2-loop}(k)\right\}\,\delta_\mathrm{D}(q-k) \nonumber\\
&+ \frac{q^3}{2\pi^2}\Bigl\{2\left[\Gamma^{(1)}_\mathrm{1-loop}(k)L^{(1)}(q,k) + 2M^{(1)}(q,k)\right]\Plin(k)+18S^{(3)}(q,k)+8Y^{(2)}(q,k)+4Q^{(2)}(q,k)\Bigr\},
\label{eq:K_SPT_2loop}
\end{align}
\end{widetext}
where the function $\Gamma_{p\mbox{-}{\rm loop}}^{(n)}$ is the $(n+1)$-point propagator at $p$-loop order. For simplification, 
we introduce the functions $L^{(1)},\ X^{(2)},\ M^{(1)},\ S^{(3)},\ Y^{(2)},$ and $Q^{(2)}$, whose expressions are summarized in Appendix~\ref{appendix:kernel_func} (see also Appendix~A of Ref.~\cite{Taruya:2012qy}). These are expressed in terms of the angle averages of the combinations of standard PT kernels $F_{\rm sym}^{(n)}$ and the linear power spectrum.

\begin{figure}[!ht]
   \leftline{\includegraphics[width=2.6cm]{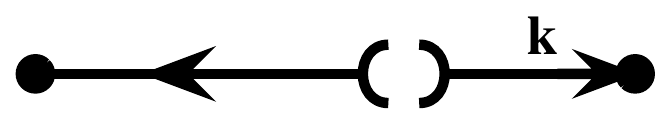}}
   \leftline{\includegraphics[width=8.4cm]{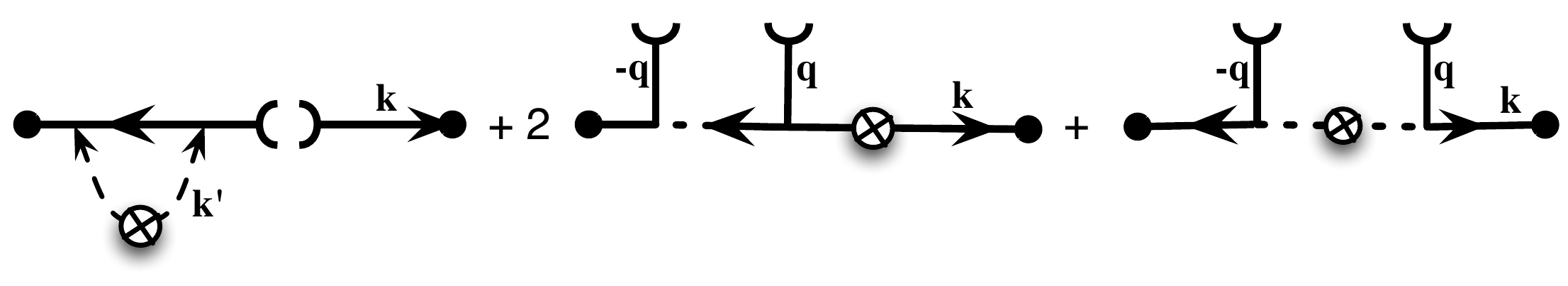}}
   \caption{Diagrammatic representation of contributions to the response function in standard PT expansion, at linear order (top row)  and at one-loop order (bottom row). The expression of such diagrams follow  rules presented in previous papers with  external lines that correspond to the propagation of wave-modes $\bfq$ and $-\bfq$. The first diagram of each row corresponds to a contribution proportional to $\Dirac(q-k)$.}
   \label{fig:Cont_0loop_1loop}
\end{figure}

\begin{figure}
\centering
\leftline{\includegraphics[width=8.2cm]{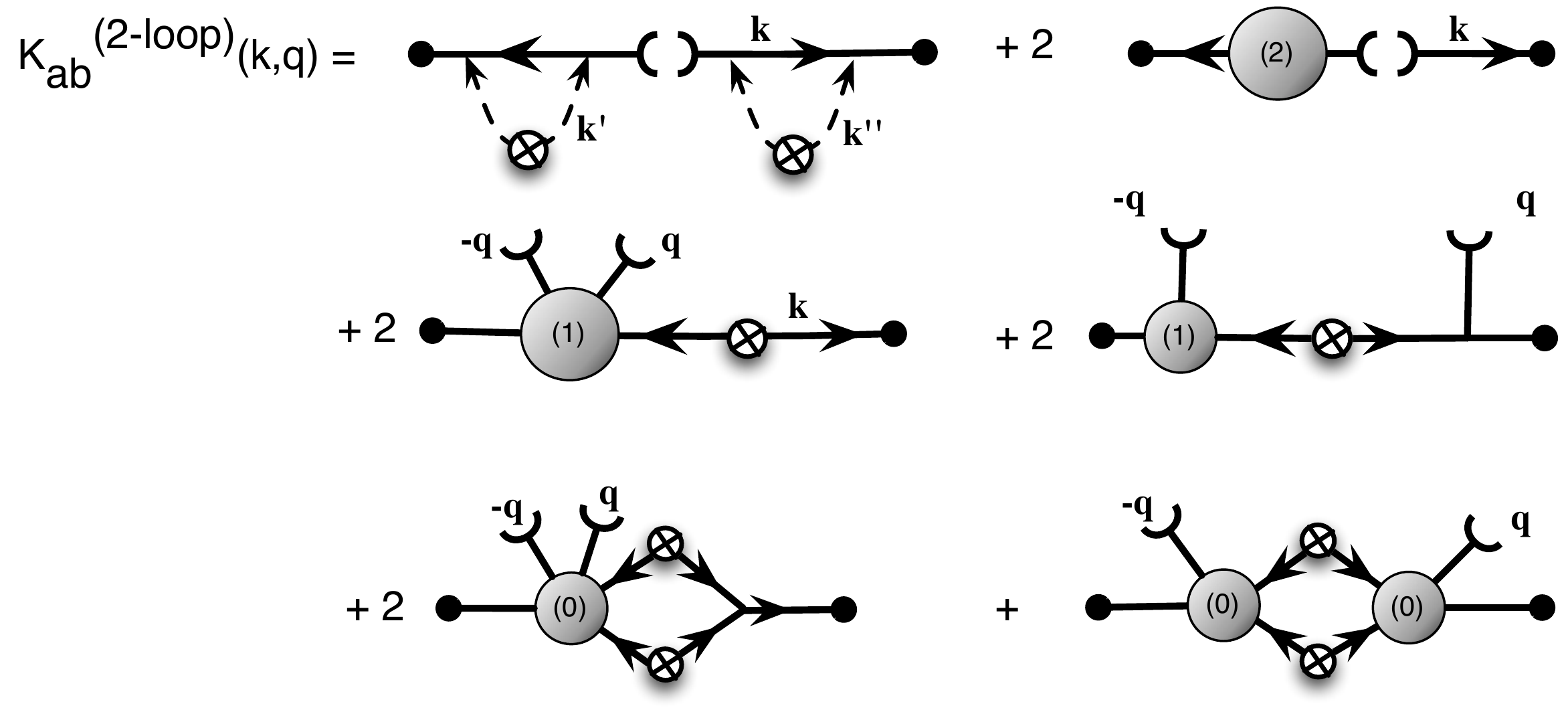}} 
\caption{Contributions of the two-loop order to the response function in standard PT. The shaded area marked with $(p)$ correspond to $\Gamma$ expression taken at $p$ loop order irrespectively of the number of lines it is connected to. 
The first line corresponds to diagrams that are proportional to $\Dirac(k-q)$. }
\label{Cont-2loop}
\end{figure}

It would then be interesting to see a diagrammatic representation of each of such contributions. Up to the one-loop order, there are four of such diagrams that are presented in Fig.~\ref{fig:Cont_0loop_1loop}. Many more can be considered at higher order (See Fig.~\ref{Cont-2loop} for the two-loop diagrams).

We can see that there are two types of contributions to the response function. The first one is those proportional to the Dirac delta function, which is already present at the linear order. These contributions tell us how much impact remains at the scale where we put a small initial perturbation in the linear power spectrum. These contributions can be found in Figs.~\ref{fig:Cont_0loop_1loop} and \ref{Cont-2loop} as the diagrams whose principal horizontal line is disconnected. The other terms are all describing the pure mode transfer effect between different scales. The diagrams in Figs.~\ref{fig:Cont_0loop_1loop} and \ref{Cont-2loop} describe the propagation of power injected at the ends of the two external lines (wave modes $\bfq$ and $-\bfq$ depicted by the semicircles) to the left and right ends expressed by the dots.

\subsubsection{The RegPT expression}

Similarly, the response function can be computed based on RegPT. Following the definition in Eq.~(\ref{eq:response_def1}) or (\ref{eq:response_def2}), the expression for the response function is derived in Appendix~\ref{appendix:derivation_PT}. The resultant expression relevant at the two-loop order is summarized as
\begin{align}
 K^\mathrm{RegPT}(k,q)=K^\mathrm{RegPT}_{\tree}(k,q)+K^\mathrm{RegPT}_{1\mbox{-}{\rm loop}}(k,q)+K^\mathrm{RegPT}_{2\mbox{-}{\rm loop}}(k,q)
\label{eq:K_RegPT}
\end{align}
with
\begin{widetext}
\begin{align}
 K^\mathrm{RegPT}_{\tree}(k,q) &=
q\,\Bigl\{\Gamma^{(1)}_{\rm reg}(k)\Bigr\}^2\,\Dirac(k-q)
+\frac{q^3}{\pi^2}\,\,
\,\Gamma^{(1)}_{\reg}(k)\bigl\{ L^{(1)}(q,k)(1+\alpha_k) +2 M^{(1)}(q,k)\bigr\}
\Plin(k)\, e^{-\alpha_k}
\nonumber
\\
&\qquad+ q\,\frac{k^2}{6\pi^2} 
\Bigl[\bigl\{1+\Gamma^{(1)}_{\oneloop}(k)+\alpha_k\bigr\}\,
e^{-\alpha_k}\,\Gamma_{\rm reg}^{(1)}(k)\,\Plin(k)
-P^{\rm RegPT}_{\rm tree}(k)\,\Bigr],
\label{eq:K_RegPT_tree}
\\
 K^\mathrm{RegPT}_{\oneloop}(k,q) &= 
\frac{2\,q^3}{\pi^2}\,\,
\Bigl[\,X^{(2)}(q,k)(1+\alpha_k)^2+\Bigl\{2Y^{(2)}(q,k)+Q^{(2)}(q,k)\Bigr\}(1+\alpha_k)
+ Z^{(2)}(q,k)+R^{(2)}(q,k)\,\Bigr] e^{-2\alpha_k}
\nonumber\\
&+ q\,\frac{k^2}{6\pi^2} 
\Bigl[\,\Bigl\{(1+\alpha_k)\,P_{\rm corr}^{(2){\tree\mbox{-}\tree}}(k) + \frac{1}{2}
P_{\rm corr}^{(2){\tree\mbox{-}\oneloop}}(k)\Bigr\}e^{-2\alpha_k}
-P^{\rm RegPT}_{1\mbox{-}{\rm loop}}(k)\,
\Bigr],
\label{eq:K_RegPT_1loop}
\\
 K^\mathrm{RegPT}_{\twoloop}(k,q) &=\frac{9\,q^3}{\pi^2}\,
S^{(3)}(q,k)\, e^{-2\alpha_k} 
- q\,\frac{k^2}{6\pi^2} \,P^{\rm RegPT}_{2\mbox{-}{\rm loop}}(k).
\label{eq:K_RegPT_2loop}
\end{align}
\end{widetext}
The quantities $P^{\rm RegPT}_{\rm tree}$ and $P^{\rm RegPT}_{\rm 1,\,2\mbox{-}loop}$ are the power spectrum in RegPT at a given order, whose explicit expressions are given in Eqs.~(\ref{eq:pkRegPT_tree})--(\ref{eq:pkRegPT_2loop}). Again, we introduced the kernels, $Z^{(2)}$ and $R^{(2)}$, as well as the spectra, $P_{\rm corr}^{(2){\tree\mbox{-}\tree}}$ and $P_{\rm corr}^{(2){\tree\mbox{-}\oneloop}}$, in addition to those introduced in $K^{\rm SPT}_{\rm 1,2\mbox{-}loop}$. These expressions are all summarized in Appendix \ref{appendix:kernel_func} (and also given in Ref.~\cite{Taruya:2012qy}).

One notable point in the RegPT expression is that as a result of the reorganized PT prescription, even the tree- and one-loop terms contain the contributions coming from the two-loop order in the standard PT. Thus, in marked contrast to the standard PT case, the leading-order response function $K^{\rm RegPT}_{\rm tree}$ has a non-vanishing support at $k\ne q$. Another interesting feature is that all the contributions proportional to the Dirac delta function in the standard PT expression are reorganized and are summarized in a single term with the regularized propagator. We can also find other terms involving regularized quantities, either the propagator or the power spectrum, which would give it some non-perturbative effects. Finally, all the terms including the one with the regularized quantities are suppressed by $\exp(-2\alpha_k)$. We will see how these features change the predictions and discuss the agreement with the simulation data shortly below.

\subsection{Comparison and considerations in different regimes}
\label{subsec:comparison}

Now we are in a position to see how different analytical models compare with the simulation measurement. We will give physical interpretations to the features seen in the simulated response function based on the comparison on different regimes.

\subsubsection{The $q\ll k$ regime}

\begin{figure}[!ht]
   \centering
   \includegraphics[width=7.3cm]{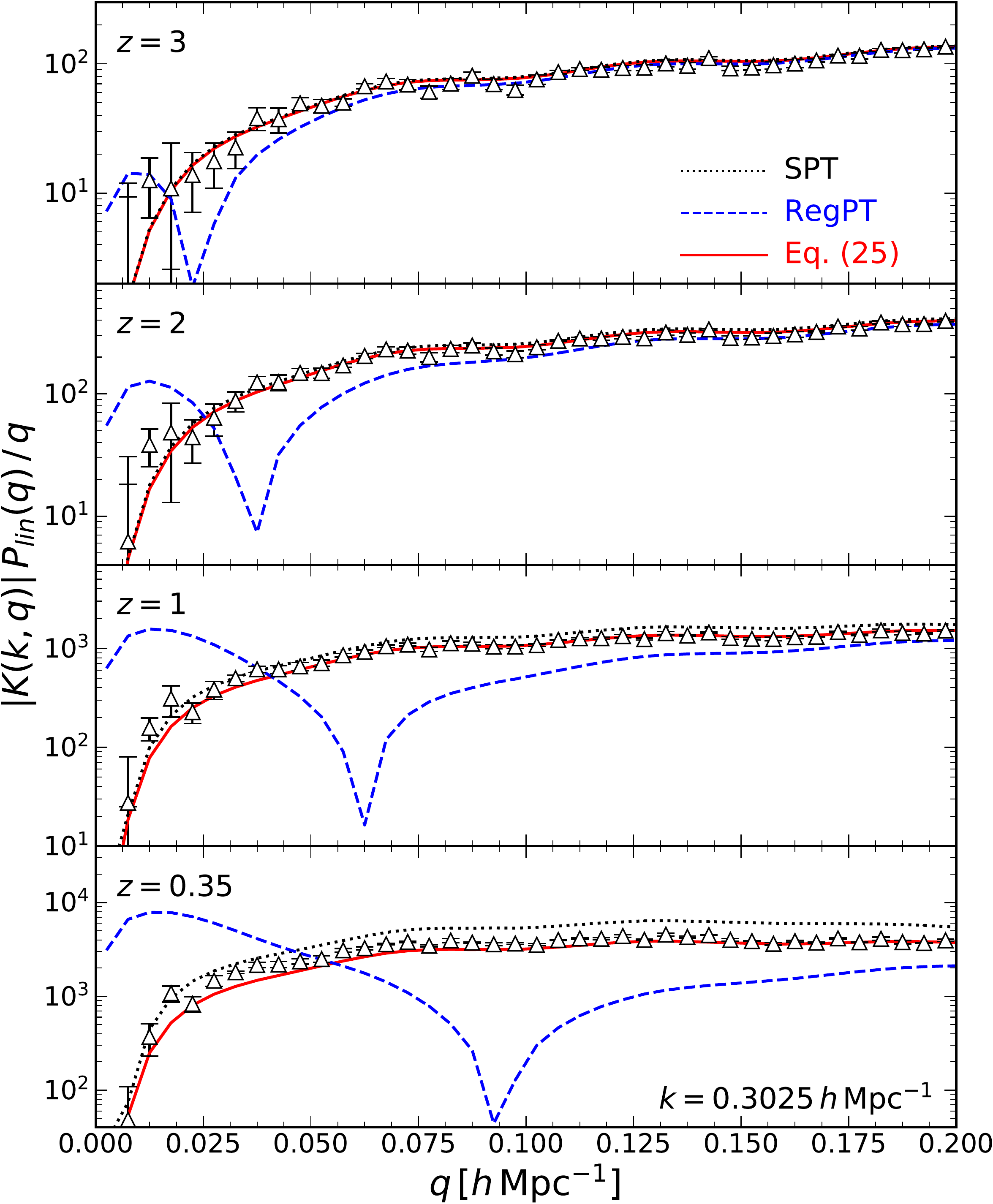}
   \caption{Response function in the $q\ll k$ regime. We fix the wavenumber $k$ to be $0.3025\hMpci$ and plot the combination $|K(k,q)|\Plin(q)/q$ as a function of $q$. The standard PT, RegPT and new prediction are shown by the dotted, dashed and solid line, respectively. The simulation measurements are plotted by the triangles with error bars showing the $1$-$\sigma$ uncertainty.}
   \label{fig:response_lowq}
\end{figure}

First, we show a comparison on $q < k$. Figure~\ref{fig:response_lowq} depicts the response function at $k\sim0.3\hMpci$ as a function of $q$. All the model predictions are computed at the two-loop level. Results at different redshifts are shown in different panels. In the small $q$ limit shown here, we expect that our response function approaches to zero due to the cancellations of the infrared contributions as expected from
the extended Galilean invariance satisfied by this system (see for instance \cite{2013JCAP...05..031P}). This seems to be the case, indeed, when we look at the measured values from the $N$-body simulations (symbols with error bars) within the quoted error level. The analytical estimate based on standard PT does reproduce this behavior. However, the RegPT prediction (dashed) clearly breaks the asymptote especially at low redshifts. At very small $q$, this model sees a zero-crossing (at $q\sim0.03, 0.04, 0.06$ and $0.1\,\hMpci$, respectively for $z=3, 2, 1$ and $0.35$), below which it gives a wrong sign.

\subsubsection{The $q\approx k$ regime}

\begin{figure}[!ht]
   \centering
   \includegraphics[width=7.3cm]{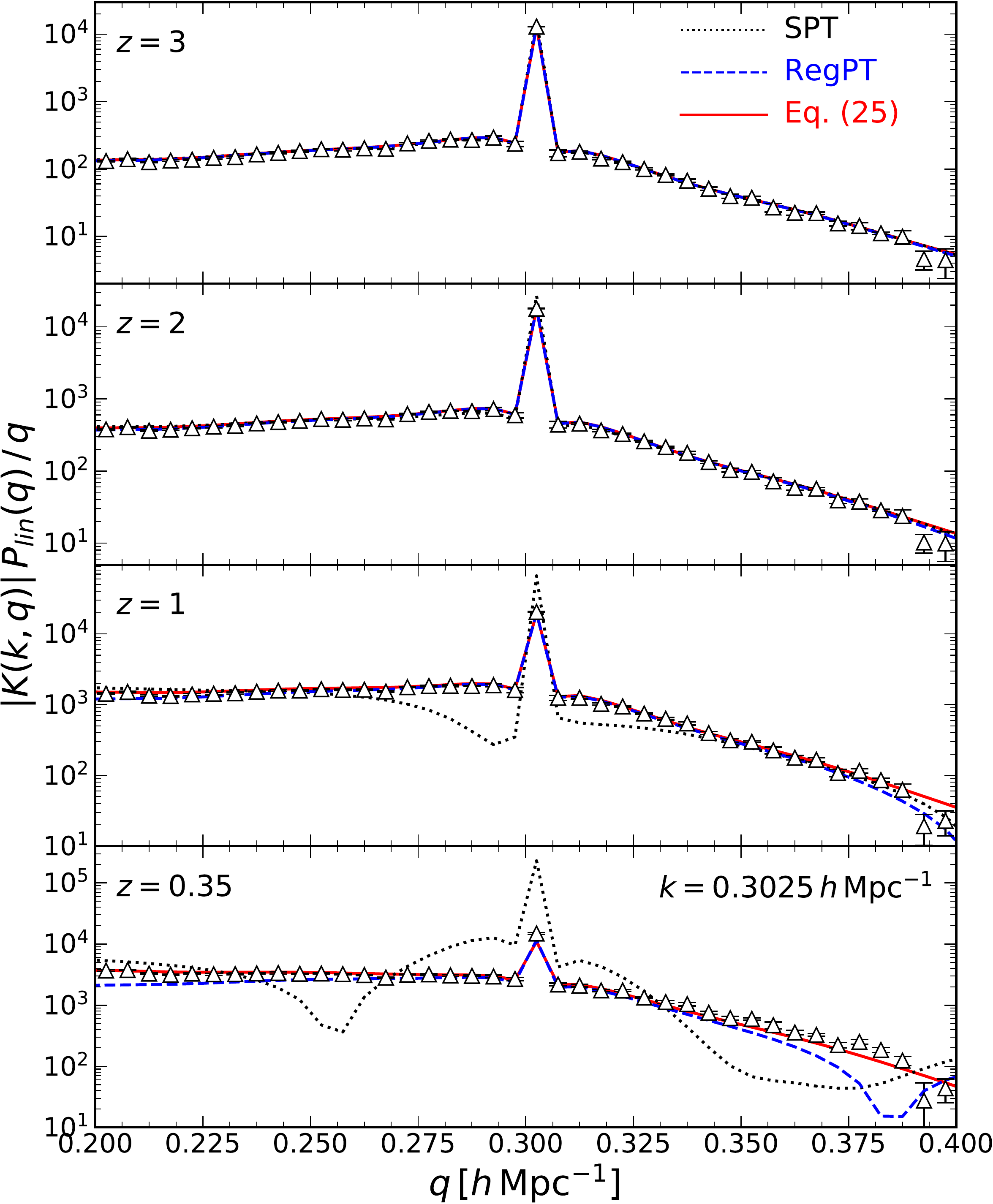}
   \caption{Same as Fig.~\ref{fig:response_lowq}, but in the $q\approx k$ regime. We again fix $k=0.3025\hMpci$} and shift the $q$ range toward smaller scales.
   \label{fig:response_midq}
\end{figure}

\begin{figure*}[!ht]
   \centerline{\includegraphics[width=14cm]{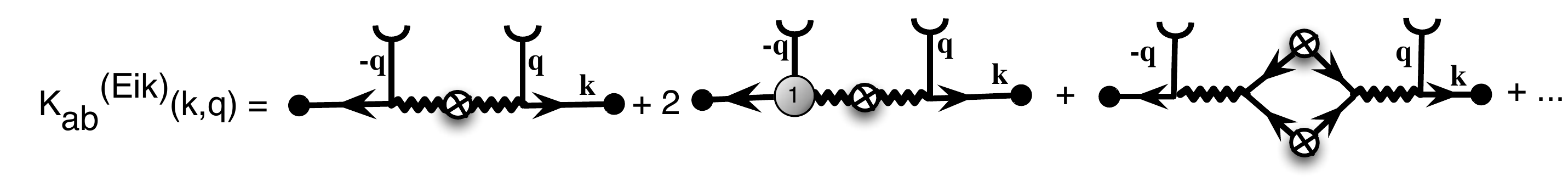}}
   \caption{Diagrammatic representation of the leading diagrams in the eikonal limit. The wiggle lines represent the soft modes one can encounter when $q\sim k$. In the Eikonal approximation the vertex values scale like $k/\vert\bfk-\bfq\vert$ and can be arbitrarily large.}
      \label{fig:Cont_Eik}
\end{figure*}

We then explore here the $q\approx k$ regime. A comparison among models and simulations can be found in Fig.~\ref{fig:response_midq}. Unlike the low-$q$ limit,
we can see that the RegPT prediction agrees with the simulation data at all the four redshifts shown here, while a clear breakdown of the standard PT calculation can be found at low redshifts\footnote{This discrepancy was not seen in \cite{Nishimichi16} due to the wider wavenumber bins in the computation of the response function.}. This can be interpreted as follows.

In this regime, configurations can be reached where modes $\bfk$
and $\bfq$ are close enough so that $\vert \bfk-\bfq\vert$ is negligible compared to $q$ and $k$. As a consequence, the vertex 
joining such a line to the rest of the diagram becomes large as it scales like $k/\vert\bfk-\bfq\vert$ (see \cite{Bernardeau12,Bernardeau13,2013arXiv1311.2724B}). This makes the resulting expression for the response function inaccurate in the standard PT expression as it affects the expansion series.

From this observation, it is, however, possible to sort diagrams 
with respect to the number of such vertices they exhibit taking advantage of the multipoint-propagator expansions that precisely organize the series expansion in this way. At the one-loop order in the standard PT, the leading contribution comes from the diagram involving
$F_{\rm sym}^{(2)}\,F_{\rm sym}^{(2)}$
as it incorporates naturally 2 of such large vertices (in Fig. \ref{fig:Cont_Eik} there are those which the wiggle lines are connected to). Moreover this diagram is indeed the dominant contributor at the one-loop order in the region $q\approx k$ (see \cite{Nishimichi16}). 
Similar un-regularized diagrams can be found in subsequent orders. Diagrams up to the two-loop order that present the same kind of divergences are presented in Fig.~\ref{fig:Cont_Eik}. It is not difficult to build a whole set of such diagrams, and the resummation~(\ref{eq:gamma_highk}) precisely accounts for all such diagrams at the high-$k$ limit.

\begin{figure*}[!ht]
   \centerline{\includegraphics[width=14cm]{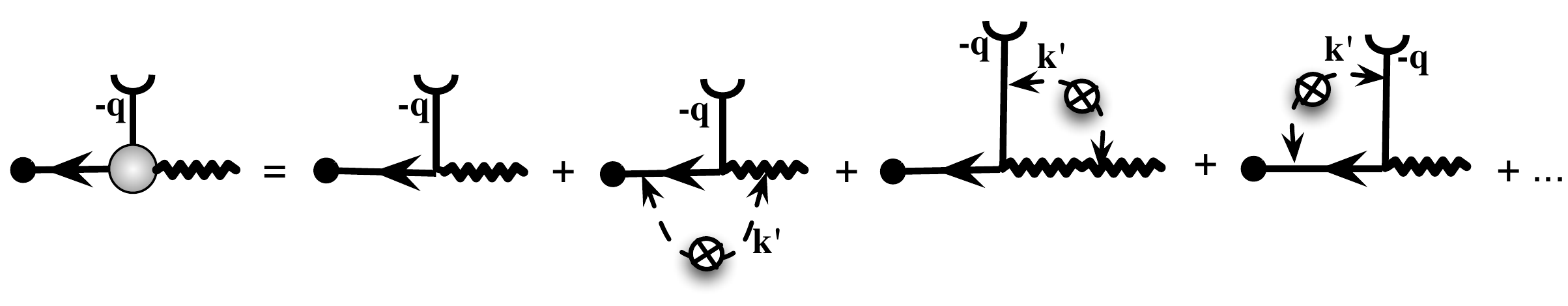}}
   \caption{Diagrammatic representation of the two-loop contribution to the resummed multipoint propagators.}
      \label{fig:Cont_G2_Renorm}
\end{figure*}

What these results suggest, however, is that $\Gamma^{(2)}(\bfq,\bfk-\bfq)$ is expected to be effectively given by an expression of the form, 
$\exp(-\alpha_k)F_\mathrm{sym}^{(2)}(\bfq,\bfk-\bfq)$ (see Fig.~\ref{fig:Cont_G2_Renorm} for diagrams contributing this resummation up to the two-loop order). The successful agreement between RegPT and simulations at the $q\sim k$ regime at low redshifts, where the standard PT does not behave nicely, justifies the usefulness of such a resummation. This is the basis of the phenomenological prescription that we will propose in Sec.~\ref{subsec:our_model}.

\subsubsection{The large $q$ regime}

\begin{figure*}[!ht]
   \centering
   \includegraphics[width=7.3cm]{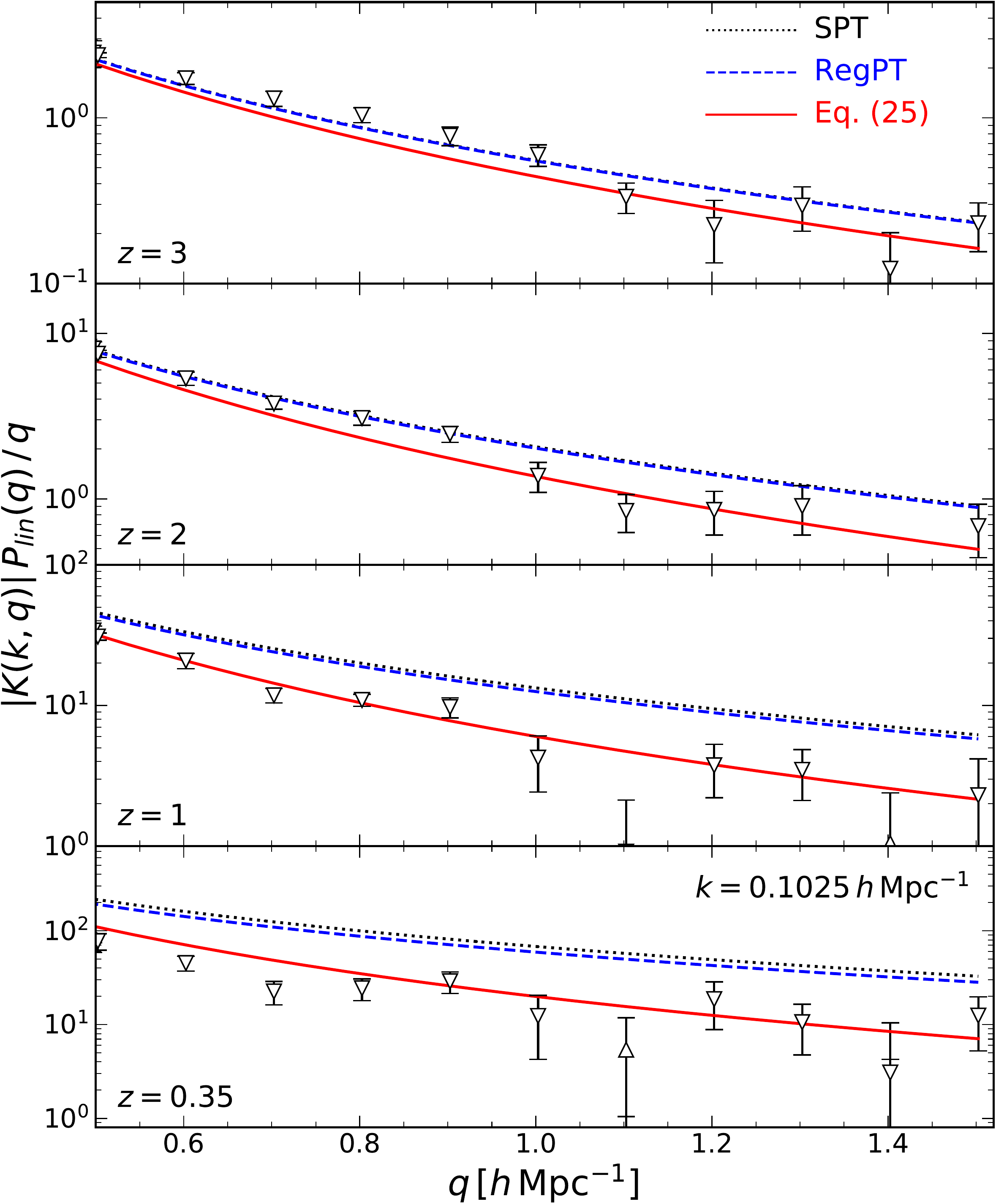}
   \includegraphics[width=7.3cm]{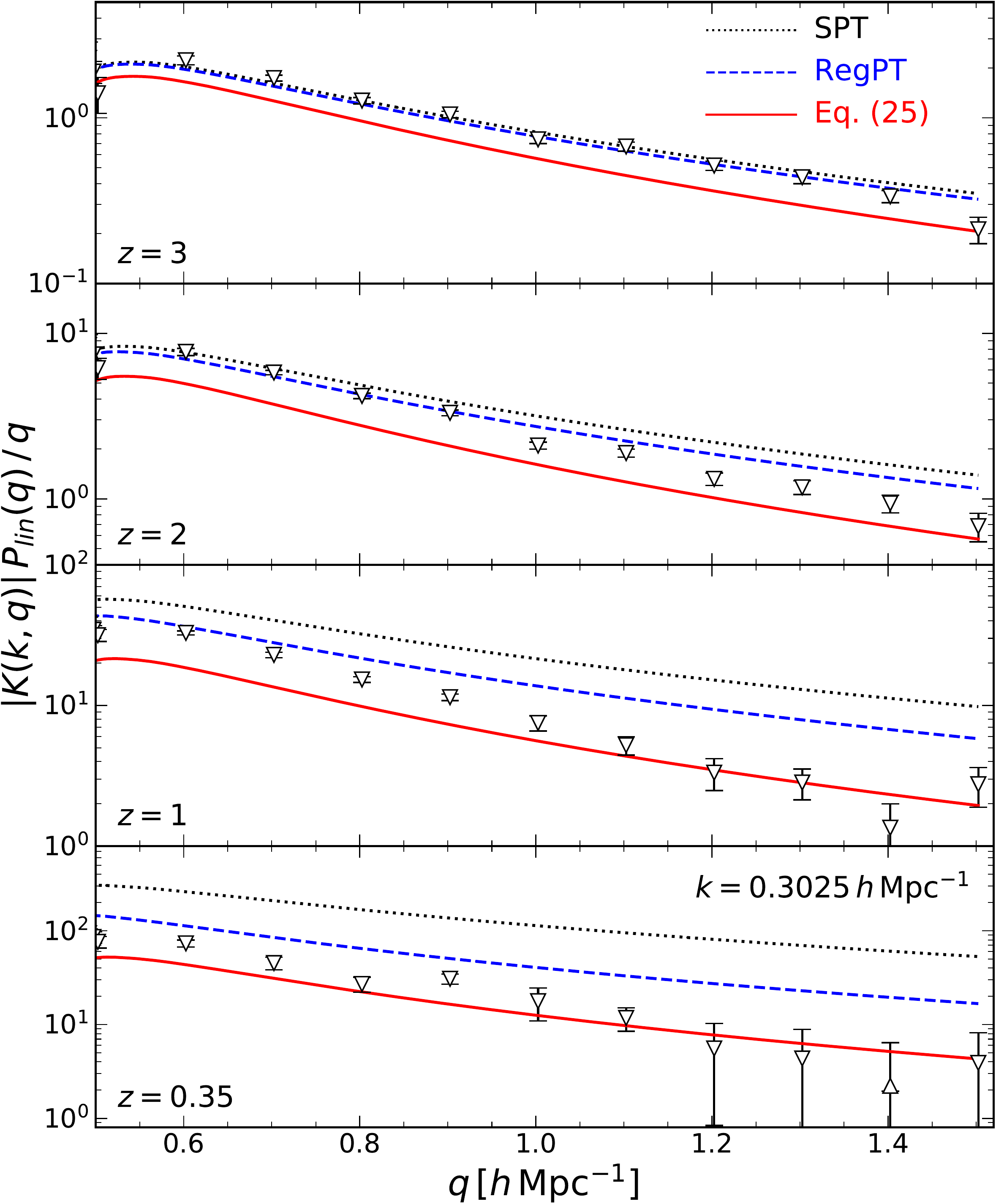}
   \caption{Same as Fig.~\ref{fig:response_lowq}, but in the $q\gg k$ regime. We now show the response function at two different values of wavenumber $k$, $0.1025\hMpci$ (left) and $0.3025\hMpci$ (right).}
   \label{fig:response_highq}
\end{figure*}

While perturbative calculations in general behave poorly on strongly nonlinear regimes, one might naively expect that a reliable prediction is still possible if one restricts oneself to sufficiently large scales, $k<k_\mathrm{max}$, which depends on the redshift. However, as recent studies have suggested, higher-loop corrections can ruin the seemingly successful predictions of the low-order calculations near $z=0$ even at very large scales, say $k=0.1\,\hMpci$ \cite{Blas:2013qy}.
It was numerically shown in Ref.~\cite{Nishimichi16} that this is because of the too strong corrections from smaller scales which PT calculations bring, compared to the numerical measurement which clearly manifests a signature of ``screening'' to prevent such a mode transfer.

We reconsider this regime, and propose a phenomenological model inspired from the trend seen in the simulation data. Before that, let us discuss if or not the RegPT prescription is a solution on this limit. We show in Fig.~\ref{fig:response_highq} the response function at two different values of $k$ (left: $0.1025\hMpci$, right: $0.3025\hMpci$), as a function of $q$.
The right panel indeed shows a significant suppression of the response in the RegPT prediction (dashed) compared to the standard PT (dotted), but the suppression is not sufficient to explain the simulation data at low redshifts. When we turn to the left panel, which shows the same comparison at a smaller $k$, the suppression is very little and almost ineffective. This is because the construction of the well-behaved propagators in RegPT respects only the $k$ dependence, and it is not designed to care the mode-transfer structure at different scales as discussed here. Thus, the resultant response is suppressed only when the wavenumber $k$ is large, and this does not give a correct $q$ dependence.

The final piece of our prescription developed here is the introduction of a damping function to take account this behavior along the $q$ directions. As shown in \cite{Nishimichi16}, the decay of the response has a Lorentzian form as a function of $q$. We will borrow this form to have an analytical model that gives the broadband shape correct in what follows.

\subsection{The proposed model}
\label{subsec:our_model}

Now we are in a position to construct a phenomenological model that respects all the findings above in different regimes. First, in order to recover the galilean invariance at the low-$q$ limit, we impose the condition to have the same asymptote as standard PT. Then following the way of our construction of the regularized propagator~(\ref{eq:gamma1_reg}), we introduce counterterms and an overall damping factor to the perturbative calculation of the response function. Unlike the exponential damping factor in RegPT, our new damping factor is designed to explain the damping both in the high $k$ and high $q$ regimes. At the two-loop order, our model reads
\begin{eqnarray}
&&K_\mathrm{model}(k,q) = \Bigl[\bigl(1+\beta_{k,q}+\frac{1}{2}\beta^2_{k,q}\bigr)K_\mathrm{tree}^\mathrm{SPT}(k,q) \nonumber\\ &&\quad+\left(1+\beta_{k,q}\right)K_\mathrm{\oneloop}^\mathrm{SPT}(k,q) + K_\twoloop^\mathrm{SPT}(k,q)\Bigr]D(\beta_{k,q})\nonumber,
\\
\label{eq:our_model}
\end{eqnarray}
where $\beta_{k,q} = \alpha_k + \alpha_q$ with $\alpha$ given by Eq~(\ref{eq:def_alpha}). Note that in the brace we have terms that depend on the $\beta_{k,q}$ factor, which we call counterterms, in addition to the standard PT responses. We design the damping function $D$ as
\begin{eqnarray}
D(x) = \left\{
\begin{array}{ll}
\exp\left(-x\right), & \mathrm{if}\,\,K_\mathrm{model}(k,q) > 0,\\
\displaystyle
\frac{1}{1+x},& \mathrm{if}\,\,K_\mathrm{model}(k,q) < 0.
\end{array}
\right.
\label{eq:damping}
\end{eqnarray}
Note that the domain for which $K_\mathrm{model}(k,q) > 0$ encompasses the $q \approx k $ regime.
We also note that in this model
the $\Dirac(k-q)$ part of the response function is very similar to that for the RegPT prescription, $q\,[\Gamma^{(1)}_\mathrm{reg}(k)]^2\,\Dirac(k-q)$, due to a similar construction. These expressions are exactly the same up to the terms of order $\mathcal{O}(\alpha_k^2)$, and both decays as $\exp(-2\alpha_k)$ in the high-$k$ limit.
We switch from the standard Gaussian damping to the Lorentzian damping after the response function sees a zero crossing to have a negative high-$q$ tail. Since the response function is positive on low $q$, we choose $\exp(-\beta_{k,q})$ there. One can see that this cancels with the counterterms in the brace in the low-$q$ limit, leaving only the standard PT expression.

The model~(\ref{eq:our_model}) is shown in Fig.~\ref{fig:response_2d} (bottom left panel), as well as in Figs.~\ref{fig:response_qdep},~\ref{fig:response_lowq},~\ref{fig:response_midq} and~\ref{fig:response_highq} (solid line). One can see that the model curve behaves well over all the regimes shown here: 
on low $q$ (Fig.~\ref{fig:response_lowq}), it behaves similarly to the standard PT prediction approaching to zero, and further, the solid line somehow keeps to fit well to the simulation data down to $z=0.35$ where the standard PT prediction starts to overpredict the response as compared with the simulation data. Then Fig.~\ref{fig:response_midq} shows that the new model (solid) behaves basically the same as the RegPT (dahsed) in the $q\sim k$ regime, both of which show an excellent agreement with the simulation data. Finally, the model prediction again fits well to the simulation data in the $k\ll q$ regime at low redshifts thanks to the Lorentzian damping factor (Fig.~\ref{fig:response_highq}). At high redshifts where the standard PT gives a reasonably agreement with simulation, however, the model predicts a bit stronger suppression. Note that even though our prescription is not always very accurate on this regime, we do not care for the reconstruction as long as the response is suppressed such that the small-to-large scale mode transfer is a minor effect.

\subsection{Validity range of the model}
\label{subsec:validity}
\begin{figure*}[!ht]
   \centering
   \includegraphics[width=7.3cm]{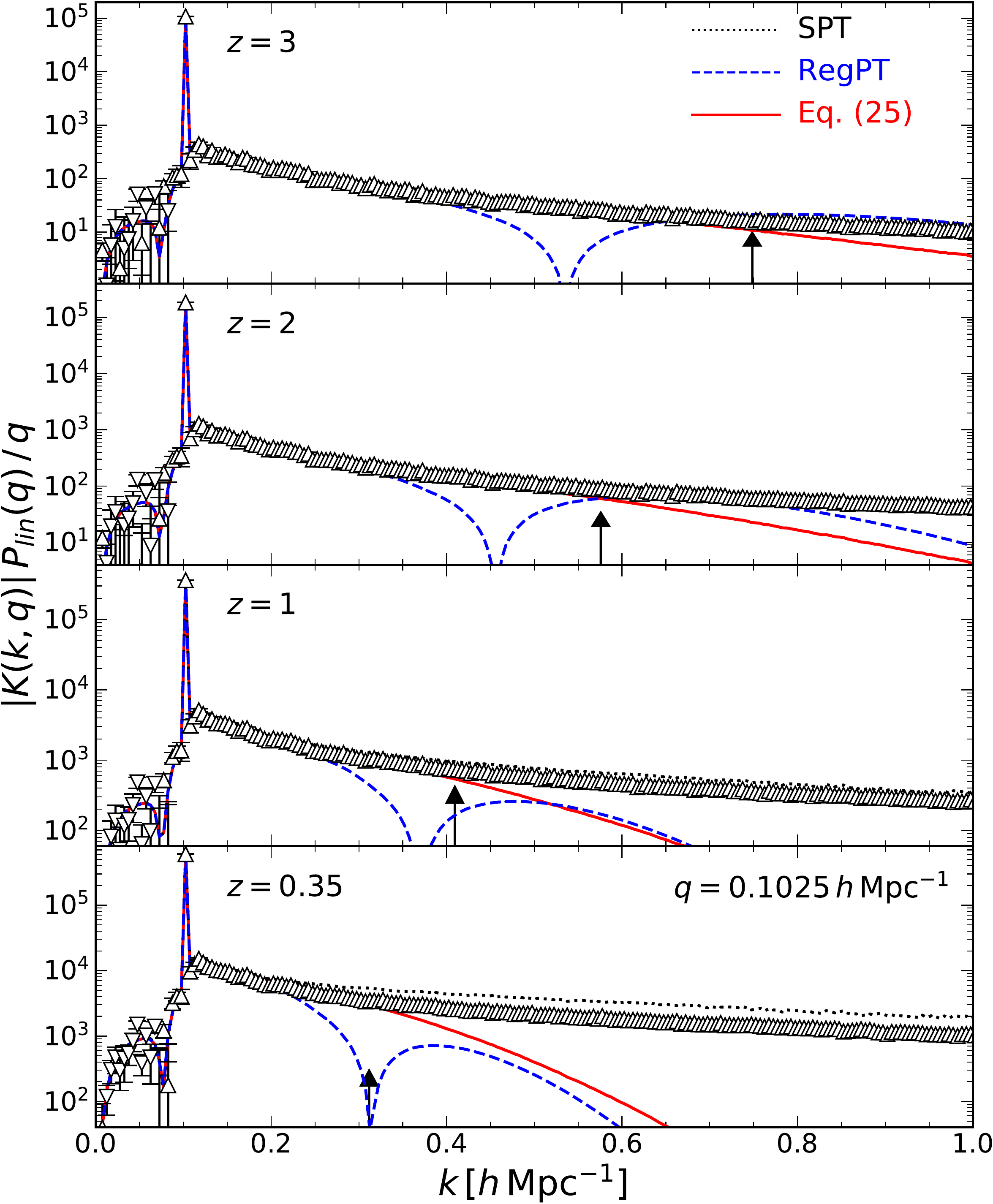}
   \includegraphics[width=7.3cm]{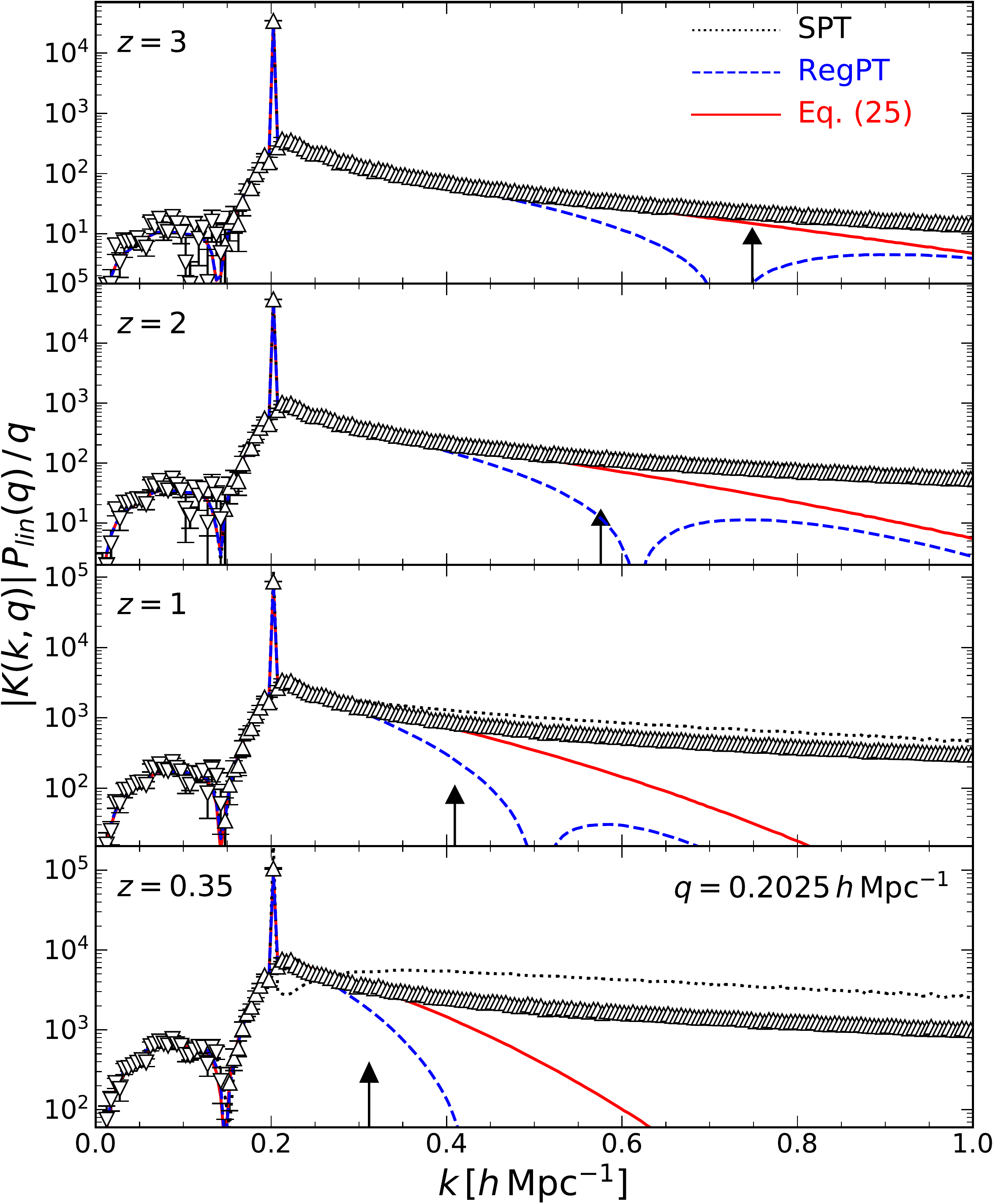}
   \caption{Response function as a function of $k$ for two fixed $q$ values in the quasi linear regime at different redshifts as indicated in the figure legend. The vertical arrows show rough estimates of the maximum wavenumber $k_\mathrm{max}$ below which the solid curve agrees with the simulation data (it is computed with $\alpha_\mathrm{max}=1$; see the main text for more detail).}
   \label{fig:response_kdep}
\end{figure*}

We have shown that the new model~(\ref{eq:our_model}) works reasonably well over a wide range of wavenumber $q$ capturing several characteristic features in different regimes. Based on a perturbative calculation, however, the model is naturally expected to show a breakdown when nonlinearity gets very strong. To see this, we investigate the dependence of the response function on the wavenumber $k$, instead of $q$. As shown in Fig.~\ref{fig:response_kdep}, the model prediction indeed gets worse as increasing $k$ for fixed values of $q$ in quasilinear scales. The scale of breakdown seems to depend on the value of $q$ very little (compare left and right).

Since the key quantity that defines the regularization of the analytical response function along the $k$-direction is $\alpha_k$ in Eq.~(\ref{eq:def_alpha}), we expect that the predicted response is suppressed too strongly when $\alpha_k$ gets large (of order unity, roughly). We introduce a number $\alpha_\mathrm{max}$ that determines the maximum wavenumber $k_\mathrm{max}$ simply via the condition, $\alpha_{k_\mathrm{max}} = \alpha_\mathrm{max}$. We show in Fig.~\ref{fig:response_kdep} the locations of $k_\mathrm{max}$ with $\alpha_\mathrm{max} = 1$ by the vertical arrow in each panel. Despite the quite different values of $q$ ($\sim0.1$ and $0.2\hMpci$ for the left and right panel), $\alpha_\mathrm{max}=1$ gives a good estimate of the maximum wavenumber for a successful prediction of the response function in both panels and at various redshifts. 

In the following section, we will discuss the accuracy of the nonlinear power spectrum reconstructed using the response function computed with Eq.~(\ref{eq:our_model}). There, it will be shown that the wavenumber $k_\mathrm{max}$ determined with $\alpha_\mathrm{max}=1$, corresponding to the arrows in the figure, is a conservative estimate of the validity range of the reconstruction with typically one-percent accuracy in the reconstructed spectrum.  In our code, $\alpha_\mathrm{max} = 1$ is thus chosen as the default value to estimate $k_\mathrm{max}$. We will come back to this point shortly.

\section{The reconstruction procedure}
\label{sec:recon}

In this section, given a well-behaved model prescription for the response function in hands, we consider a practical problem of prediction or reconstruction of the nonlinear matter power spectrum from the linear power spectrum. We begin by briefly describing the basic idea and method in Sec.~\ref{subsec:basic_idea}. The proposed methodology uses a precision power spectrum for a given fiducial cosmological model, and in Sec.~\ref{subsec:highres}, we perform a set of high-resolution simulations. Then, subsequent sections \ref{subsec:single_step} and \ref{subsec:multi_step} present a detailed implementation of the reconstruction method. Sec.~\ref{subsec:single_step} describes the simplest case based on a single-step reconstruction, and Sec.~\ref{subsec:multi_step} generalizes this procedure employing multiple steps, which allows us to predict the nonlinear power spectrum over a rather broader range of cosmological parameters. All the ideas and procedures described in this section are implemented in the python package, \texttt{RESPRESSO}, which is available at \url{http://www-utap.phys.s.u-tokyo.ac.jp/~nishimichi/public_codes/respresso/index.html}.

\subsection{The basic idea}
\label{subsec:basic_idea}

Suppose that a very well-calibrated nonlinear matter power spectrum template is available at one (or a small number of) cosmological model(s) based on a large set of simulations. Then, the question is how we can accurately estimate the nonlinear power spectrum for different cosmological parameters without performing any extra $N$-body simulations. If the difference between the two cosmological models is small enough, the concept of perturbative approach should be applied, and it would be sufficient to just predict a small correction to the baseline result. In this context, the key quantity is the response function. From the definition of the response function in Eq.~(\ref{eq:response_def1}), we have
\begin{eqnarray}
&&{\Pnl}(k;\bfp_1)\approx {\Pnl}(k;\bfp_0)+\int\dd \ln q\,K(k,q)\,\nonumber\\
&&\qquad\qquad\times\left[{\Plin}(q;\bfp_1)-{\Plin}(q;\bfp_0)\right],
\label{eq:recon1}
\end{eqnarray}
where $\bfp_i$ denotes the $i$-th cosmological parameter set. This relation tells us that when we have an accurate power spectrum template for the cosmological parameters $\bfp_0$ (a fiducial model), we can predict the nonlinear power spectrum for the parameters $\bfp_1$ (a target model) using the response function. One notable point is that provided a reliable template for the power spectrum and response function, Eq.~(\ref{eq:recon1}) can be applied even when the density field itself is in a strongly nonlinear regime. In this respect, the proposed method is said to be non-perturbative. Although Eq.~(\ref{eq:recon1}) assumes that the difference between the linear power spectra, ${\Plin}(q;\bfp_1)-{\Plin}(q;\bfp_0)$, is small, we will later introduce a multi-step reconstruction that allows us to relax this assumption, and thus the method can be applied to a broader parameter range over the $\Lambda$CDM-like models.

\subsection{Power spectrum template from high-resolution simulations}
\label{subsec:highres}

For practical implementation of the idea in Sec.~\ref{subsec:basic_idea},   
we first need to prepare an accurate power spectrum template for a fiducial cosmological model.  For this purpose, we perform another set of simulations, and tabulate the measured power spectra at different redshifts. This table serves as the initial guess of the nonlinear power spectrum for models with different cosmological parameters. 

\begin{figure}[!ht]
   \centering
   \includegraphics[width=8.2cm]{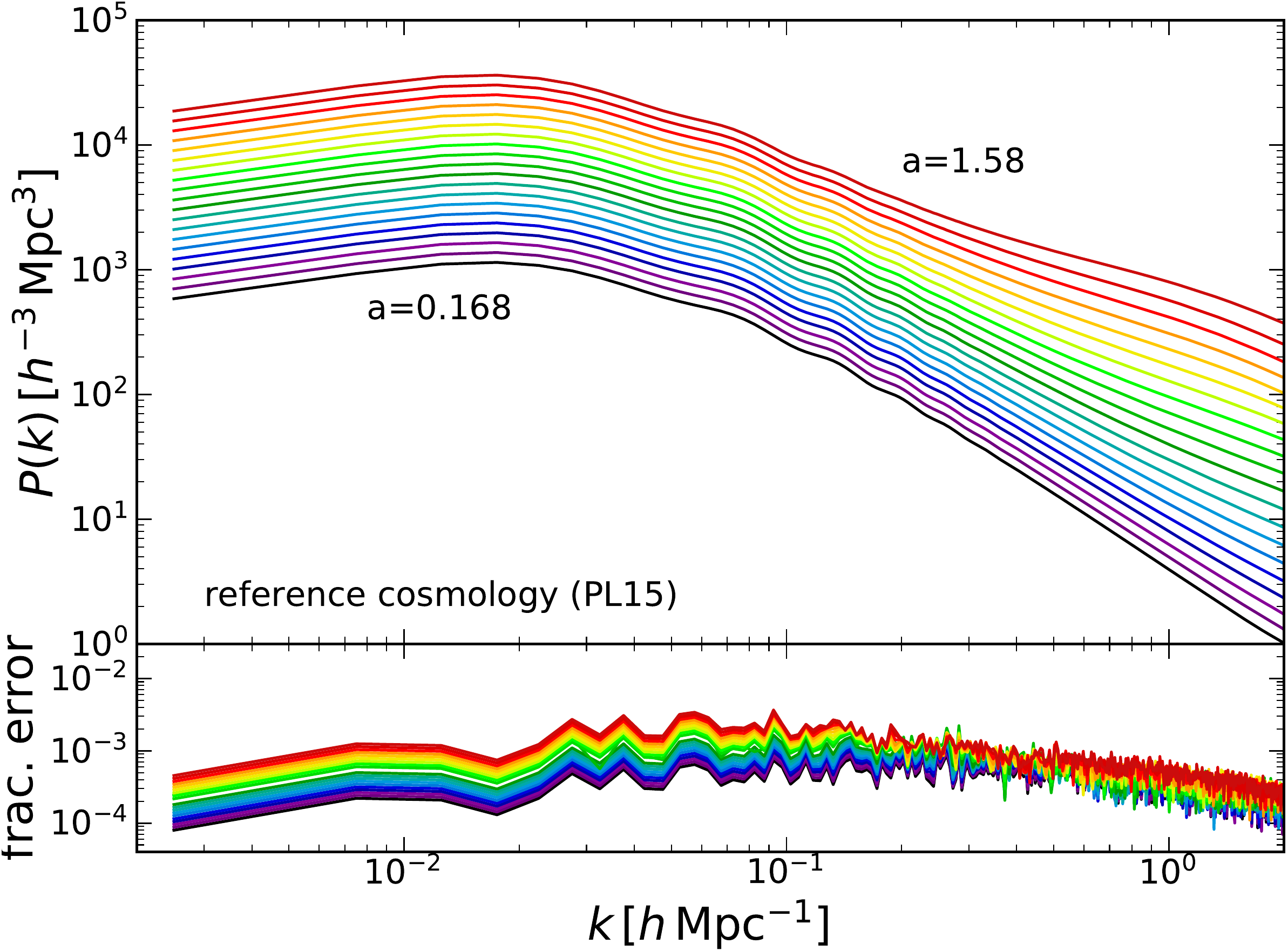}
   \caption{Nonlinear power spectra for the reference cosmological model, \texttt{PL15}. Upper panel shows the spectra at the 20 output epochs, while the lower panel shows the fractional error estimated from the scatter among realizations. The data are stored in a table, and we use them by interpolating in two dimensions, wavenumber $k$ and scale factor $a$, for the reconstruction.}
   \label{fig:pnl_ref}
\end{figure}

The set of simulations employed here uses $2,048^3$ particles in comoving cubes with a side length of $2,048\,\hiMpc$. We follow the ``paired-and-fixed'' method proposed by \cite{Angulo16} [the Angulo-Pontzen (AP) method, hereafter] that fixes the amplitude of the initial density contrast in Fourier space to the expectation value and then takes an average of a pair of such simulations with reversed initial phases to suppress the cosmic variance. We create five pairs of such simulations for the PLANCK 2015 cosmology (\cite{PLANCK15}; see \texttt{PL15} in Table~\ref{tab:cosmos} for the exact values of the cosmological parameters) and roughly estimate the statistical accuracy from the scatter among the ten (i.e., five AP pairs) simulations. Since the AP method is designed such that the difference within an AP pair efficiently cancels (this cancellation is exact at the next-to-leading perturbative order), the scatter gives an upper bound of the true error level. The error level estimated this way is indeed very small: it is always less than $1\%$ of the signal over all the redshifts up to wavenumber $k=1\hMpci$, with a weak increasing trend with time (see the lower panel of Fig.~\ref{fig:pnl_ref}).

The simulation parameters, other than the number of particles and the box size, are scaled appropriately from the low-resolution suit used to measure the response function: the initial redshift is chosen to keep the rms displacement in unit of the mean inter-particle separation ($25\%$), and the softening scale is $5\%$ of the same distance. We save $20$ snapshots for every factor $1.2$ in linear growth factor squared starting at $a=0.168$ (at which $D_+(a)=1.2^{-17/2}=0.212$, where $D_+$ is normalized to be unity at $a=1$).  The simulations are run until we reach $a=1.58$ ($D_+(a)=1.2$) in order to provide the reference power spectra for reconstruction in Sec.~\ref{sec:recon}. These outputs after $a=1$ are indeed needed when we perform a reconstruction of spectra for higher-amplitude (i.e., $\sigma_8$) models. 

The relatively dense sampling in time, together with the smooth time evolution, allows us to interpolate the measured nonlinear power spectrum over time and scale quite easily and accurately. We adopt the cubic spline interpolation over wavenumber $k$ and scale factor $a$ to provide a model template at any scale and time covered by our reference simulations (see the top panel of Fig.~\ref{fig:pnl_ref}).  Indeed we test the interpolation accuracy by using only every two outputs to perform the same interpolation and see the agreement with the unused data points. The interpolation error is found to be at most $0.2\%$ level. 

We also perform simulations for other cosmological models as listed in Table~\ref{tab:cosmos}. We generate only one pair of AP realizations for each of these cosmologies, and the snapshots are dumped only at $z=3, 2, 1, 0.5$ and $0$ (we store only at $z=0.5$ for the runs, \texttt{low-ns} and \texttt{high-ns}). These simulations are used to test the accuracy of the reconstruction starting from the PLANCK 2015 cosmology.

In the measurement of the power spectrum, we apply the aliasing correction based on interlacing \cite{Sefusatti16} to
provide a better accuracy control around the Nyquist wavenumber. With a CIC density assignment on $1280^3$ grid points, the
alias-corrected spectrum at $z=0$ agrees with simulations with even higher resolution started at a 
higher redshift ($N=2048^3$, $L=1000\hiMpc$ and $z_\mathrm{in}=59$; Nishimichi et al. in preparation)
within $1\%$ down to the Nyquist frequency $\sim1.96\hMpci$.

\subsection{Single-step reconstruction}
\label{subsec:single_step}

\subsubsection{Method}
\label{subsubsec:single_step_method}

In Eq.~(\ref{eq:recon1}), one further needs to specify the cosmological model at which we compute the response function $K(k,q)$. Roughly speaking, it should be evaluated at a cosmological model somewhere in between the fiducial and the target model. As one of the simplest examples, we consider that the response function is given at the fiducial model $\bfp_0$ (i.e., \texttt{PL15}), and always use this irrespective of the target model $\bfp_1$ in this subsection.

Another point to be clarified is the redshift at which we evaluate the fiducial nonlinear power spectrum. One does not necessarily evaluate the spectrum at the same redshift as the one for the target model. Moreover, one might be able to find a better redshift at which the two spectra are closer. The guideline to find the optimal redshift is thus to minimize the difference of the two linear power spectra, $\Plin(q,z_1;\bfp_1)-\Plin(q,z_0;\bfp_0)$. 

\begin{figure*}[!ht]
   \centering
   \includegraphics[width=15.1cm]{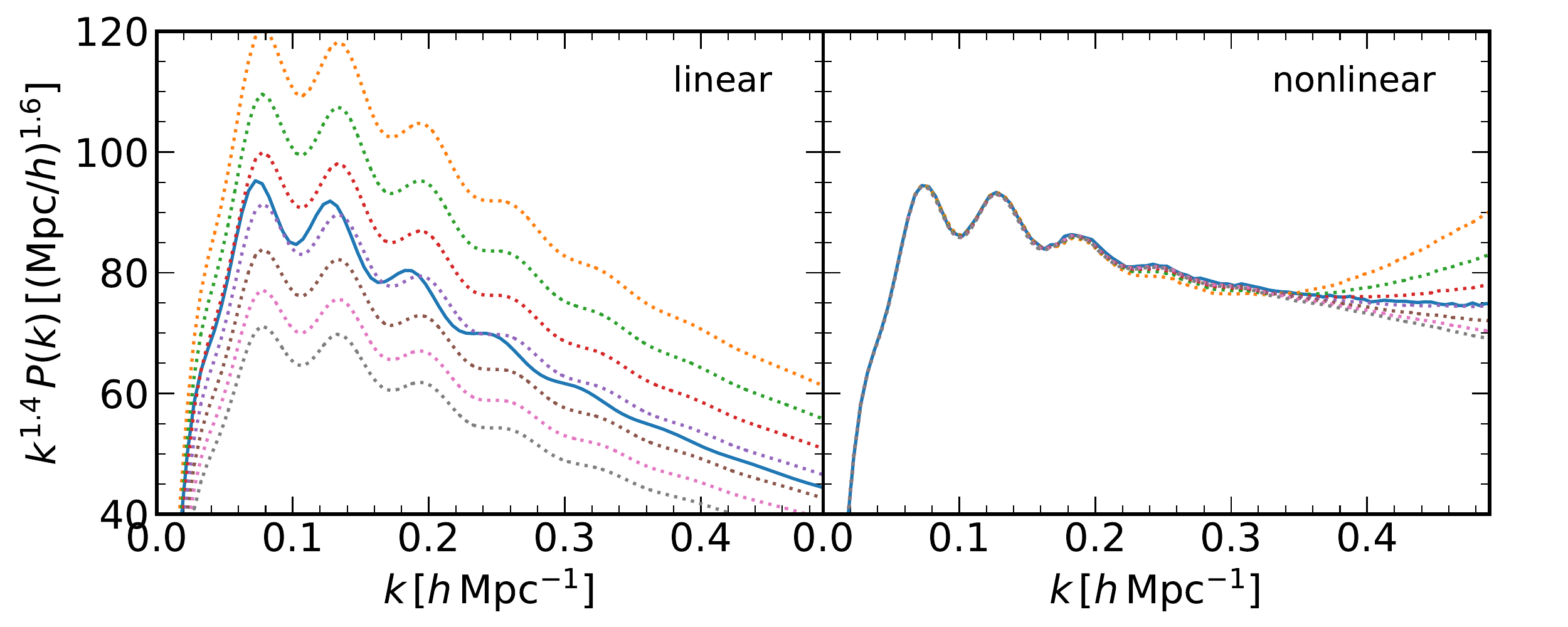}
   \caption{Dependence of the reconstructed power spectrum on the redshift at which the fiducial spectrum template is evaluated. Left: linear power spectra for the target (solid; \texttt{WM5} at $z=1$) and the fiducial (dotted; \texttt{PL15} from $z=0.7$ to $1.3$ in every $0.1$ from top to bottom) model. Right: nonlinear power spectra for the target cosmology, reconstructed (dotted; starting from the fiducial model shown in the left panel) and the direct simulation result (solid). The order of the reconstructed spectra on high $k$ is the same as in the left panel.}
   \label{fig:optimal_redshift}
\end{figure*}

Figure~\ref{fig:optimal_redshift} shows how the different choice of redshift $z_0$ affects the reconstruction. Quite notably, a wide range of $z_0$ (from $0.7$ to $1.3$; note the variety in amplitude of the linear power spectrum in the left panel) gives successful reconstruction on large scales (i.e., up to $k\sim 0.3\,\hMpci$). The accuracy of the reconstruction is rather insensitive to the choice of redshift on such large scales. As expected, the closest linear spectrum (evaluated at $z_0=1$; the fourth from the top among the seven dotted lines in the left panel) leads to the best result in the right panel (almost on top of the solid curve).

To obtain a reasonable choice of $z_0$ for various cosmological models, we implement this point as follows. We first define the matching wavenumber, $q_\mathrm{mat}$, up to which we match the fiducial and target linear power spectra. We define this as the upper bound in the integral
\begin{eqnarray}
\int_0^{q_\mathrm{mat}} \Delta_{\mathrm{lin}}^2(q)\,\mathrm{d}\ln q = C,
\label{eq:qmatch}
\end{eqnarray}
where $\Delta_{\mathrm{lin}}^2(q) = q^3\Plin(q)/(2\pi^2)$ is the dimensionless linear power spectrum.
This integral gives us the amplitude of power on scales larger than $q_{\mathrm{mat}}$. As is already clear from the measured response function, the dominant mode transfer appears from larger to smaller scales. We try to fix the amount of the transfer from scales up to a certain wavenumber to be a constant $C$ both in the fiducial and the target model to determine $z_0$.

In practice, we first perform this integral for the \textit{target} linear power spectrum (evaluated at the redshift of evaluation, $z_1$) to determine $q_\mathrm{mat}$ for a given value of $C$. Then, with the same value of $q_\mathrm{mat}$, we search for the redshift $z_0$ at which the integral evaluated for the \textit{fiducial} linear power spectrum exactly gives $C$. For the value of $C$, we have made several tests in different cosmological models and at redshifts, and find that $C=0.25$ is a reasonable choice, with which the reconstruction method gives an accurate prediction for all of the cases we have examined (see what follows).

The reconstruction procedure is now unique. That is, once $z_0$ is determined with the procedure above, we then calculate the fiducial nonlinear power spectrum, $\Pnl(k,z_0;\bfp_0)$, from the spline interpolator of the simulation template. The response function, $K(k, q)$ is then computed at the same fiducial model according to the PT-based prescription in Eq.~(\ref{eq:our_model}) at the same redshift $z_0$. Plugging these two quantities into Eq.~(\ref{eq:recon1}), the rest is to perform the one-dimensional integration, which can be done quickly, to finally obtain the nonlinear power spectrum at the target cosmology, $\Pnl(k,z_1;\bfp_1)$. 

\subsubsection{Results}
\label{subsubsec:single_step_results}

\begin{figure*}[!ht]
   \centering
   \includegraphics[width=8.2cm]{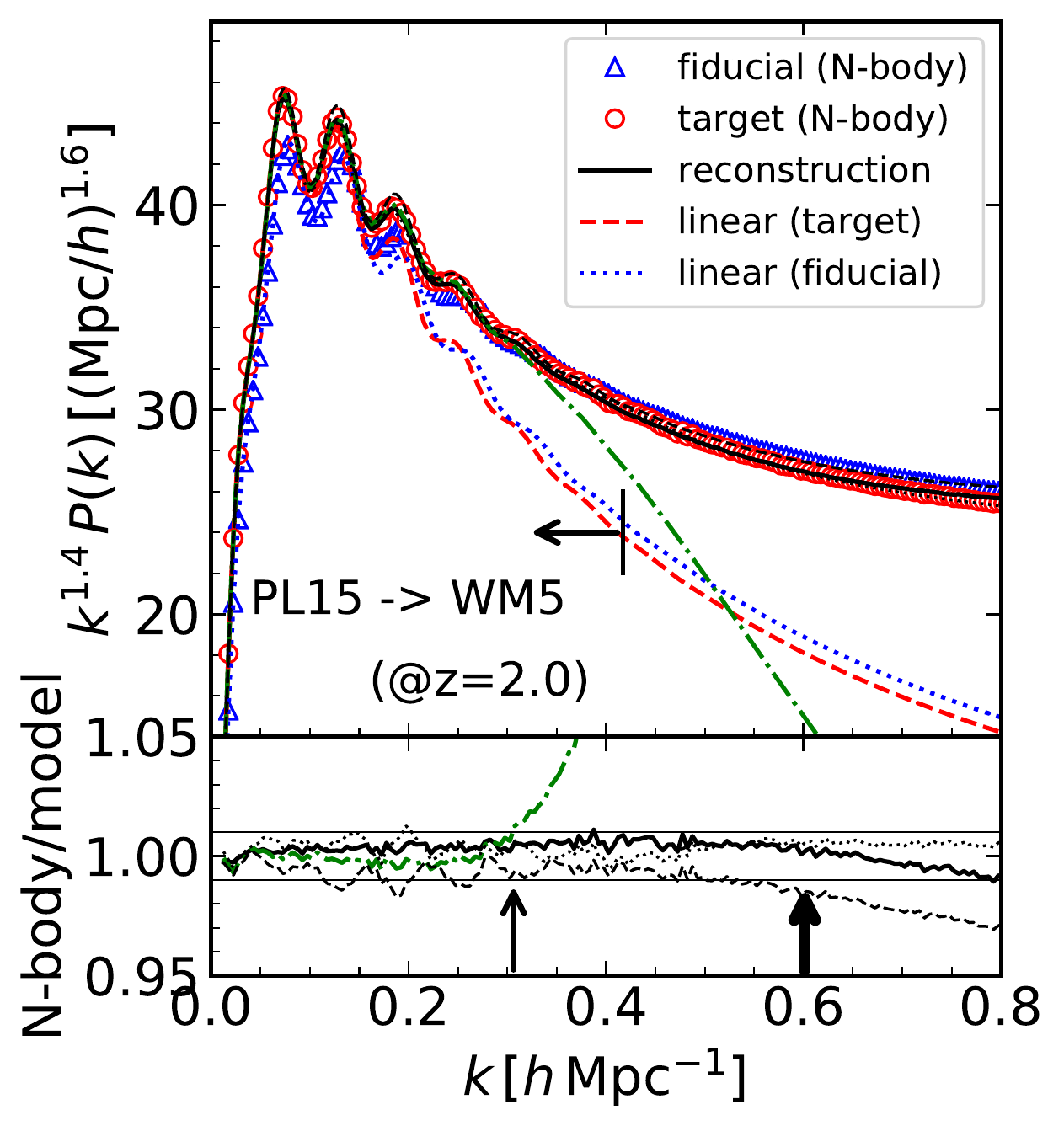}
   \includegraphics[width=8.2cm]{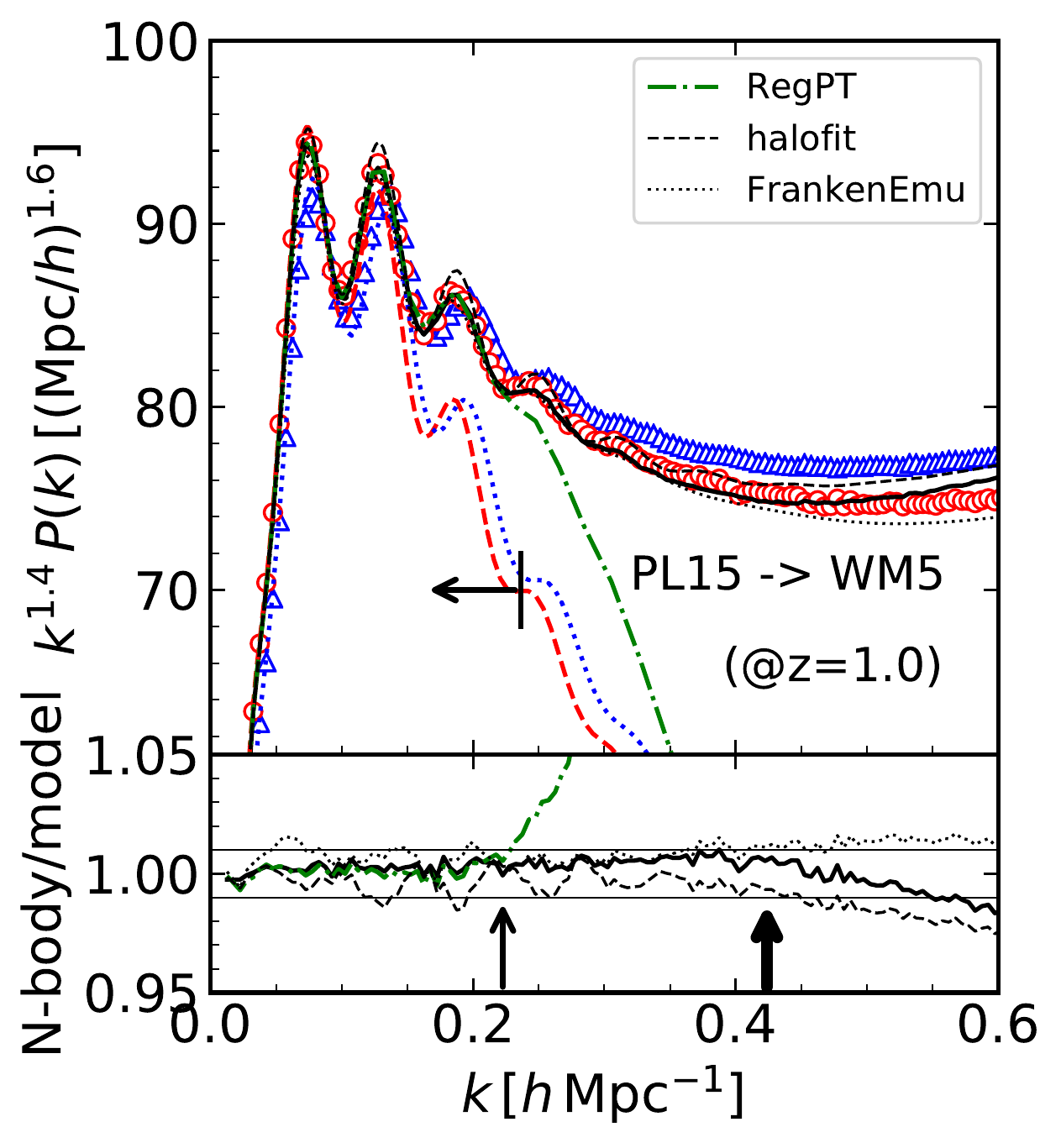}
   \includegraphics[width=8.2cm]{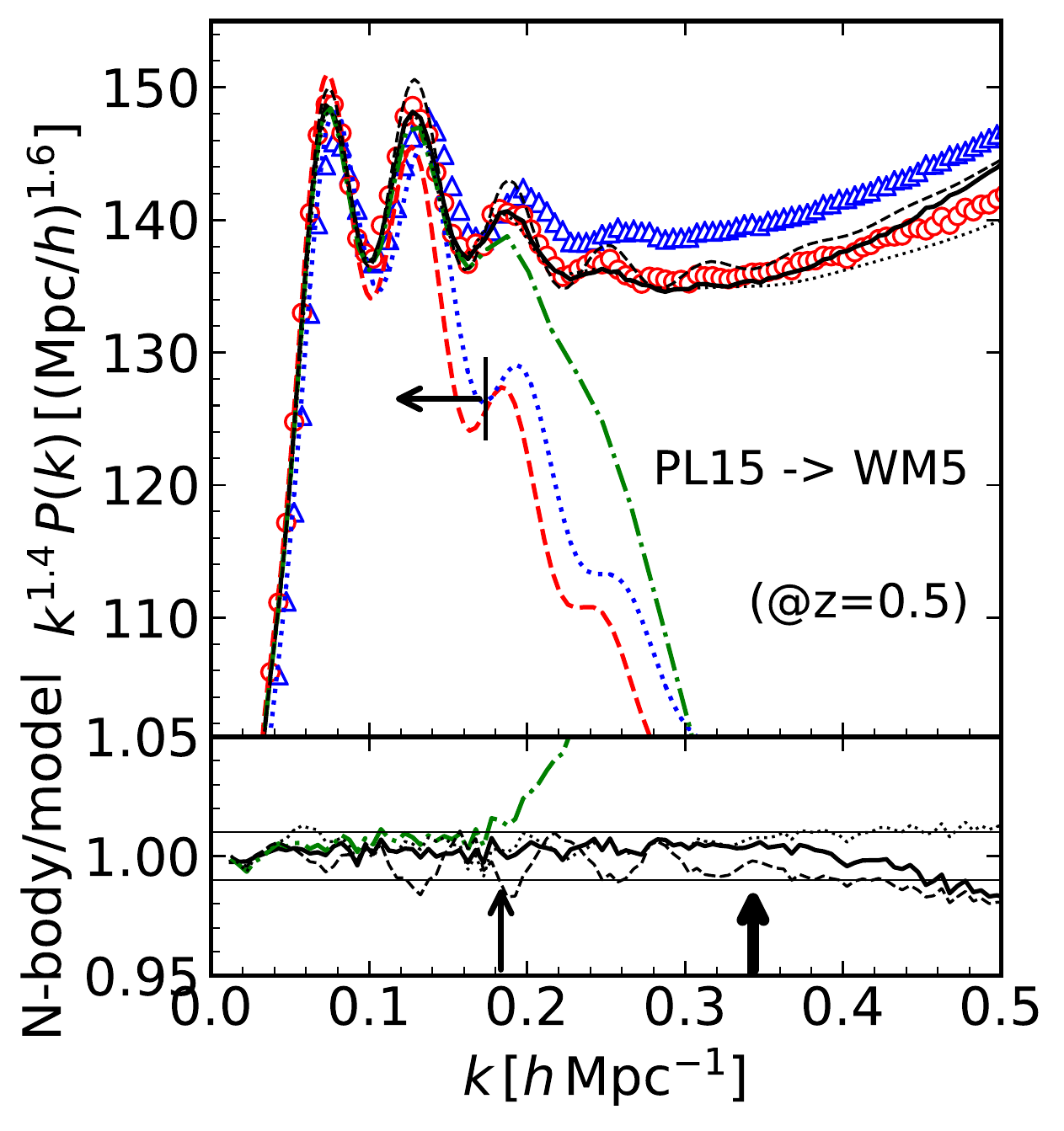}
   \includegraphics[width=8.2cm]{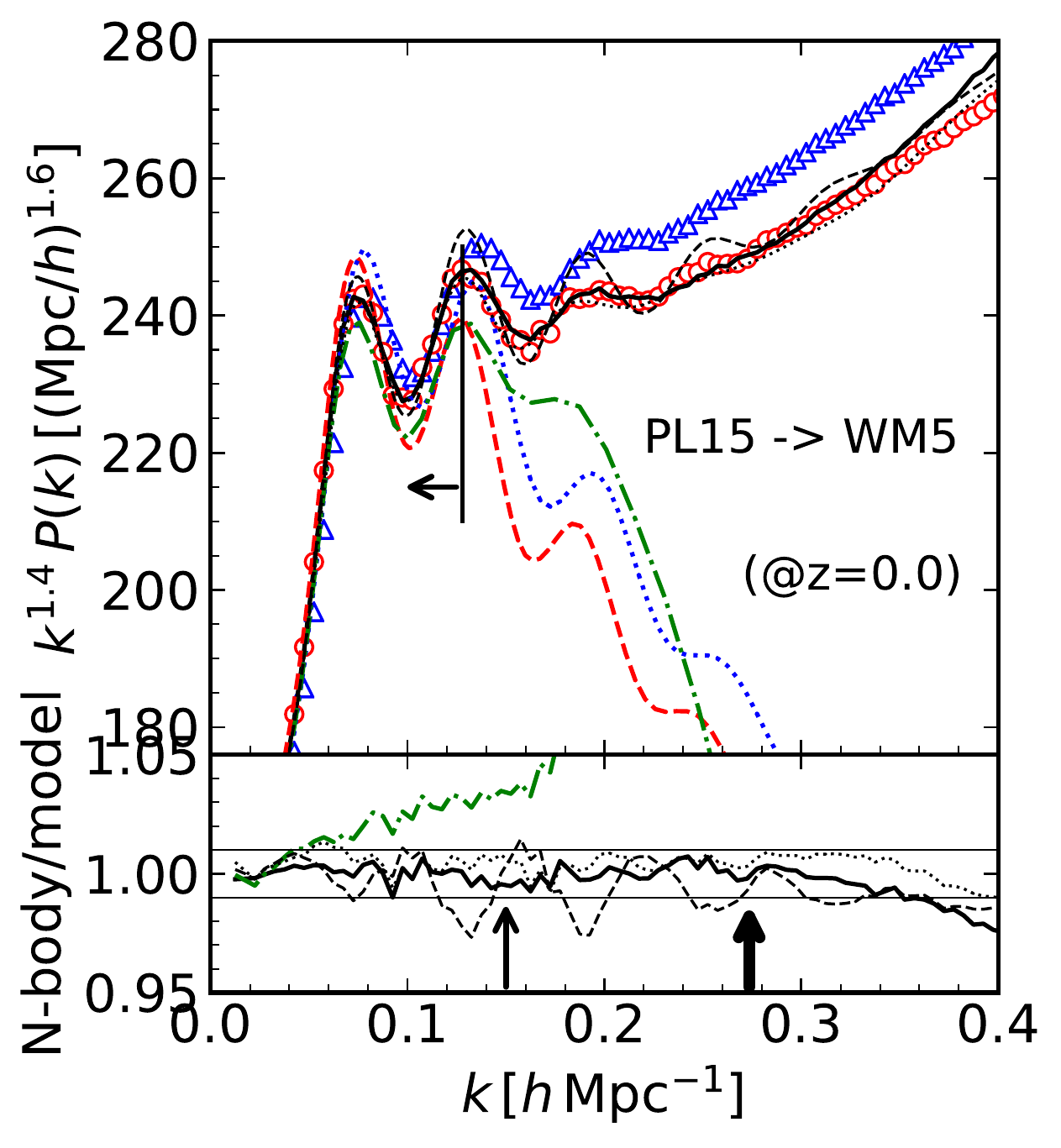}
   \caption{Reconstruction of the nonlinear power spectrum for a nearby cosmology. Upper: we start from \texttt{PL15} cosmological model as the fiducial cosmology (triangles; see Table~\ref{tab:cosmos}) and perform the reconstruction of the nonlinear power spectra for \texttt{WM5} model (target model) at different redshifts ($z=2, 1, 0.5$ and $0$ from top left to bottom right). The reconstructed spectrum is plotted by the solid curve, while the direct simulation result at the target model is shown by the circles. We also show the RegPT prediction for the target power spectrum at the two-loop order (dot-dashed).
The linear power spectra for the fiducial (dashed) and target (dotted) models are also plotted. The horizontal arrow indicates the wavenumber range over which the amplitude of the linear power spectra is matched [see Eq.~(\ref{eq:qmatch}) and text for detail]. Lower: we show the ratio of the measured power spectrum to the model predictions at the target cosmology. The two vertical arrows show the location of the estimated maximum reliable wavenumber for the model predictions, RegPT (thin; $\alpha_\mathrm{max}=0.25$) and reconstruction (thick; $\alpha_\mathrm{max}=1$).  In addition to the predictions based on the reconstruction and the RegPT, we show the ratio to other nonlinear prescriptions in the literature by the thin lines in this panel: halofit (dashed) and FrankenEmu (dotted). The two horizontal solid lines mark the $\pm 1\%$ accuracy range.}
   \label{fig:reconstruction_wmap5}
\end{figure*}

\begin{figure*}[!ht]
   \centering
   \includegraphics[width=8.2cm]{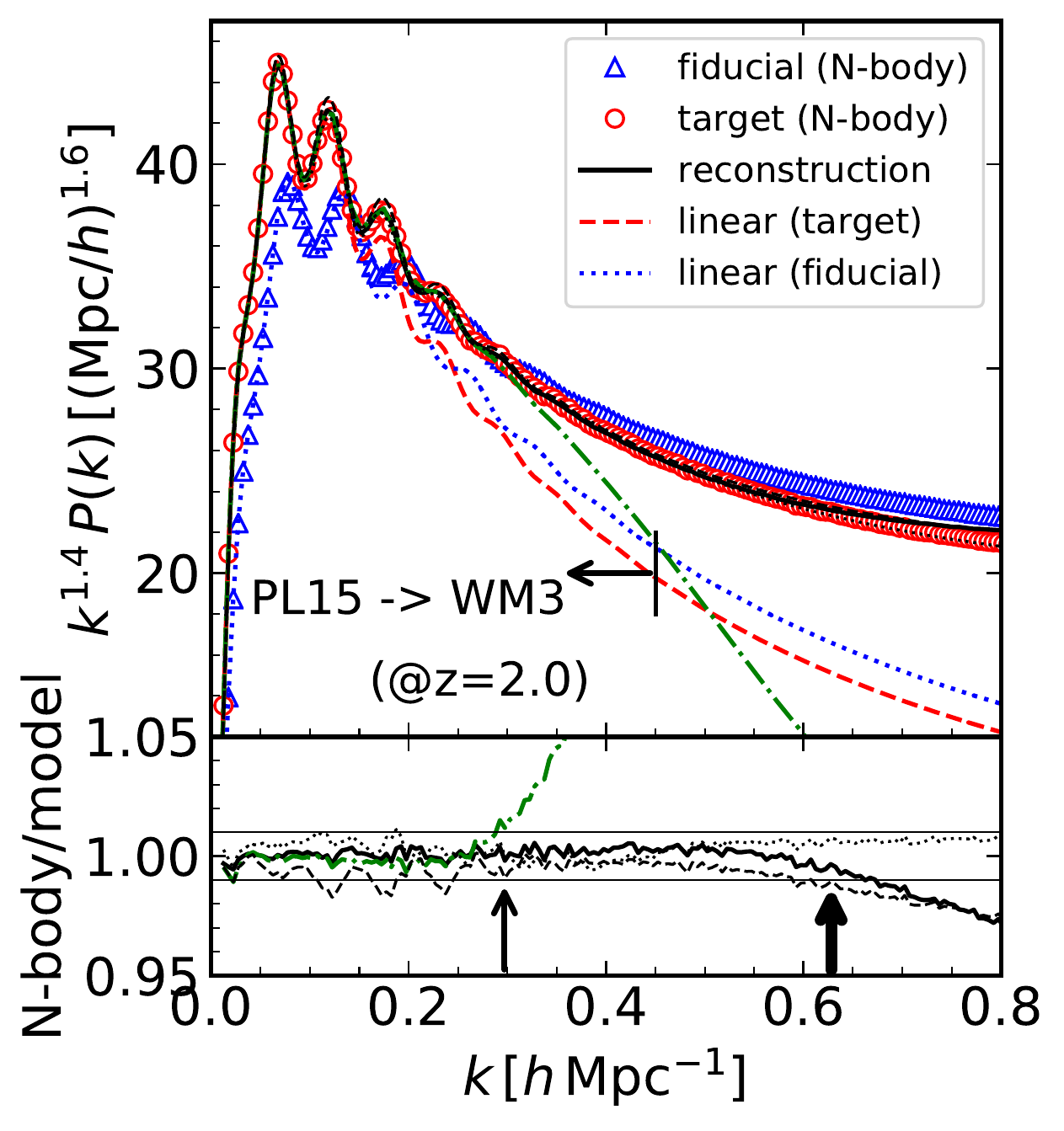}
   \includegraphics[width=8.2cm]{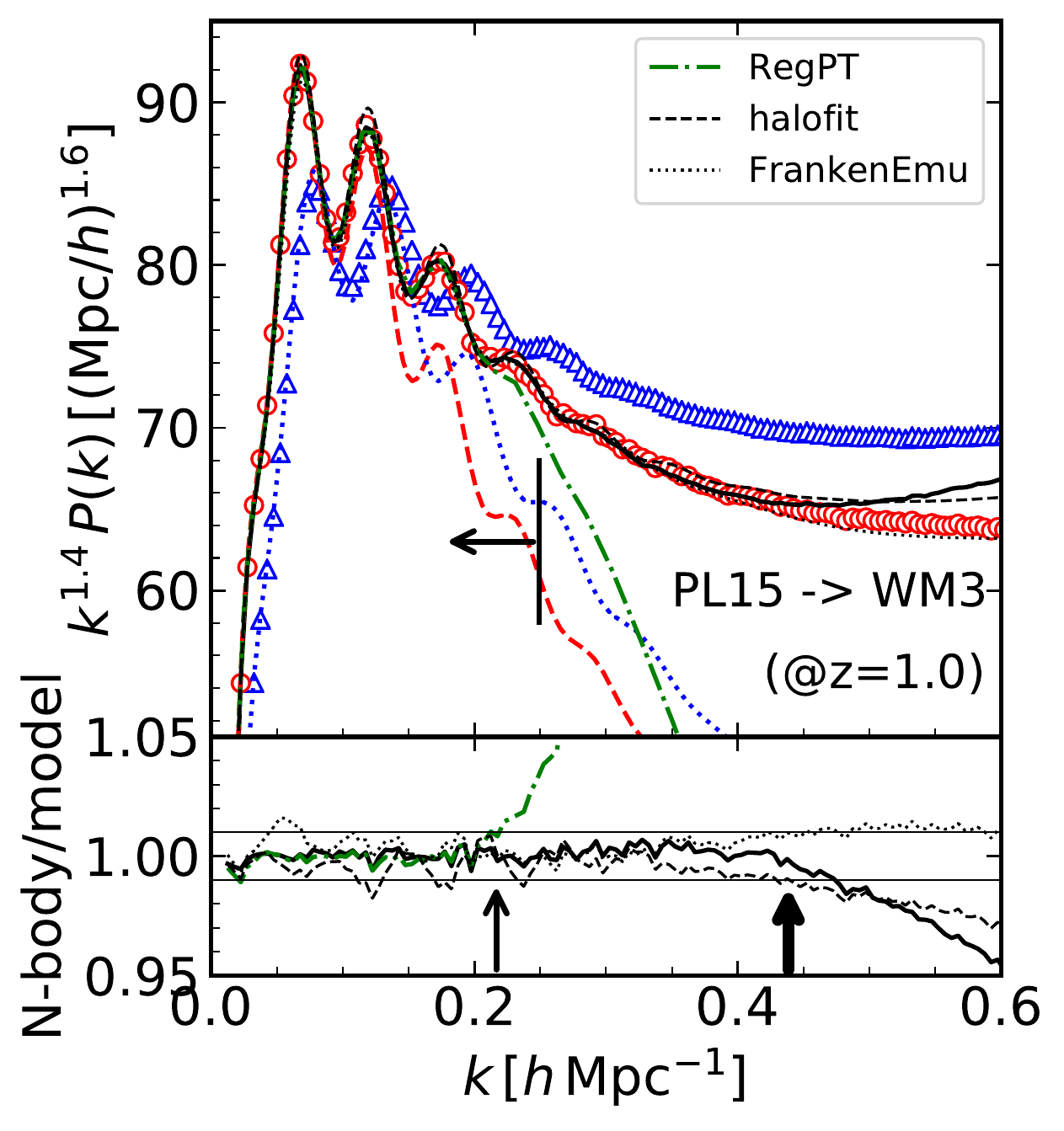}
   \includegraphics[width=8.2cm]{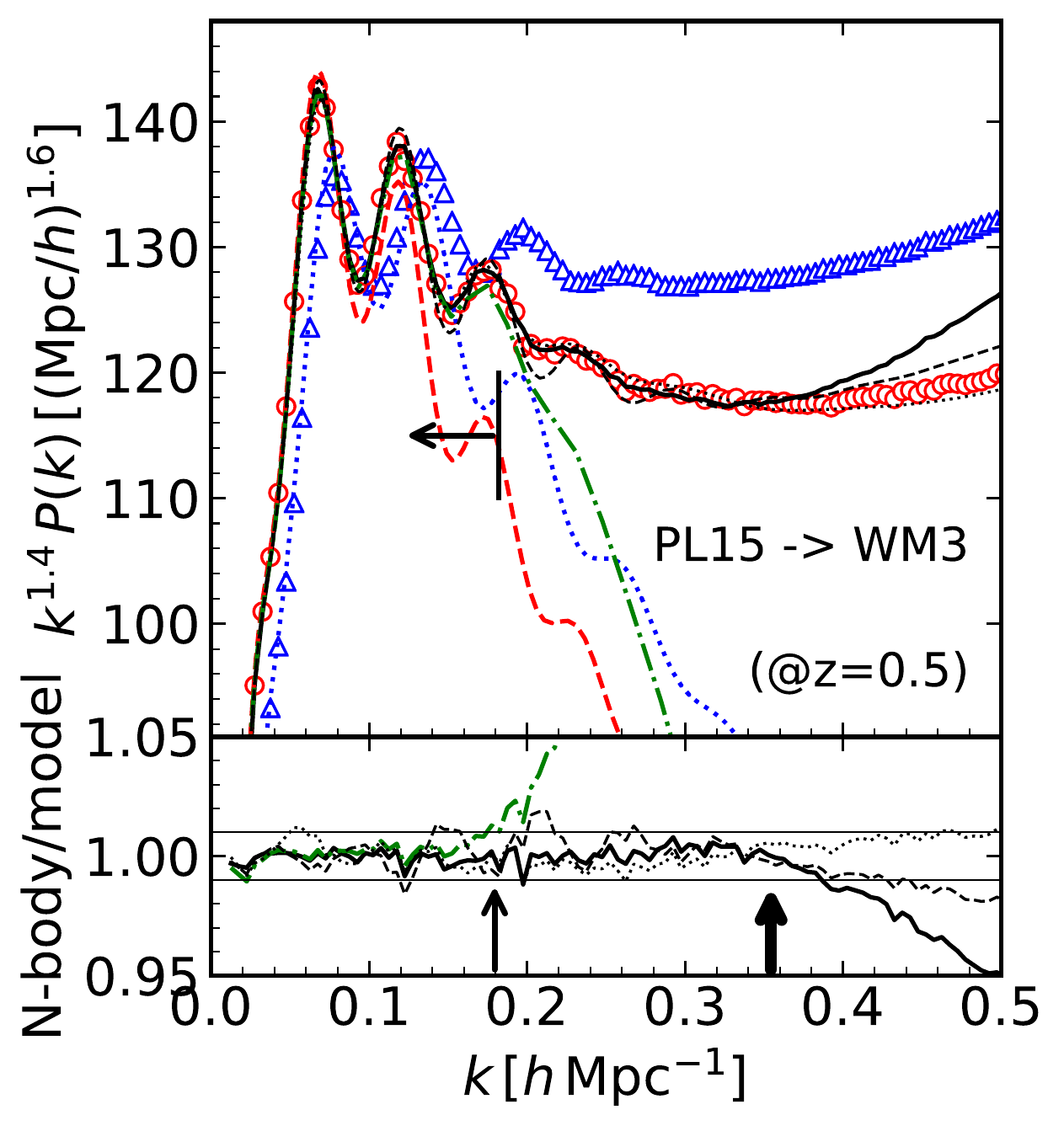}
   \includegraphics[width=8.2cm]{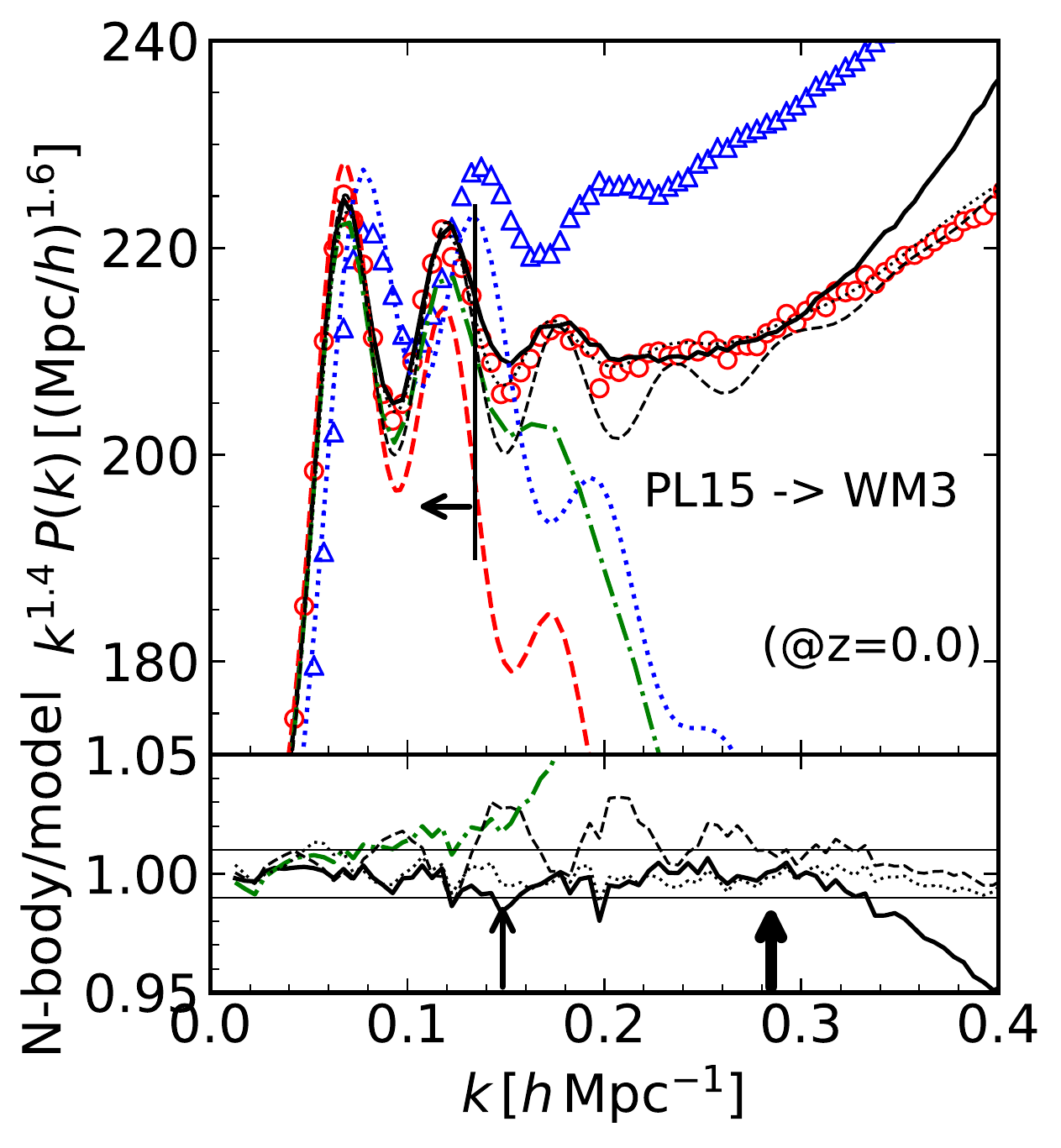}
   \caption{Same as in Fig.~\ref{fig:reconstruction_wmap5}, but for \texttt{WM3} cosmology as the target model.}
   \label{fig:reconstruction_wmap3}
\end{figure*}

We now perform the reconstruction based on the procedures explained in \S\ref{subsubsec:single_step_method}. The results are shown for two different cosmological models in Fig.~\ref{fig:reconstruction_wmap5} (\texttt{PL15} to \texttt{WM5}; see Table~\ref{tab:cosmos} for their parameters) and Fig.~\ref{fig:reconstruction_wmap3} (\texttt{PL15} to \texttt{WM3}). 

One important point that we should address is to what extent we can trust the results of reconstructed power spectrum. This involves two different aspects: first, the two linear power spectra should be close enough such that the higher-order corrections are negligible. This is already visible in Fig.~\ref{fig:optimal_redshift}, in which we have tested the normalization of the fiducial model. When the fiducial and the target linear spectra are closer, the reconstructed spectrum is in a better agreement with the direct simulation result up to a larger wavenumber. While we can find $z_1$ that gives a quite similar linear power spectrum to the target in this particular example, this is not always the case especially when the cosmological parameters are significantly different. The second is that the accuracy of the model~(\ref{eq:our_model}) is worse on higher $k$ and lower $z$ as discussed in \S~\ref{subsec:validity}. We postpone the discussion on the reconstruction error originating from the former to the subsequent subsection and only discuss the latter point here.

The limiting factor for the accurate reconstruction of nonlinear power spectrum is in the prescription of the response function. To be precise, the accuracy of the response function relies on the phenomenologically introduced damping factor $D(\beta_{k,q})$.  While this helps us to have a well-behaved response function on intermediated scales, it eventually erases the response structure in the high $k$ regime. This situation is exactly the same as the prediction of the nonlinear power spectrum based on the multi-point propagator expansion; the exponential damping in the propagators leads to an unrealistic suppression of power on small scales. 

As discussed in \S~\ref{subsec:validity}, the condition, $\alpha_{k_\mathrm{max}}=\alpha_\mathrm{max}$ with $\alpha_\mathrm{max}$ order unity provides a reasonable guess of $k_\mathrm{max}$ for a successful prediction of the response function. We compute the value of $k_\mathrm{max}$ for the target linear power spectrum and check the reconstruction accuracy around this $k_\mathrm{max}$. We find that $k_\mathrm{max}$ with $\alpha_\mathrm{max}=1$ is a conservative estimate for the reconstructed spectra to have $\sim1\%$ accuracy. The thick vertical arrows in the lower panel of Figs.~\ref{fig:reconstruction_wmap5} and \ref{fig:reconstruction_wmap3} show the locations of this wavenumber. While the ratio is within the one-percent band even beyond $k_\mathrm{max}$ thus estimated for the \texttt{WM5} model, it quickly goes away from the band for the \texttt{WM3} model soon after $k$ exceeds $k_\mathrm{max}$. This is because the former is closer to the fiducial \texttt{PL15} model than the latter, and we need to compute only a very small correction to the fiducial nonlinear power spectrum around that wavenumber range.

In Figs.~\ref{fig:reconstruction_wmap5} and \ref{fig:reconstruction_wmap3}, we also plot by the dot-dashed line the direct RegPT calculation at the two-loop level without any reconstruction involved. One can see that the prediction starts to depart from the simulation data much earlier than the reconstruction method at a lower value of $k$. We show by the thin vertical arrows the values of $k_\mathrm{max}$ obtained with $\alpha_\mathrm{max}=0.25$, which serves as a good indicator of the breakdown of RegPT except at $z=0$ where the prediction gets worse at a wavenumber lower than this. The gain of the reconstruction over RegPT is about twice in the maximum wavenumber $k_\mathrm{max}$. 

Further, we show in the bottom panel of Figs.~\ref{fig:reconstruction_wmap5}~and~\ref{fig:reconstruction_wmap3} the ratio of the $N$-body simulations to two other prescriptions for the nonlinear power spectrum. They are halofit \cite{2003MNRAS.341.1311S} with the parameters recently refined by \cite{2012ApJ...761..152T} (dashed) and FrankenEmu \cite{FrankenEmu} (dotted).　While these models broadly agree with our simulation data, mostly within the $1\%$ band shown by the two horizontal solid lines, our model exhibits some preferable feature over these models: it shows a more stable performance on larger scales and at higher redshifts. This is thanks to the fact that the perturbative approach gets better when the fluctuations are smaller and our phenomenological correction plays a minor role. Therefore, our model is more useful when one tries to model the clustering signal on, e.g., BAO scale at a high redshift, while the other two might be more suitable to model the broadband shape up to larger wavenumbers deep in the non-perturbative regime.

To further test the validity of the new scheme presented here, we also compare the reconstructed power spectrum to the simulations, \texttt{low-ns} and \texttt{high-ns}, which have the spectrum index $n_s$ apart by $\pm 0.05$ from the fiducial simulations. The amplitude of the primordial spectrum for these models are set to keep the present-day linear amplitude, $\sigma_8$. Note that the value of $n_s$ has already been constrained very tightly by CMB experiments, and its uncertainty is one order of magnitude smaller, $\sim 0.005$, from Planck \cite{PLANCK15} ($68\%$ C.L., assuming a flat $\Lambda$CDM cosmology). Thus, we are testing the scheme for a bit extreme cases here.

\begin{figure}[!ht]
   \centering
   \includegraphics[width=8.2cm]{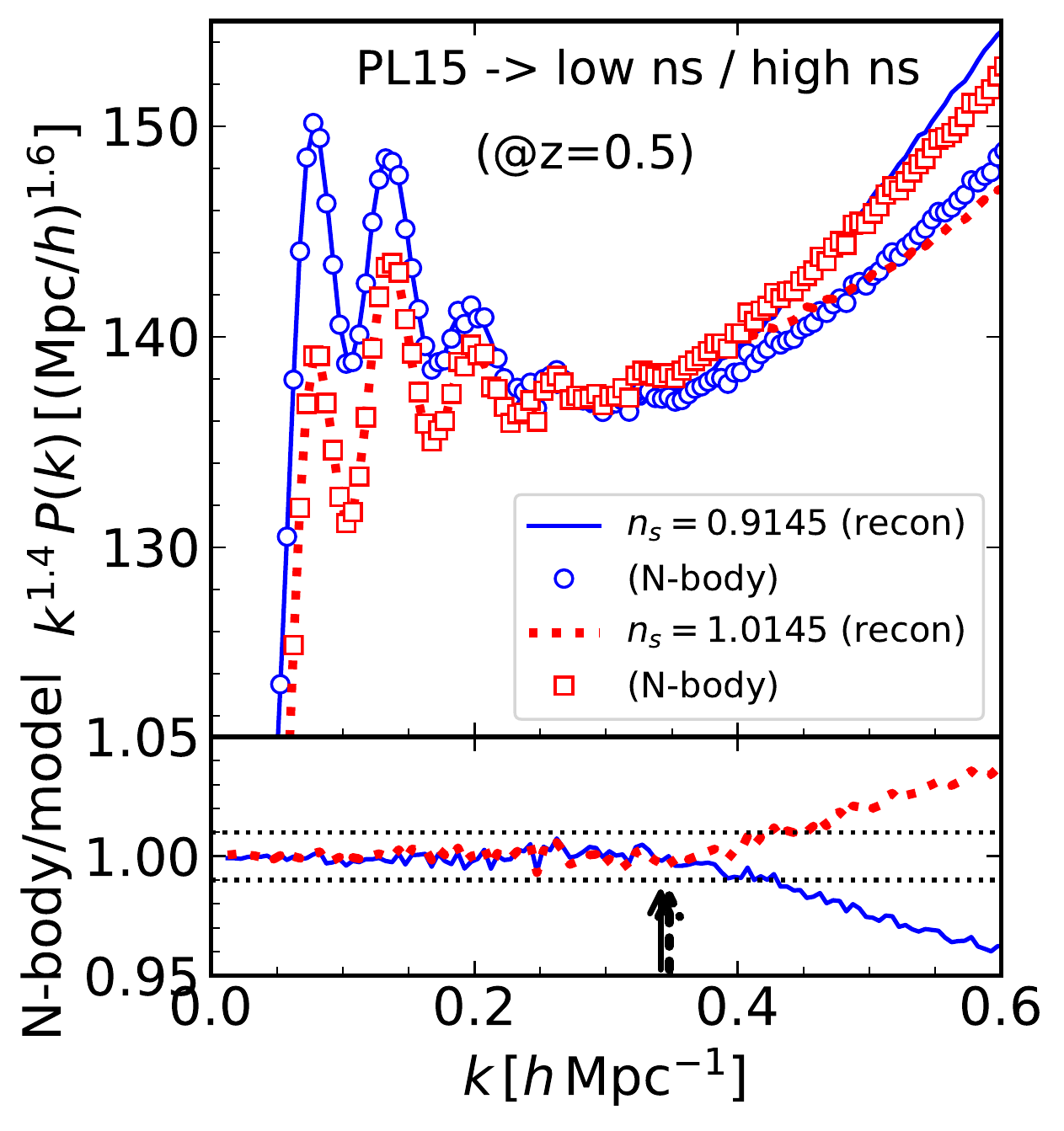}
   \caption{A single-step reconstruction to the models with different spectral indices (\texttt{low-ns} and \texttt{high-ns}) from the fiducial \texttt{PL15} model at $z=0.5$. This time, we only show the reconstructed and the simulated nonlinear power spectra of the two target models for clarity in the top panel: the solid and the circles for \texttt{low-ns}, the dashed line and squares for \texttt{high-ns}. The bottom panel is the same as before, but the ratio for the two target models are shown together with the same line type as in the top panel. The estimated limiting wavenumber is also shown by the arrows (they are almost on top of each other).}
   \label{fig:ns_varied}
\end{figure}

The comparison is made at $z=0.5$ and is shown in Fig.~\ref{fig:ns_varied}. The performance of the reconstruction is as good as the previous examples despite the rather different overall tilt. Note that we do not need, for these models, to resort to the multi-step reconstruction scheme laid out in the next subsection, and the criterion introduced there actually tells that the number of steps should be one for them. In this sense, the current uncertainty of the spectral index is such that the possible change in the linear power spectrum is well within the reach of the reconstruction scheme, which is expected to work for ``nearby'' cosmologies.

\subsection{Multi-step reconstruction with precomputed response table}
\label{subsec:multi_step}

\subsubsection{Method}
\label{subsubsec:multi_step_method}

\begin{figure*}[!ht]
   \centering
   \includegraphics[width=12cm]{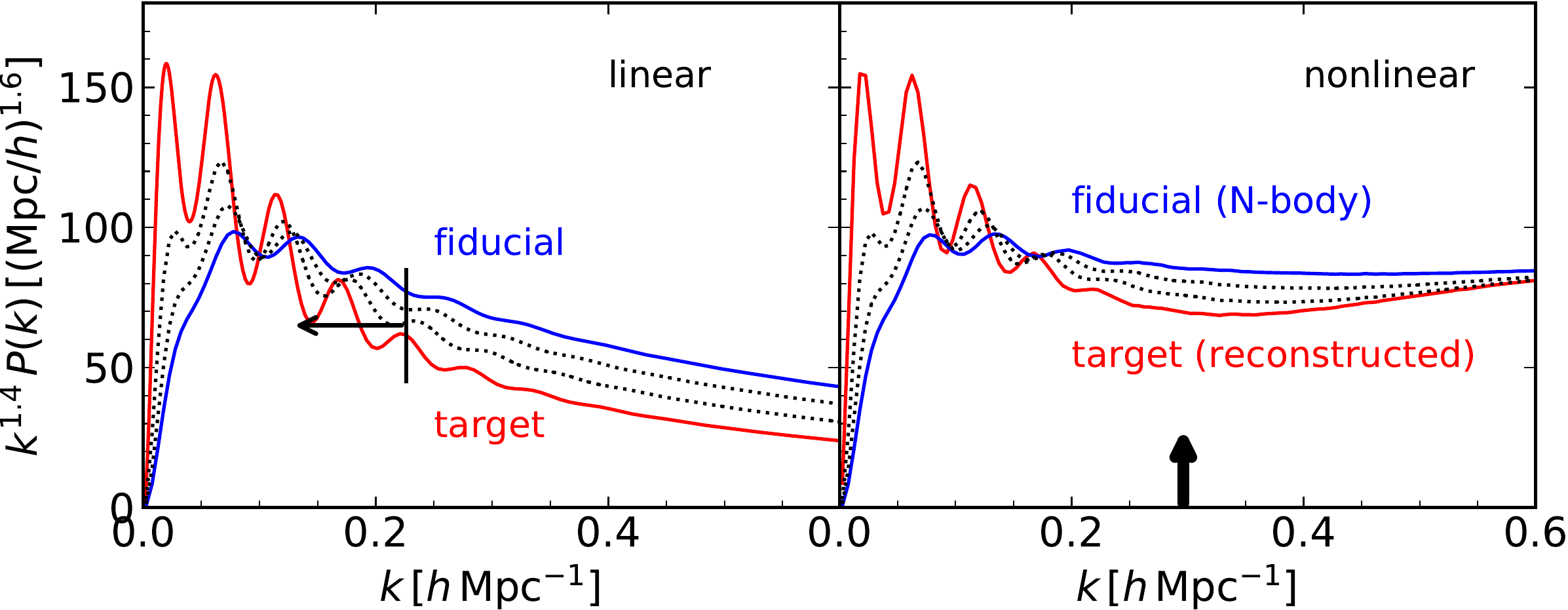}
   \caption{Linear (left) and nonlinear (right) power spectra in the multi-step reconstruction.  The two intermediate steps are depicted by the dotted curves in both panels. We match the integral~(\ref{eq:qmatch}) up to the wavenumber specified by the end point of the horizontal arrow in the left panel for all the steps. The estimated maximum wavenumber $k_\mathrm{max}$ for a successful reconstruction is marked by the vertical arrow in the right panel (see text for detail).}
   \label{fig:multi_step}
\end{figure*}

\begin{figure}[!h]
   \centering
   \includegraphics[width=8.2cm]{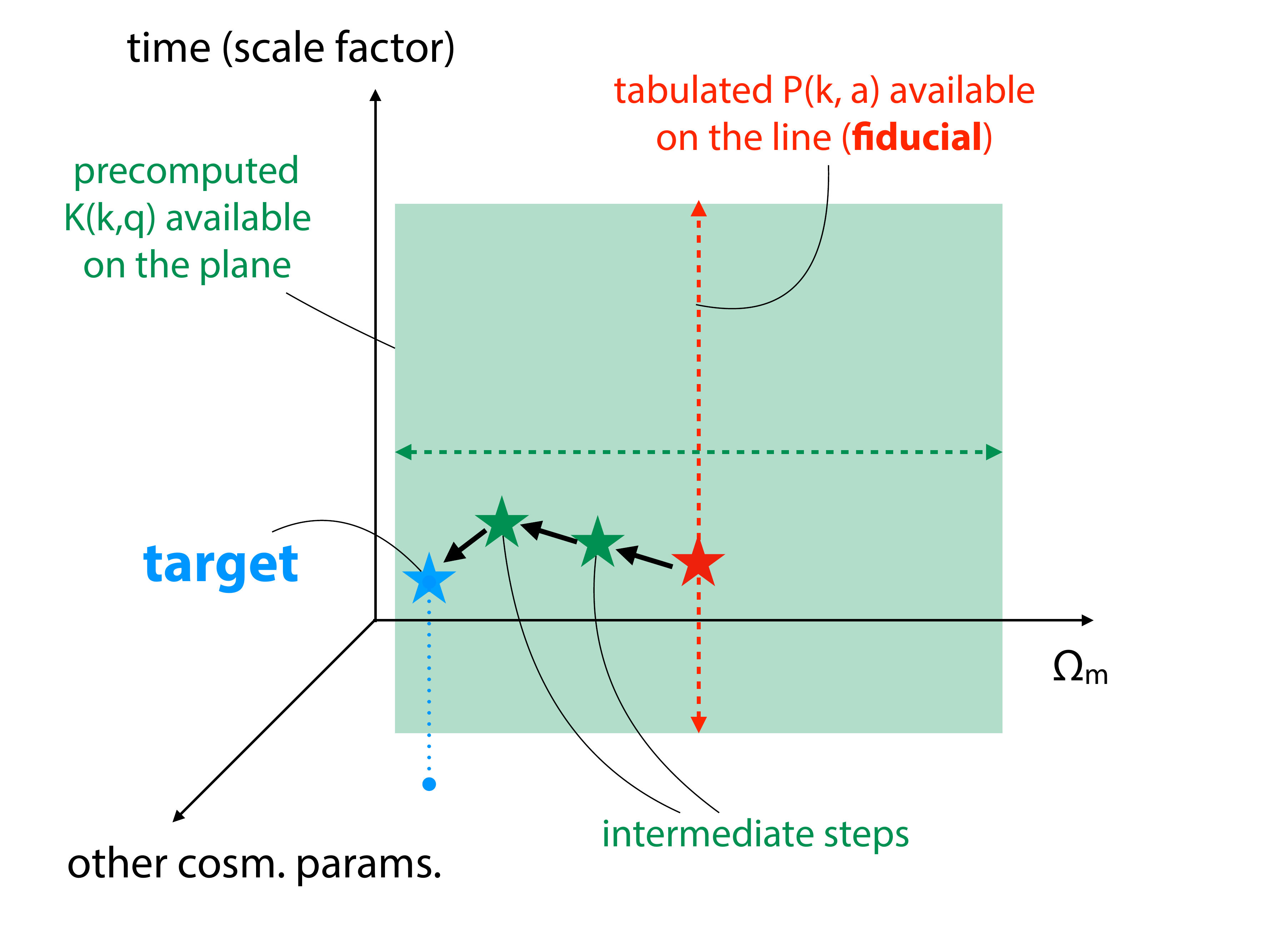}
   \caption{Schematic figure illustrating the multi-step reconstruction. We first determine the scale factor for the fiducial cosmological model from which we start the reconstruction (the rightmost star symbol in the figure), and then how many steps we take according to the distance to the target model (the leftmost star). In this example, we take two intermediate steps on the $a$ -- $\Omega_\mathrm{m}$ plane passing through the fiducial model, where the precomputed response function is available.}
   \label{fig:schematic}
\end{figure}

While the main usage of our reconstruction method is for cosmological models close to the reference \texttt{PL15} cosmology, it is still of interest to understand and extend its applicable range. Since Eq.~(\ref{eq:recon1}) is an approximation to omit the higher-order functional derivatives, we expect to see the signature of the breakdown when the difference between the fiducial and the target cosmologies get large. Indeed, we have not discussed how close the two models should be for a successful reconstruction in the previous sections. It is then useful if we can develop a method to reduce the error caused by the higher-order effects.

Here, we try to address this by employing a \textit{multiple-step} reconstruction scheme, in which we repeatedly use Eq.~(\ref{eq:recon1}) for nearby cosmologies to gradually make a transition from the fiducial to the target cosmology. 
For this purpose, we prepare tabulated files for the kernel functions relevant for the calculation of our model, such as $L^{(1)}$ or $X^{(2)}$, for different values of $\Omega_\mathrm{m}$ in the range $[0.1,0.5]$ at every $0.01$, while the other parameters are kept fixed to the values to the \texttt{PL15} model. 
These tables are interpolated again by a cubic spline function to give a smooth prediction over the above range. 
We focus on the change in $\Omega_\mathrm{m}$ because this is one of the least constrained parameters for instance in weak lensing surveys, and it can alter the shape of the power spectrum significantly. 
Typically, one obtains a much better constraint on some combination of $\sigma_8$ and $\Omega_\mathrm{m}$, and the error ellipse tend to be elongated for each of these parameters. 
Since the dependence on the redshift or the amplitude parameter in the analytical formula of the response function~(\ref{eq:our_model}) is quite trivial, our table can serve as a template for a range of cosmologies (and also redshifts), whose linear power spectrum is different in amplitude and shape, for such observational projects.

Given the tabulated response function, our next task is then to find a reasonable prescription to define the ``path'' from the fiducial to the target cosmology for the multi-step reconstruction. Figure~\ref{fig:schematic} illustrates how the multi-step reconstruction works. Here the fiducial model is shown by the vertical dashed arrow, and the starting model $\bfp_0$ for the reconstruction is picked on this line at a scale factor $a_0$ by the method described in the previous section (the rightmost star symbol in the schematic figure). Then the precomputed kernel template is provided on the plane of fixed cosmological parameters other than $\Omega_\mathrm{m}$, on which we locate some intermediate steps (we denote them as $\bfp_1$, ..., $\bfp_n$ at scale factor $a_1$, ..., $a_n$, and are shown by the two star symbols in the middle in the schematic figure). We aim to predict the nonlinear power spectrum at the target cosmology $\bfp_{n+1}$ at $a_{n+1}$ shown by the leftmost star, which is not necessarily on the aforementioned plane.

In practice, we first determine the cosmological model $\bfp_{n}$ on the plane (second star symbol from the left in the schematic figure; the last intermediate step) that is the closest to the target cosmology $\bfp_{n+1}$. In doing this, we minimize the following distance indicator:
\begin{eqnarray}
d(\bfp_{i},\bfp_{j}) = \frac{1}{q_\mathrm{mat}}\int_0^{q_\mathrm{mat}}\mathrm{d}q\left\{\frac{\Plin(q;\bfp_j) - \Plin(q;\bfp_i)}{[\Plin(q;\bfp_j) + \Plin(q;\bfp_i)] /2}\right\}^2,
\nonumber\\
\label{eq:distance_def}
\end{eqnarray}
with $q_{\mathrm{mat}}$ defined in Eq.~(\ref{eq:qmatch}). Namely, we quantify the fractional difference of the linear power spectra in the wavenumber range where we try to match the amplitude.
Then, we locate the other steps, $\bfp_1$, ..., $\bfp_{n-1}$, in between $\bfp_{n}$ and $\bfp_{0}$, on the same plane with a constant interval in $\Omega_\mathrm{m}$. Note that the scale factor $a_i$ ($i=0, ..., n$) are all determined to give the same integral~(\ref{eq:qmatch}) of the linear power spectrum as the target model $\bfp_{n+1}$ up to the wavenumber $q_\mathrm{mat}$. 

After we fix the path, we then evaluate the response function at the models between the step, $\bfp_{i+1/2}$, at which $\Omega_\mathrm{m}$ equals to the average of those at $\bfp_i$ and $\bfp_{i+1}$ when we make a step from $\bfp_i$ to $\bfp_{i+1}$ except for $i=n$. Since we have the kernel template only on a plane shown in Fig.~\ref{fig:schematic}, we simply use the response function evaluated at $\bfp_n$ for the last step from $\bfp_n$ to $\bfp_{n+1}$. The whole procedure can be summarized as
\begin{eqnarray}
&&{\Pnl}(k;\bfp_{i+1})= {\Pnl}(k;\bfp_i)+\int\dd \ln q\,K(k,q;\bfp_{i+1/2})\,\nonumber\\
&&\qquad\qquad\qquad\qquad\times\left[{\Plin}(q;\bfp_{i+1})-{\Plin}(q;\bfp_i)\right],\nonumber\\
&&\qquad\qquad\qquad\qquad\qquad\qquad (\mathrm{for}\,\,i=0,\dots,n-1)
\label{eq:recon_multi1}
\\
&&{\Pnl}(k;\bfp_{n+1})= {\Pnl}(k;\bfp_n)+\int\dd \ln q\,K(k,q;\bfp_{n})\,\nonumber\\
&&\qquad\qquad\qquad\qquad\times\left[{\Plin}(q;\bfp_{n+1})-{\Plin}(q;\bfp_n)\right].
\label{eq:recon_multi2}
\end{eqnarray}

\subsubsection{Results}
\label{subsubsec:multi_step_results}

An example three-step reconstruction can be found in Fig.~\ref{fig:multi_step}, where the linear and the nonlinear spectra are at each step are shown on the left and right panel, respectively. Here the target model (\texttt{EXT015}, see Table~\ref{tab:cosmos}) and the fiducial model (\texttt{PL15}) are plotted by the solid lines, and the two intermediate steps are by the dotted lines. On the left panel, we show by the horizontal arrow the location of the wavenumber $k_\mathrm{mat}$ below which we adjust the amplitude of the linear spectra. On the other hand, we show the expected maximum wavenumber $k_\mathrm{max}$ with $\alpha_\mathrm{max}=1$ on the right panel (see later discussion on how we determine $k_\mathrm{max}$ in the multi-step reconstruction in more detail).

\begin{figure}[ht]
   \centering
   \includegraphics[width=7.8cm]{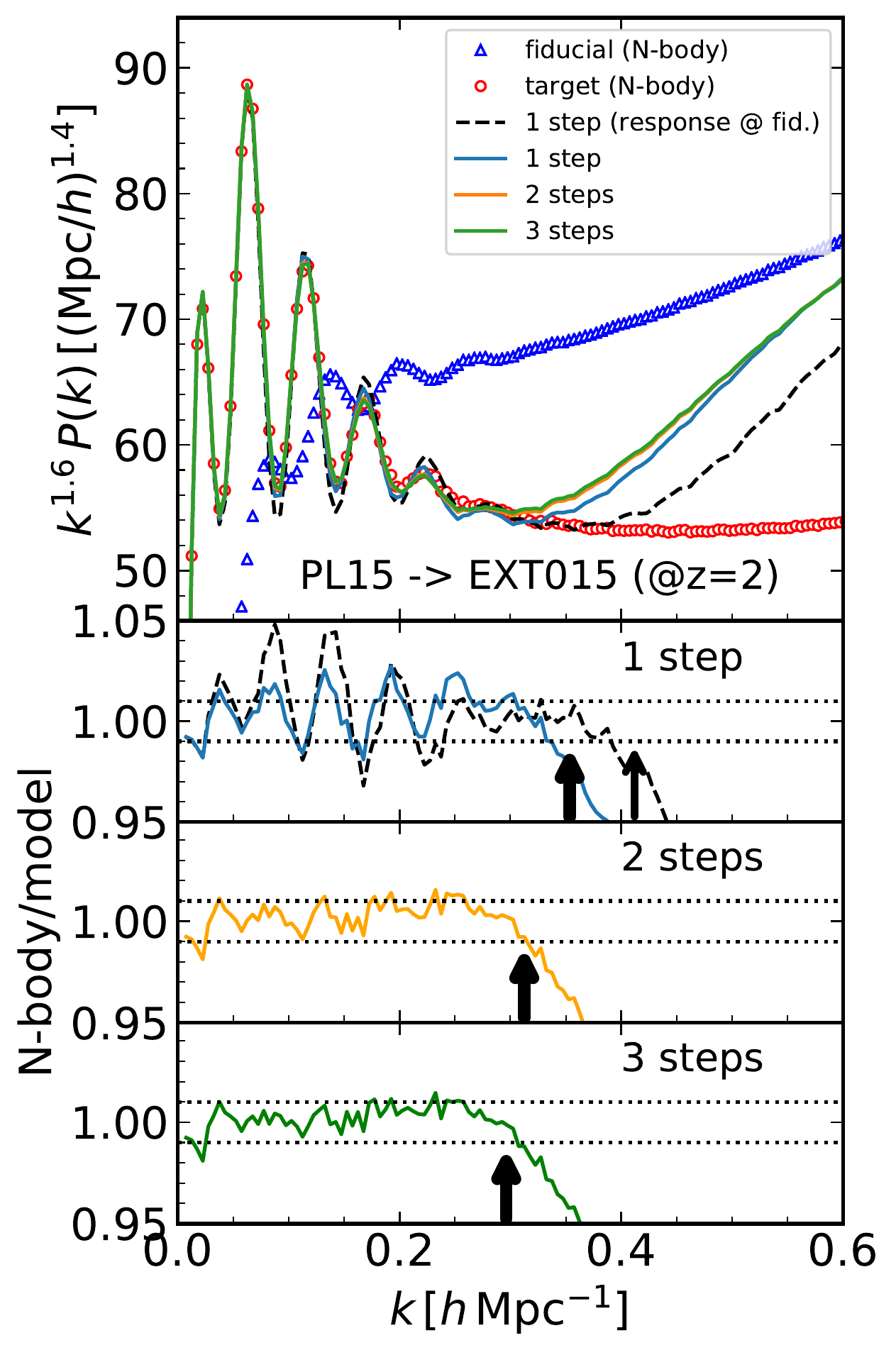}
   \caption{Multi-step reconstruction from \texttt{PL15} to \texttt{EXT015} at $z=2$ ($\sigma_8$ scaled to this redshift is $0.51$ for this model). We show the fiducial and the target nonlinear power spectra by symbols in the top panel. The results of the reconstruction with different number of steps are also shown by lines. These are almost on top of each other (i.e., solid curves), except the one with a single step and the response is evaluated at the fiducial model (dashed). We show in the lower panels the ratio of the simulated and reconstructed power spectra at the target cosmology. The horizontal dotted lines mark the $\pm 1\%$ accuracy interval.}
   \label{fig:multi_step2}
\end{figure}

We compare the reconstructed power spectra with different number of steps in Fig.~\ref{fig:multi_step2}. We consider a reconstruction from the \texttt{PL15} model to the \texttt{EXT015} model in this example, which are quite apart in the linear power spectra. In the top panel, the simplest single-step reconstruction described in the previous section is plotted by the dashed line, which should be compared to the red symbols obtained directly from simulations performed for the target cosmology. We also show in the top panel the results based on Eq.~(\ref{eq:recon_multi1}) with different number of steps (solid, almost on top of each other). Note that we do not have to perform the last step from $\bfp_n$ to $\bfp_{n+1}$ in Eq.~(\ref{eq:recon_multi2}), since this particular target model is located on the plane where the precomputed kernel template is available. Thus, we use only Eq.~(\ref{eq:recon_multi1}) to obtain the curves in this figure. The ratio of the reconstructed spectra and the direct simulation result are shown in the lower three panels with the corresponding line types.

Unlike the previous example, the ratio exhibit an oscillatory feature around unity with the amplitude reaching to $\sim 5\%$. With the response function evaluated at the intermediate cosmological model, even the single-step reconstruction works better than the previous procedure (compare the solid and the dashed curves in the second panel of Fig.~\ref{fig:multi_step2}). The oscillatory feature in the ratio in the bottom panels is significantly suppressed already by choosing a more appropriate cosmological model at which the response function is evaluated. The result gets improved with two steps but is almost the same when we further increase the number of steps, suggesting the stability of our procedure against number of steps.

We evaluate the analytical response function multiple times in this procedure, and the estimated maximum wavenumber $k_\mathrm{max}$ for an accurate prediction of the function can vary at different steps. To be conservative, we identify the final estimate of $k_\mathrm{max}$ for a successful reconstruction to the smallest one among those evaluated at every reconstruction step. In the example of Fig.~\ref{fig:multi_step2}, the location of $k_\mathrm{max}$ estimated with $\alpha_\mathrm{max}=1$ is shown by the vertical arrow in each of the lower panels. The estimated $k_\mathrm{max}$ gets smaller with increasing the number of steps in this example [note that we distinguish $k_\mathrm{max}$ for the two cases in the second panel by the thin arrow and the thick arrow, respectively for the simplest reconstruction in \S~\ref{subsec:single_step} and for the one in in this subsection with Eq.~(\ref{eq:recon_multi1})]. In all the cases, the estimated maximum wavenumber $k_\mathrm{max}$ well represent the wavenumber around which the ratio start to deviate from unity except for the oscillatory feature in the single-step reconstruction.

\begin{figure}[ht]
   \centering
   \includegraphics[width=7.8cm]{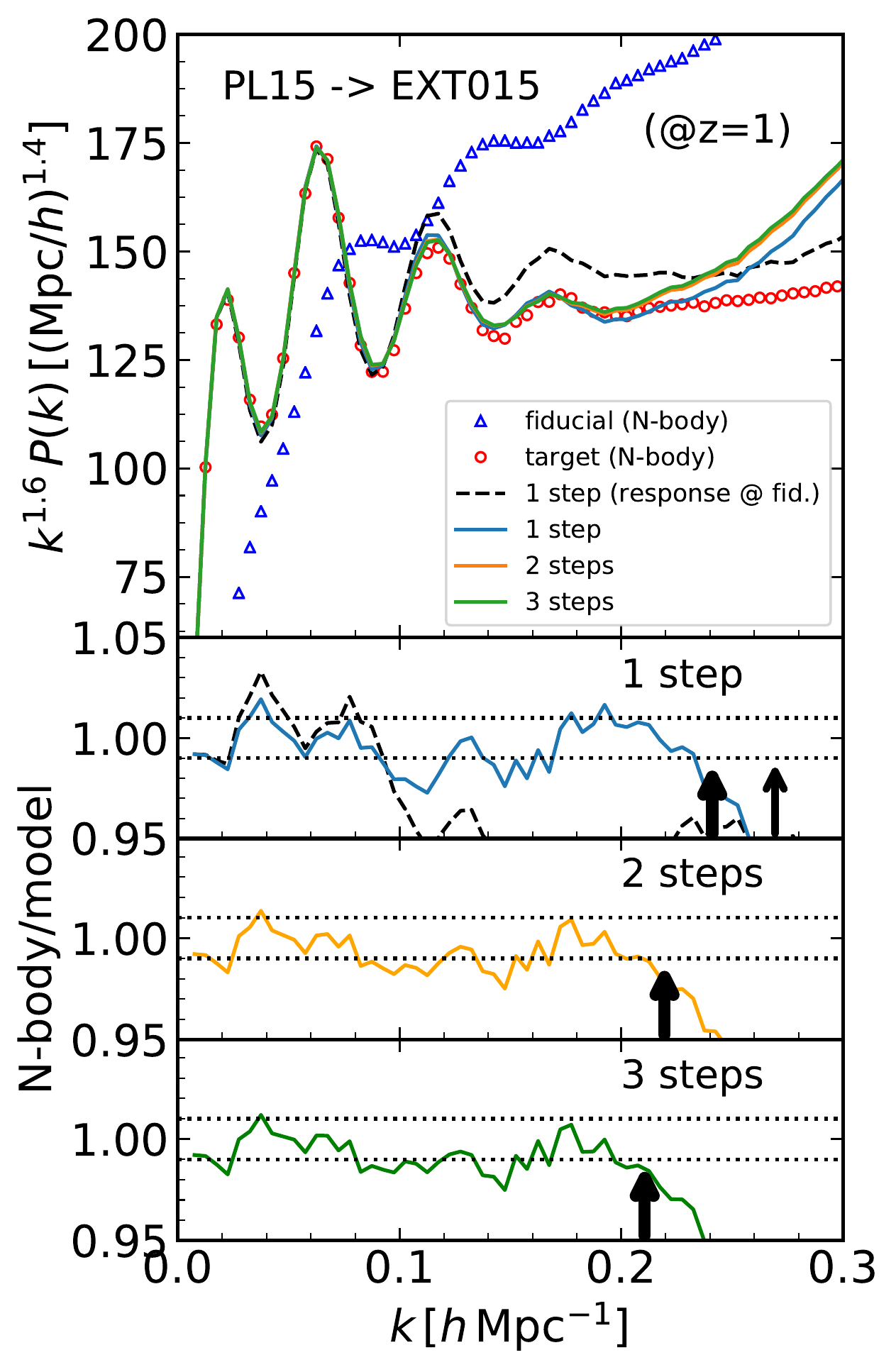}
   \caption{Same as Fig.~\ref{fig:multi_step2}, but at $z=1$. Note that the amplitude parameter of the fluctuation, $\sigma_8$, scaled to this redshift already reaches $0.71$ for this model.}
   \label{fig:multi_step2-2}
\end{figure}

We also perform the same exercise at a lower redshift, $z=1$ (see Fig.~\ref{fig:multi_step2-2}, note that the value of $\sigma_8$ scaled to this redshift is already quite high, $0.71$). The reconstruction result is almost converged with two steps as before. Now, however, the accuracy of the reconstruction is somewhat degraded: $\sim 2\%$ level up to the wavenumbers marked by the vertical arrows. Although not shown here, the reconstruction accuracy for this model gets much worse at even lower redshifts, $z=0.5$ or $0$, corresponding to $\sigma_8(z) = 0.86$ and $1.03$, respectively. 

We consider that the reason for the larger discrepancy is the following. Although our phenomenological model for the response function is calibrated to reproduce the global trend quite well down to $z=0.35$ with simulations for the \texttt{WM5} model, the validity is not tested at lower redshifts or for models with larger amplitudes where the system is in the strongly non-perturbative regime. It might work poorly in such cases. Then, the error in the response function propagates to the reconstructed power spectrum in a way that depends on the distance to the target model. The reconstruction works by definition perfectly when the target model is identical to the fiducial model no matter how large the amplitude of the fluctuations is. A large error in the response function might not affect the performance of reconstruction to a close enough target model because the correction itself is small. Indeed, the reconstruction works with $\sim 1\%$ accuracy at $z=0$ for \texttt{WM5} and \texttt{WM3}, where the $\sigma_8(z)$ is even higher, around $0.8$, as already shown in Figs.~\ref{fig:reconstruction_wmap5} and \ref{fig:reconstruction_wmap3}. After some tests, we find that the lower performance is seen when $\sigma_8(z)$ is larger than $\sim0.7$ and one needs a multi-step reconstruction. Note also that, the simulation data used as the target cosmology is noisier in such cases because of the imperfect cancellation of the cosmic-variance error with the AP method\footnote{The cosmic variance is canceled perfectly at the second order in the linear density contrast using the Angulo-Pontzen method.}. The noisy feature in the ratio of the simulated and to the reconstructed power spectra for the \texttt{EXT015} model at $z=1$ is due to this effect. The same can be seen to a lesser extent in the right bottom panel of Fig.~\ref{fig:reconstruction_wmap3} for the \texttt{WM3} model at $z=0$.

Following the results discussed so far, we wish to optimize the number of steps such that it is large enough to erase the mismatch between the reconstructed and the true spectra appeared as the oscillatory feature in the ratio, and small enough not to unnecessarily increase the computational cost and potentially narrow the reliable range expressed by $k_\mathrm{max}$. For this purpose, we use the distance previously defined in Eq.~(\ref{eq:distance_def}), now between the fiducial model $\bfp_0$ and the model closest to the target model within the plane, $\bfp_n$.
In the example above, this distance is calculated to be $d\simeq0.216$. Following this result and the other example to be presented shortly, we decide to compute the number of intermediate steps by taking the integer part of $(d/0.08)$ for simplicity. This gives 2 steps in this particular example.

\begin{figure}[!ht]
   \centering
   \includegraphics[width=7.8cm]{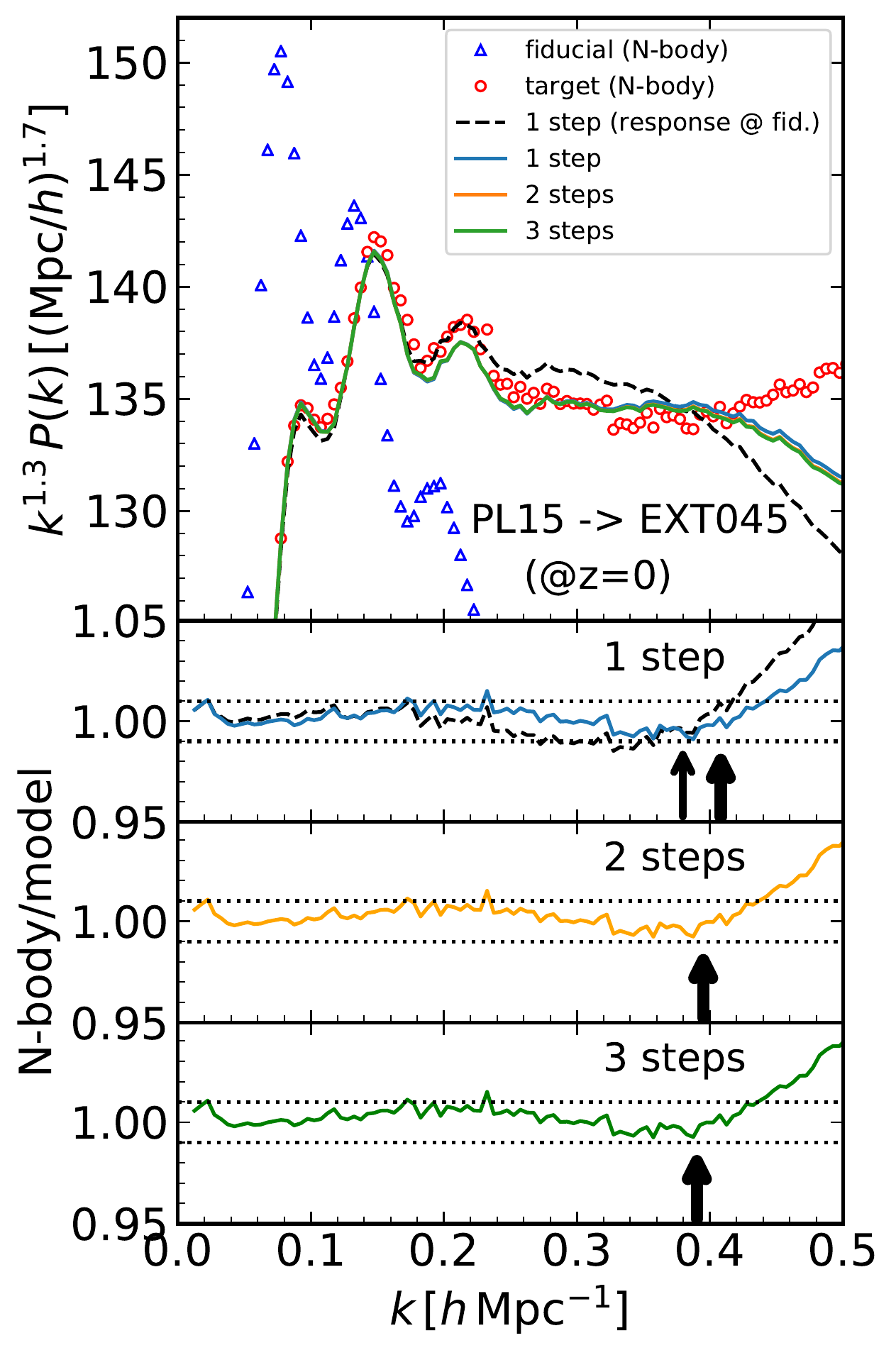}
   \caption{Same as Fig.~\ref{fig:multi_step2}, but from \texttt{PL15} to \texttt{EXT045} at $z=0$ ($\sigma_8 = 0.60$).}
   \label{fig:multi_step2-3}
\end{figure}
We show another example in Fig.~\ref{fig:multi_step2-3}, where we show a reconstruction from \texttt{PL15} to \texttt{EXT045}. Now, the distance between the fiducial and the target models defined in Eq.~(\ref{eq:distance_def}) is $0.108$, and our code chooses to take one intermediate step. This indeed gives a converged result against the number of steps, as clearly seen in the lower panels. It is promising that we can see again that the $k_\mathrm{max}$ calculated with the default value of $\alpha_\mathrm{max}=1$, depicted by the vertical arrows in the lower panels, is a good indicator of the reliable range of reconstruction with a $\sim1\%$ accuracy. In our code, users can easily modify the value of $\alpha_\mathrm{max}$ for more aggressive or conservative estimate of $k_\mathrm{max}$ depending on the required accuracy of the problem.
\section{Conclusions}
In this paper, we have investigated the response function defined as the functional derivative of the nonlinear power spectrum with respect to the linear power spectrum with detailed accuracy as a function of two wavenumbers $q$ and $k$. We take advantage of a large set of low-resolution simulations to measure the response functions with unprecedented accuracy revealing its detailed dependence with the wave modes.
Those results are then compared to the standard and renormalized analytical calculations based on perturbation theory. Important features, such as the cancellation of terms in the low-$q$ limit, suppression of the peak structure around $k\sim q$, as well as the damped high-$q$ tail are investigated in detail. Based on the result, we propose a phenomenological model that smoothly connect the behavior of the response function in different regimes.

Using the phenomenological model, the second half of the paper focuses on the reconstruction of the nonlinear power spectrum from one fiducial cosmology for which a higher-resolution simulation template is available. We show the method works well for target cosmological models near the fiducial one for the standard implementation with a single reconstruction step. Quite naturally, the reliable range of the reconstruction is roughly consistent with the scale up to which our phenomenological model successfully predicts the response function. Typically, we can reach to wavenumbers twice larger than that from the direct calculation of the spectrum based on the renormalized PT to have the prediction error controlled to lower than $\sim1\%$. Further, it is shown that employing multiple steps, we can extend the applicable range -- in terms of cosmological parameters -- of the method and reach to some extreme models such as those with $\Omega_\mathrm{m}=0.15$ or $0.45$ starting from the recent Planck cosmology. The reconstruction procedures presented here is implemented in a python code, and together with pre-computed data, the package \texttt{RESPRESSO} is publicly available.

Although we focused here the use of the response function to a regime where it can be described from expressions motivated by Perturbation Theory results -- hence limiting its range of applicability -- we believe it can be used for much wider cosmological parameters.
We note in particular that the response function is very well-behaved and smooth on scales where the perturbation-theory motivated model fails. It suggests that it should be possible to find a reasonable analytical prescription to account for such scales making possible to extend the reconstruction method proposed in this paper into such a non-perturbative regime. Another line of extension is to consider non-standard cosmological models, such as under the presence of massive neutrinos. Such a generalization might be possible as long as the sum of the neutrino masses is small such that the linear power spectrum is close to the base-line linear power spectrum for massless neutrinos and the extra history dependence of the nonlinear power spectrum is well understood. We postpone these studies for future research.

\label{sec:conclusions}

\acknowledgments
To complete this work, discussions during the workshop, YITP-T-17-03, held at Yukawa Institute for Theoretical Physics at Kyoto University were useful. TN acknowledges financial support from Japan Society for the Promotion of Science (JSPS) KAKENHI Grant Number 17K14273 and Japan Science and Technology Agency (JST) CREST Grant Number JPMJCR1414. This works is supported in part by grant ANR-12-BS05-0002 of the French Agence Nationale de la Recherche (FB), and MEXT/JSPS KAKENHI Grant Numbers JP15H05889 and JP16H03977 (AT). Numerical calculations for the present work have been carried out on Cray XC30 at Center for Computational Astrophysics, CfCA, of National Astronomical Observatory of Japan, and partly at the Yukawa Institute Computer Facility.


\bibliography{lssref}

\appendix
\section{Derivation of the analytical response functions}
\label{appendix:derivation_PT}

In this Appendix, we outline the derivation of the analytic expressions of the response function based on the perturbation theory calculations at the two-loop order in standard PT and RegPT. 

First consider the standard PT. We begin by writing down the explicit expression for the power spectrum of density field. At the two-loop order, we have 
\begin{widetext}
\begin{align}
 P^{\rm SPT}(k)&=P_{\rm lin}(k)+P^{\rm SPT}_{\rm 1\mbox{-}loop}(k)+ P^{\rm SPT}_{\rm 2\mbox{-}loop}(k)\,;
\label{eq:pk_SPT}
\\
P^{\rm SPT}_{\rm 1\mbox{-}loop}(k)&= 2\,P_{\rm lin}(k)\,\Gamma_{\rm 1\mbox{-}loop}^{(1)}(k)
+2\int\frac{d^3\bfq}{(2\pi)^3}\,\{F_{\rm sym}^{(2)}(\bfq,\bfk-\bfq)\}^2\,P_{\rm lin}(|\bfk-\bfq|)P_{\rm lin}(q),
\label{eq:pk_SPT_1loop}
\\
P^{\rm SPT}_{\rm 2\mbox{-}loop}(k)&= 
\left\{\,\left[\,\Gamma^{(1)}_{1\mbox{-}{\rm loop}}(k)\,\right]^2 
+2\,\Gamma^{(1)}_{2\mbox{-}{\rm loop}}(k)\,\right\}\,P_{\rm lin}(k)
+6\,\int\frac{d^3\bfp d^3\bfq}{(2\pi)^6}
\left\{F_{\rm sym}^{(3)}(\bfp,\bfq,\bfk-\bfp-\bfq)\right\}^2\nonumber\\ 
&\qquad \times P_{\rm lin}(|\bfk-\bfp-\bfq|)P_{\rm lin}(p)P_{\rm lin}(q)
+4\int\frac{d^3\bfq}{(2\pi)^3}\,F_{\rm sym}^{(2)}(\bfq,\bfk-\bfq)
\Gamma^{(2)}_{1\mbox{-}{\rm loop}}(\bfq,\bfk-\bfq)
\,P_{\rm lin}(|\bfk-\bfq|)P_{\rm lin}(q),
\label{eq:pk_SPT_2loop}
\end{align}
\end{widetext}
Here, to simplify the expression, we used the function $\Gamma^{(n)}_{p\mbox{-}{\rm loop}}$, which is the $(n+1)$-point propagator at $p$-loop order computed with standard PT [see Eq.~(\ref{eq:def_Gamma^n_p-loop})]. 

With the expressions given above, we consider a small variation of linear power spectrum around the fiducial model spectrum $P_{\rm lin,{\rm fid}}$:
\begin{align}
P_{\rm lin}(k)=P_{\rm lin, {\rm fid}}(k) + \delta\,P_{\rm lin}(k).
\label{eq:var_pklin}
\end{align}
Assuming $\delta P_{\rm lin}\ll P_{\rm lin,{\rm fid}}$, we expand 
the standard PT power spectrum around the fiducial power spectrum 
$P_{\rm lin,{\rm fid}}$. We then have
\begin{align}
P^{\rm SPT}(k)  \simeq  P^{\rm SPT}_{\rm fid}(k)+ \delta P^{\rm SPT}(k), 
\end{align}
where the term $\delta P^{\rm SPT}(k)$ summarizes the first-order variation of the power spectrum, $\delta P_{\rm lin,{\rm fid}}$. Using the symmetric property of the kernel over its arguments, we obtain
\begin{widetext}
\begin{align}
 \delta P^{\rm SPT}(k)&=\delta P^{\rm SPT}_{\rm tree}(k)+ \delta P^{\rm SPT}_{\rm1\mbox{-}loop }(k) + \delta P^{\rm SPT}_{\rm1\mbox{-}2oop }(k);
\nonumber
\\
\delta P^{\rm SPT}_{\rm tree}(k) &= \delta P_{\rm lin}(k),
\label{eq:delpkSPT_tree}
\\
\delta P^{\rm SPT}_{\rm 1\mbox{-}loop}(k) &= 2 \left\{\delta P_{\rm lin}(k) \,\Gamma^{(1)}_{\rm 1\mbox{-}loop}(k) + P_{\rm lin}(k)\,\delta\Gamma^{(1)}_{\rm 1\mbox{-}loop}(k) \right\}
 + 4 \int\frac{d^3\bfq}{(2\pi)^3}\,\{F_{\rm sym}^{(2)}(\bfq,\bfk-\bfq)\}^2\,P_{\rm lin}(|\bfk-\bfq|)\,\delta P_{\rm lin}(q),
\label{eq:delpkSPT_1loop}
\\
\delta P^{\rm SPT}_{\rm 2\mbox{-}loop}(k) &= 
\left\{\,\left[\,\Gamma^{(1)}_{1\mbox{-}{\rm loop}}(k)\,\right]^2 
+2\,\Gamma^{(1)}_{2\mbox{-}{\rm loop}}(k)\,\right\}\,\delta P_{\rm lin}(k)
+\left\{\,2\,\Gamma^{(1)}_{1\mbox{-}{\rm loop}}(k)\,\delta\Gamma^{(1)}_{1\mbox{-}{\rm loop}}(k)
+2\,\delta\Gamma^{(1)}_{2\mbox{-}{\rm loop}}(k)\,\right\}\,P_{\rm lin}(k)
\nonumber\\
&\qquad\qquad 
+18\,\int\frac{d^3\bfp d^3\bfq}{(2\pi)^6}
\left\{F_{\rm sym}^{(3)}(\bfp,\bfq,\bfk-\bfp-\bfq)\right\}^2 
P_{\rm lin}(|\bfk-\bfp-\bfq|)P_{\rm lin}(p)\,\delta P_{\rm lin}(q)
\nonumber\\
&\qquad\qquad+4\int\frac{d^3\bfq}{(2\pi)^3}\,F_{\rm sym}^{(2)}(\bfq,\bfk-\bfq)
\Gamma^{(2)}_{1\mbox{-}{\rm loop}}(\bfq,\bfk-\bfq)
\,P_{\rm lin}(|\bfk-\bfq|)\,\delta P_{\rm lin}(q),
\nonumber\\
&\qquad\qquad+4\int\frac{d^3\bfq}{(2\pi)^3}\,F_{\rm sym}^{(2)}(\bfq,\bfk-\bfq)\,
\delta\Gamma^{(2)}_{1\mbox{-}{\rm loop}}(\bfq,\bfk-\bfq)
\,P_{\rm lin}(|\bfk-\bfq|)\,P_{\rm lin}(q).
\label{eq:delpkSPT_2loop}
\end{align}
\end{widetext}
In the above, we have shortly denoted $P_{\rm lin,fid}$ by $P_{\rm lin}$. The quantity, $\delta \Gamma^{(n)}_{p\mbox{-}{\rm loop}}$, is the first-order variation of the propagator, and the resultant expression is given by
\begin{widetext}
\begin{align}
\delta \Gamma^{(n)}_{p\mbox{-}{\rm loop}}(\bfk_1,\cdots,\bfk_n)&=p\,c^{(n)}_p
\int\frac{d^3\bfq_1\cdots d^3\bfq_p}{(2\pi)^{3p}}\,F_{\rm sym}^{(n+2p)}(\bfq_1,-\bfp_1,\cdots,\bfq_p,-\bfq_p,\bfk_1,\cdots,\bfk_n)\,
P_{\rm lin}(q_1)\cdots P_{\rm lin}(q_{p-1})\,\delta P_{\rm lin}(q_p).
\label{eq:delta_Gamma^n_p-loop}
\end{align}
\end{widetext}
Substituing Eq.~(\ref{eq:delta_Gamma^n_p-loop}) into the above, we obtain the following integral form: 
\begin{align}
\delta P^{\rm SPT}(k)=&\int d\ln q\left\{K_{\rm tree}(k,q)+K_{\rm 1\mbox{-}loop}(k,q)\right.\nonumber\\
&\qquad\left.+K_{\rm 2\mbox{-}loop}(k,q)\right\} \delta P_{\rm lin}(q),   
\end{align}
which finally gives Eqs.~(\ref{eq:K_SPT_tree}), (\ref{eq:K_SPT_1loop}), and (\ref{eq:K_SPT_2loop}).

Next consider the response function in RegPT.  Again, we write down the expression of the power spectrum. Using the regularized propagators given at Eqs.~(\ref{eq:gamma1_reg})--(\ref{eq:gamma3_reg}), the expression relevant at two-loop order is 
\begin{widetext}
\begin{align}
&P^{\rm RegPT}(k)= P^{\rm RegPT}_{\rm tree}(k) + P^{\rm RegPT}_{\rm 1\mbox{-}loop}(k) + P^{\rm RegPT}_{\rm 2\mbox{-}loop}(k)\,;
\nonumber
\\
&
P^{\rm RegPT}_{\rm tree}(k) = [\Gamma^{(1)}_{\rm reg}(k)]^2\,P_{\rm lin}(k),
\label{eq:pkRegPT_tree}
\\
&
P^{\rm RegPT}_{\rm 1\mbox{-}loop}(k) =
2\,\int\frac{d^3\bfq}{(2\pi)^3}\,\{\Gamma^{(2)}_{\rm reg}
(\bfq,\bfk-\bfq)\}^2\,P_{\rm lin}(|\bfk-\bfq|) P_{\rm lin}(q),
\label{eq:pkRegPT_1loop}
\\
&
P^{\rm RegPT}_{\rm 2\mbox{-}loop}(k) =
6\,\int\frac{d^3\bfp d^3\bfq}{(2\pi)^6}\,\{\Gamma^{(3)}_{\rm reg}
(\bfp,\bfq,\bfk-\bfp-\bfq)\}^2\,P_{\rm lin}(|\bfk-\bfp-\bfq|)P_{\rm lin}(p)P_{\rm lin}(q).
\label{eq:pkRegPT_2loop}
\end{align}
\end{widetext}
Applying similarly the small variation of linear power spectrum given at Eq.~(\ref{eq:var_pklin}) to the above,  one obtains
\begin{align}
P^{\rm RegPT}(k)\simeq P^{\rm RegPT}_{\rm fid}(k)+ \delta P^{\rm RegPT}(k)
\end{align}
with the variation $\delta P^{\rm RegPT}(k)$ given by
\begin{widetext}
\begin{align}
\delta P^{\rm RegPT}(k) =& \delta P^{\rm RegPT}_{\rm tree}(k) + 
\delta P^{\rm RegPT}_{\rm 1\mbox{-}loop}(k) + \delta P^{\rm RegPT}_{\rm 2\mbox{-}loop}(k)\,;
\nonumber
\\
\delta P^{\rm RegPT}_{\rm tree}(k) =& 2\Bigl\{ 
[\Gamma^{(1)}_{\rm reg}(k)]^2\,\delta P_{\rm lin}(k)
+ 2\Gamma^{(1)}_{\rm reg}(k)\,\delta \Gamma^{(1)}_{\rm reg}(k)\,P_{\rm lin}(k) 
\Bigr\},
\nonumber
\\
\delta P^{\rm RegPT}_{\rm 1\mbox{-}loop}(k) =& 
4\,\int\frac{d^3\bfq}{(2\pi)^3}\,\{\Gamma^{(2)}_{\rm reg}
(\bfq,\bfk-\bfq)\}^2\, 
\delta P_{\rm lin}(|\bfk-\bfq|)\,P_{\rm lin}(q)
\nonumber\\
&+
4\,\int\frac{d^3\bfq}{(2\pi)^3}\,\Gamma^{(2)}_{\rm reg}
(\bfq,\bfk-\bfq)\,\delta \Gamma^{(2)}_{\rm reg}(\bfq,\bfk-\bfq)\, P_{\rm lin}(|\bfk-\bfq|)P_{\rm lin}(q),
\nonumber
\\
\delta P^{\rm RegPT}_{\rm 2\mbox{-}loop}(k) =&
18\,\int\frac{d^3\bfp d^3\bfq}{(2\pi)^6}\,\{\Gamma^{(3)}_{\rm reg}
(\bfp,\bfq,\bfk-\bfp-\bfq)\}^2\,
P_{\rm lin}(|\bfk-\bfp-\bfq|)P_{\rm lin}(p)\,\delta P_{\rm lin}(q)
\nonumber
\\
& + 
12\,\int\frac{d^3\bfp d^3\bfq}{(2\pi)^6}\,\Gamma^{(3)}_{\rm reg}
(\bfp,\bfq,\bfk-\bfp-\bfq)\,\delta \Gamma^{(3)}_{\rm reg}
(\bfp,\bfq,\bfk-\bfp-\bfq)\,P_{\rm lin}(|\bfk-\bfp-\bfq|)P_{\rm lin}(p)P_{\rm lin}(q).
\nonumber
\end{align}
\end{widetext}
In the above, the variation of the regularized propagators, $\delta \Gamma^{(n)}_{\rm reg}$, must be evaluated taking account of the fact that the factor $\alpha_k$ [see Eq.~(\ref{eq:def_alpha}) for definition] also depends on the linear power spectrum. Then, we have
\begin{align}
& \delta\Gamma^{(1)}_{\rm reg}(k) = 
\Bigl\{ \delta\Gamma^{(1)}_{1\mbox{-}{\rm loop}}(k)(1+\alpha_k) + 
\delta\Gamma^{(1)}_{2\mbox{-}{\rm loop}}(k) \Bigr\}\,e^{-\alpha_k}
\nonumber\\
&\qquad +\frac{k^2}{2}\Bigl\{\left(1+\Gamma^{(1)}_{1\mbox{-}{\rm loop}}(k)+
\alpha_k\right)\,e^{-\alpha_k}-\Gamma^{(1)}_{\rm reg}(k)\,\Bigr\}\,
\delta\sigma_{\rm d}^2
\nonumber
\\
&\delta\Gamma^{(2)}_{\rm reg}(\bfq,\bfk-\bfq) = 
\delta\Gamma^{(2)}_{1\mbox{-}{\rm loop}}(\bfq,\bfk-\bfq)\,e^{-\alpha_k}
\nonumber\\
&\qquad 
+\frac{k^2}{2}\left\{
F^{(2)}_{\rm sym}(\bfq,\bfk-\bfq)\,e^{-\alpha_k}-\Gamma^{(2)}_{\rm reg}(\bfq,\bfk-\bfq)
\right\} \delta\sigma_{\rm d}^2,
\nonumber\\
&\delta\Gamma^{(3)}_{\rm reg}(\bfp,\bfq,\bfk-\bfp-\bfq) = -\frac{k^2}{2}\Gamma^{(3)}_{\rm reg}(\bfp,\bfq,\bfk-\bfp-\bfq)\,\delta\sigma_{\rm d}^2,
\end{align}
with the variation $\sigma_{\rm d}$ given by
\begin{align}
 \delta \sigma_{\rm d}^2=\int \frac{dq}{6\pi^2}\,\delta P_{\rm lin}(q). 
\end{align}
Note here that we have omitted the subscript $_{\rm fid}$, and the quantities $\alpha_k$, $\Gamma_{\rm reg}^{(n)}$ and $\Gamma^{(n)}_{\rm m\mbox{-}loop}$ are evaluated with the fiducial power spectrum. Finally, substituting these expressions into the variation of power spectrum, after some manupilations, we obtain the following integral form: 
\begin{align}
 \delta P^{\rm RegPT}(k) =& \int d\ln q 
\left[
K^{\rm RegPT}_{\rm tree}(k, q) + 
K^{\rm RegPT}_{\rm 1\mbox{-}loop}(k, q) \right.
\nonumber\\
&\left.+ \qquad
K^{\rm RegPT}_{\rm 2\mbox{-}loop}(k, q) 
\right]\,\delta P_{\rm lin}(q) 
\end{align}
which finally leads to the expressions at Eqs.~(\ref{eq:K_RegPT_tree}), (\ref{eq:K_RegPT_1loop}), and (\ref{eq:K_RegPT_2loop}).

\section{Kernel functions}
\label{appendix:kernel_func}

Here, we summarize the explicit expressions for all the functions given in Eqs.~(\ref{eq:K_SPT}) and (\ref{eq:K_RegPT}).
\begin{widetext}
\begin{align}
&L^{(1)}(q,k)=3\int\frac{d^2\bfOmg_q}{4\pi}\,F^{(3)}_{\rm sym}(\bfq,-\bfq,\bfk),
\label{eq:L1}\\
&M^{(1)}(q,k)=15\int\frac{d^3\bfp\,d^2\bfOmg_q}{4\pi(2\pi)^3} 
F^{(5)}_{\rm sym}(\bfp,-\bfp,\bfq,-\bfq,\bfk)\,P_{\rm lin}(p) ,
\label{eq:M1}\\
&X^{(2)}(q,k)=\frac{1}{2}\int_{-1}^1d\mu_q\,
\left\{F^{(2)}_{\rm sym}(\bfq,\bfk-\bfq)\right\}^2\,
P_{\rm lin}(\sqrt{k^2-2kq\,\mu_q+q^2}),
\label{eq:X2}\\
&Y^{(2)}(q,k)=\frac{1}{2}\int_{-1}^1d\mu_q\,
F^{(2)}_{\rm sym}(\bfq,\bfk-\bfq)\,\Gamma^{(2)}_{\rm 1\mbox{-}loop}
(\bfq,\bfk-\bfq)\,
P_{\rm lin}(\sqrt{k^2-2kq\,\mu_q+q^2}),
\label{eq:Y2}\\
&Z^{(2)}(q,k)=\frac{1}{2}\int_{-1}^1d\mu_q\,
\left\{\Gamma^{(2)}_{\rm 1\mbox{-}loop}
(\bfq,\bfk-\bfq)\right\}^2\,
P_{\rm lin}(\sqrt{k^2-2kq\,\mu_q+q^2}),
\label{eq:Z2}\\
&Q^{(2)}(p,k)=\int\frac{d^3\bfq}{(2\pi)^3}
F^{(2)}_{\rm sym}(\bfq,\bfk-\bfq)\,K(p:q,|\bfk-\bfq|,k)\,
P_{\rm lin}(|\bfk-\bfq|)\,P_{\rm lin}(q),
\label{eq:Q2}\\
&R^{(2)}(p,k)=\int\frac{d^3\bfq}{(2\pi)^3}
\Gamma^{(2)}_{\rm 1\mbox{-}loop}(\bfq,\bfk-\bfq)\,
K(p;q,|\bfk-\bfq|,k)\,
P_{\rm lin}(|\bfk-\bfq|)\,P_{\rm lin}(q),
\label{eq:R2}\\
&S^{(3)}(q,k)=\frac{1}{2}\int_{-1}^1 d\mu_q\,\int\frac{d^3\bfp}{(2\pi)^3}
\left\{F^{(3)}_{\rm sym}(\bfp,\bfq,\bfk-\bfp-\bfq)\right\}^2\,
P_{\rm lin}(|\bfk-\bfp-\bfq|)\,P_{\rm lin}(p)
\label{eq:S3}
\end{align}
\end{widetext}
with the variable $\mu_q$ being the directional cosine between $\bfk$ and $\bfq$, i.e., $\mu_q=\cos(\hat{\bfk}\cdot\hat{\bfq})$. Here, the function $K$ is defined by
\begin{align}
&K(q;k_1,k_2,k_3) = 6\int\frac{d^2\bfOmg_q}{4\pi}\,F^{(4)}_{\rm sym}(\bfq,-\bfq,\bfk_1,\bfk_2); 
\nonumber\\
&\qquad\qquad\qquad\qquad\qquad \bfk_1+\bfk_2=\bfk_3.
\end{align}
Also, the expressions for the power spectra, 
$P^{(2)\rm tree\mbox{-}tree}_{\rm corr}$ and 
$P^{(2)\rm tree\mbox{-}1loop}_{\rm corr}$, which appears at Eq.~(\ref{eq:K_RegPT_1loop}), are summarized below:
\begin{widetext}
\begin{align}
P^{(2)\rm tree\mbox{-}tree}_{\rm corr}(k) &= 
2\int\frac{d^3\bfq}{(2\pi)^3}F_{\rm sym}^{(2)}(\bfq,\bfk-\bfq)
F_{\rm sym}^{(2)}(\bfq,\bfk-\bfq)
\, P_{\rm lin}(q)P_{\rm lin}(|\bfk-\bfq|),
\label{eq:P2_tree-tree}
\\
P^{(2)\rm tree\mbox{-}1loop}_{\rm corr}(k) &= 
4\int\frac{d^3\bfq}{(2\pi)^3} 
F_{\rm sym}^{(2)}(\bfq,\bfk-\bfq)
\Gamma_{\rm 1\mbox{-}loop}^{(2)}(\bfq,\bfk-\bfq)
\, P_{\rm lin}(q)P_{\rm lin}(|\bfk-\bfq|).
\label{eq:P2_tree-1loop}
\end{align}
\end{widetext}

\end{document}